\documentclass{aa}

\usepackage{tabularx}
\newcolumntype{Y}{>{\centering\arraybackslash}X}
\usepackage{comment}
\usepackage{graphicx}
\usepackage{adjustbox}
\usepackage{colortbl}
\usepackage{txfonts}
\usepackage{subcaption}
\usepackage[format=plain,font=small,labelfont=bf]{caption}
\usepackage{xcolor}
\usepackage{float} 
\usepackage{hyperref}
\hypersetup{
    linkbordercolor=red,citebordercolor=blue}

\newcommand{\prodimo}{P{\tiny RO}D{\tiny I}M{\tiny O}\;}

\begin{document}

   \title{{The effect of metallicity on the abundances of molecules in protoplanetary disks}}

   \subtitle{}

   \author{R. Guadarrama
          \inst{1},
          Eduard I. Vorobyov\inst{1,2}, Christian Rab\inst{3,4,5}, Manuel Güdel\inst{1}
          }

   \institute{Department of Astrophysics, University of Vienna,
              T\"urkenschanzstrasse 17, A-1180 Vienna, Austria \\
             \email{rodrigo.guadarrama@univie.ac.at}
             \and
Research Institute of Physics, Southern Federal University, Rostov-on-Don, 344090 Russia
\and Max-Planck-Institut für extraterrestrische Physik, Giessenbachstrasse 1, 85748 Garching, Germany
\and Kapteyn Astronomical Institute, University of Groningen, PO Box 800, 9700 AV Groningen, The Netherlands
\and University Observatory, Faculty of Physics, Ludwig-Maximilians-Universität München, Scheinerstr. 1, 81679 Munich,
Germany
             }


 
  \abstract
   {Diverse studies have shown that it is important to consider the impact of metallicity on the chemodynamical evolution of protoplanetary disks. {It has been suggested that there are different chemistry cycles in non-solar metallicity environments or that there exists a dependence of the efficiency of mass transport in protostars and pre-main-sequence stars on the metallicity.}}   
   {{We study the influence of different metallicities on the physical, thermal, and chemical properties of protoplanetary disks, and in particular on the formation and destruction of carbon-based molecules.}}
   {With the thermo-chemical code \prodimo we investigate the impact of lower metallicities on the radiation field, disk temperature, and the abundance of different molecules (H$_2$O, CH$_4$, CO, CO$_2$, HCN, CN, HCO$^+$ and N$_2$H$^+$). We use a fiducial disk model as a reference model and produce two models with lower metallicity.  
   The resulting influence on different chemical species is studied by analyzing their abundance distribution throughout the disk and their vertical column density. Furthermore, the formation and destruction reactions of the chemical species are studied.} 
   {The results show a relation between the metallicity of the disk and the strength of the stellar radiation field inside
   the disk. As the metallicity decreases the radiation field is able to penetrate deeper regions of the disk. As a result, there is a stronger radiation field overall in the disk with lower metallicity which also heats up the disk.
   This triggers a series of changes in the chemical formation and destruction efficiencies for different chemical species.
   In most cases, the available species abundances change and have greater values compared to scaled-down abundances by constant factors. Metallicity has a clear impact on the snowline of the molecules studied here as well. As metallicity decreases the snowlines are pushed further out and existing snow rings shrink in size.} 
   {We find that the abundances of the studied molecules in lower metallicity disks cannot be understood or reproduced by scaling down the respective species abundances of the reference disk model. This is because the chemical reactions responsible for the destruction and formation of the studied molecules change as the metallicity of the disk is reduced. We found a strong overabundance (relative to scaled-down values) in the models with lower metallicity for gaseous species (CN, CO, HCO$^+$, N$_2$H$^+$) which are particularly useful in observations. This could be advantageous for future observations in low metallicity environments.  
   Further studies considering different aspects of the disk are needed to gain a deeper understanding of the relation between metallicity and disk thermochemical evolution. Future studies should consider different dust grain size distribution, different stellar radiation fields, and stellar burst scenarios among other processes.}

   \authorrunning{Guadarrama et al.}
\titlerunning{Water and carbon-based molecules in sub-solar metallicity disks}
   \maketitle
%

\section{Introduction}

Protoplanetary disks form during the gravitational collapse of rotating cloud cores and consist of molecular hydrogen, dust grains of various sizes, and chemical species of multiple complexities. While molecular hydrogen constitutes the main mass reservoir of the disk, dust plays an important role in determining the disk's thermal balance and supplying the planet formation process with building material. Chemical species, in particular water and carbon-based molecules, are the main prerequisites of life and also play a key role in the process of dust growth \citep[e.g.][]{2016ApJ...821...82O}. According to \cite{2013Sci...341..630Q}, the fronts where volatile species condense (snowlines) play an important role in the planet formation process. They define the location where dust coagulation is favoured because of an increase in the stickiness of icy grains.

The position of the snowlines for the most abundant volatiles may also have an influence on the elemental and molecular composition of planetesimals and planets. The position of the respective snowlines for different chemical species determine at what radii which molecules freeze onto dust grains. This influences the chemical composition of icy mantels of dust grains at specific regions in the disk. As the dust grains are the seeds of planet formation, the chemical composition of their icy mantels also influences the chemical composition of forming planetesimals and subsequently, planets. In the case of gas giants, the composition of the atmosphere is also linked to the position of the different snowlines, and therefore to their location on the disk. For instance, the carbon to oxygen ratio, a defining characteristic of the atmosphere, is set by the amount of gas phase oxygen and carbon a gas giant accretes from the disk. The accreted amount of oxygen and carbon-bearing molecules in the gas phase is dependent on the location on the disk where they freeze out (snowline).  
Thus, planet formation efficiencies, their core-, and atmosphere compositions are strongly linked to the snowlines of volatile species. 

Until recently, most numerical studies of protoplanetary disks were limited to the solar metallicity case. However, the detection of a growing number of exoplanets has revealed that the frequency of the giant planet occurrence depends on the metallicity of the host star \citep{2005A&A...437.1127S,2014prpl.conf..691B}. In addition, studies of protostars and pre-main-sequence stars in the Magellanic Clouds indicate that mass accretion rates are inversely proportional to metallicity \citep{2013ApJ...775...68D}. This finding seems to be corroborated by \citet{2010ApJ...723L.113Y,2016AJ....151...50Y} who inferred shorter lifetimes to low-metallicity disks in the outer regions of the Galaxy. However, this trend could also be explained by higher rates of disk photoevaporation due to a more diluted medium that allows a stronger radiation field \citep{2018ApJ...857...57N}. Another example of these studies is the work performed by \cite{2016ApJ...827..126B}. By observing the chemical abundances of HII regions in 12 nearby dwarf galaxies, the study suggests that there are different chemistry cycles in non-solar metallicity environments. 
This evidence suggests that disk evolution and planet formation in protoplanetary disks may proceed differently in low metallicity environments. 

The chemical evolution and composition of protoplanetary disks have been extensively studied in the context of solar-metallicity disks \citep[see a review by][]{2014prpl.conf..363P,2013ChRv..113.9016H}.
Among various compounds that are found in protoplanetary disks, volatile species with low sublimation temperatures are best studied because they are central to the process of planet formation. 
In this work, we use the radiation thermo-chemical code \prodimo \citep{2009A&A...501..383W} to study the impact of low metallicity on important molecules (e.g. H$_2$O, CO) on protoplanetary disks.

When we refer to metallicity in this study, we mean the values of element abundances and the gas-to-dust ratio. To simulate a lower metallicity we reduce the element abundance of metals together with the amount of dust. Then we analyze the resulting temperature structure, radiation field, and chemical abundances of various molecules in different metallicities. The formation and destruction reactions of chemical species are also analyzed. Except for ionized molecules, we track the position of the respective snowlines. For a deeper analysis of the chemical abundances, we perform a comparison of the vertical column density between the models with lower metallicity and an artificially scaled-down reference model. Although it is expected that the behaviour of the column densities as a function of the metallicity may not be linear, we use the comparison with scaled-down models to be able to define the regions of the disk where the discrepancy is largest and to quantify the effect of non-linearity. We used an evolved, externally irradiated Class II disk for our models. 
 
In Sect.\ref{method} we describe the gas and dust structure of the disk in the models as well as the chemical network and element abundances used for this work. We also explain how we applied the modifications in the disk models to simulate lower metallicities in the disks. Our results are presented in Sect.\ref{Results}. In Sect.\ref{Limitations of model} we discuss some aspects of the used model and the results and in Sect.\ref{Summary and conclusions}  we present our conclusions and describe the following steps for this work.

\section{Method}\label{method}

We employed the radiation thermo-chemical code \prodimo \citep{2009A&A...501..383W,2010A&A...510A..18K,2011MNRAS.412..711T} to model the chemical composition of young protoplanetary disks with distinct metallicities. \prodimo includes the radiative transfer of stellar and background radiation along with gas and dust thermal balance to provide the temperature structure and the local radiation field of the disk. The chemical abundances are then calculated using steady-state chemistry employing a chemical reaction network used in \cite{2017A&A...607A..41K}. Various heating and cooling processes are also considered to solve consistently for the gas temperature and the chemistry. The implementation of X-rays and X-ray chemistry is according to \cite{2011A&A...526A.163A}.

\subsection{{Metallicities and initial abundances}}\label{lower metallicity}

\begin{table}[]
    \centering
     \caption{Elements included in this study, and their abundances in model Z1 on the scale log$n_{\mathrm{H}}$ =12 with $\epsilon$ as the abundances relative to H$_2$ (middle column) and their masses in amu (right column). The values are taken from \cite{2016A&A...586A.103W}}.
    \begin{tabular}{c|c|c}
    \hline
    \hline
    element & 12+log $\epsilon $ & $m$ [amu]  \\
    
    \hline
    H &  12.00 &  1.0079 \\
He  & 10.984  & 4.0026   \\
C   & 8.14    & 12.011  \\
N   & 7.90    & 14.007  \\
O   & 8.48    & 15.999  \\
Ne  & 7.95    & 20.180  \\
Na  & 3.36    & 22.990  \\
Mg  & 4.03    & 24.305  \\
Si  & 4.24    & 28.086  \\
S   & 5.27    & 32.066  \\
Ar  & 6.08    & 39.948  \\
Fe  & 3.24    & 55.845   \\
\hline
    \end{tabular}
    \label{table:el_abun}
\end{table}

We use different element abundances and dust-to-gas ratios to study the impact of lower metallicities on the chemical composition of the disk, focusing on water and several most abundant carbon-bearing molecules. We considered three model realizations of protoplanetary disks with three distinct metallicities $Z$.
In all cases the mass of the central star, stellar effective temperature, stellar luminosity and X-ray luminosity are fixed and set equal to $M_\ast=0.7~M_\odot$, $T_\ast=4000$~K, $L_\ast=1.0~L_\odot$ and $L_X=2.6\times10^{-4}~L_\odot$.
We assume a solar metallicity value of Z$_{\odot} = 0.02$ which is in agreement with the value of Z$_{\odot} = 0.0196 \pm 0.0014$ from \cite{2016ApJ...816...13V}. The abundances for our reference model are the abundances used for standard TTauri disk in the \prodimo code \citep{2016A&A...586A.103W}. {As a result, our reference model Z1 has a metallicity value of Z1$= 0.017$ and has, therefore, a value of Z1$=0.85\, Z_{\odot}$.} The element abundances corresponding to model Z1 are shown in Table \ref{table:el_abun}. The models are named according to their disk metallicity. Model Z01 and Z001 correspond to a metallicity of 0.1 and 0.01 (10\% and 1\% of the reference metallicity Z1 respectively). Thus, the metallicities of our models lie in the $0.85\times10^{-2}- 0.85\ Z_{\odot}$ limits. In order to produce models with different metallicities, we performed two modifications to the initial setup of the models:

\begin{table}[!h]
\caption{Variations of the  model used in this study. The columns represent the disk gas mass, metallicity, and dust-to-gas ratio.}
    \label{table:metallicities}
    \centering
    \begin{tabular}{l |c|c|c}
    \hline
    \hline
    Model     &  $M_{\rm disk}$ [$M_{\sun}$] & Z [$Z_{\sun}$] & dust-to-gas ratio\\
    \hline
   
    Z1  &    $3.0\times10^{-2}$ &   $8.5\times10^{-1}$ & $1.0\times10^{-2}$ \\
    Z01 &    $3.0\times10^{-2}$ &   $8.5\times10^{-2}$ & $1.0\times10^{-3}$\\
    Z001&    $3.0\times10^{-2}$ &   $8.5\times10^{-3}$ & $1.0\times10^{-4}$\\
    
    \hline
    \end{tabular}
\end{table}
\begin{enumerate}
  \item {We reduced the initial abundances of 13 of the 15 elements (all except H and He) by a factor of 10 and 100 for the gas and the dust.}
  
  \item We reduced the dust-to-gas mass ratio in the input parameters by the same factors (models Z01 and Z001, respectively).
\end{enumerate}

This produced the above-mentioned three models (from herein referred to as Z1, Z01, and Z001) with the reference, 1/10 and 1/100 of reference metallicity, respectively. The disk mass, the metallicity relative to the solar value, and the dust-to-gas ratio are displayed in Table \ref{table:metallicities}. 

Further variations of disk and star properties with metallicity will be considered in follow-up studies. The grid resolution is 150 points in the horizontal direction and 100 in the vertical direction. The details of the initial disk's physical structure, dust content, and chemical composition are provided below.

\subsection{The disk model}\label{disk model}

The disk model we use for this study assumes certain simplifications in its physical properties such as the star, the disk geometry, dust settling, and others. The set of assumptions is taken to reproduce the most commonly observed multi-wavelength properties of Class II protoplanetary disks. The basic structural properties of the model are described in detail in ~\cite{2016A&A...586A.103W}. The chemical network used in this study is introduced and explained in \cite{2017A&A...607A..41K}. In section \ref{ssec:gas disk structure}, \ref{ssec:dust disk structure}, and \ref{ssec:chemical model}we provide a brief overview of this model.  

\subsubsection{Gas disk structure}\label{ssec:gas disk structure}

\begin{figure}[!h]
    \centering
    \includegraphics[width=0.48\textwidth]{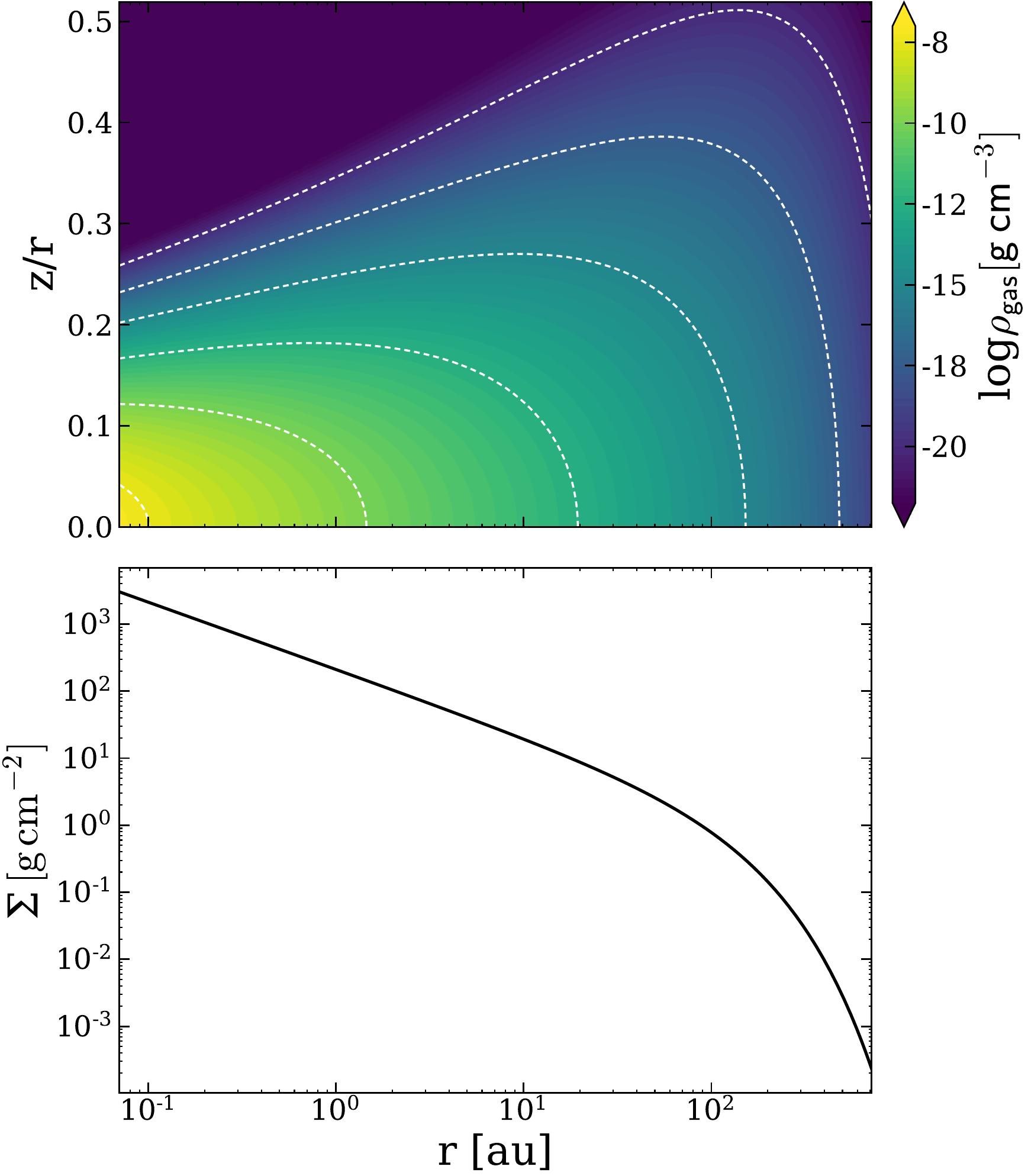}
    \caption{Gas density structure (top) and radial surface density profile (bottom) of the reference model Z1.}
    \label{fig:dens_struc_fiducial}
\end{figure}

We use a fixed parameterized density structure for the disk which is based on viscous evolution models. For further details see~\cite{2016A&A...586A.103W}.
The gas density structure is an axisymmetric flared (2D) function of the radius and the vertical height of the disk ($r$ and $h$, respectively) and it is given by 

\begin{equation}\label{eq:dens}
\rho(r,z) = \frac{\Sigma(r)}{\sqrt{2\pi}\cdot h(r)}\exp{\left (  -\frac{z^{2}}{2h(r)^{2}}\right )}\quad   [\text{g cm}^{-3}].
\end{equation} 
Here, $\Sigma(r)$ is the radial surface density profile of the disk. We assume a simple power-law distribution with a tapered outer edge

\begin{equation}\label{eq:sigma}
\Sigma(r) = \Sigma_{0}\left (\frac{r}{R_{\rm in}}\right )^{-\pmb{\lambda}}\exp{\left (-\left (\frac{r}{R_{\rm tap}}\right )^{2-\pmb{\lambda}}\right )}\quad  [\text{g cm}^{-2}].
\end{equation}
Here, $R_{\rm in}$ is the inner disk radius and $R_{\rm tap}$ is the characteristic radius. The constant $\Sigma_0$ is determined from the specified disk mass $M_{\mathrm{disk}} = 2\pi\int \Sigma (r)r\,dr$ and Eq.\ref{eq:sigma}. The vertical scale height $h(r)$ is given by a radial power law

\begin{equation}\label{eq:scaleh}
h(r) = H_{100}\left (\frac{r}{100\ \mathrm{au}}\right )^{\beta},
\end{equation}
where $H_{100}$ 
is the disk scale height at $r = 100$~au and has the value of 10 au and $\beta$ is the flaring power index. Fig.\ref{fig:dens_struc_fiducial} shows the density structure and the radial surface density. The parameters for the disk density structure are listed in Table \ref{table:1}.  

\subsubsection{Dust disk structure}\label{ssec:dust disk structure}

For the fiducial model Z1 with the reference metallicity of Z1$=0.85\,Z_{\odot}$ we assumed a dust-to-gas mass ratio of 0.01 in the disk. The respective gaseous element abundances are listed in Table \ref{table:el_abun}and are taken from \cite{2017A&A...607A..41K}.
We take dust growth into account by using a dust size distribution with a minimum and  maximum dust grain sizes of $a_{\rm min}=0.05\:\rm{\mu m}$ and $a_{\rm max}=3000\:\rm{\mu m}$, respectively. We use a simple power-law for the dust size distribution $f(a) \propto a^{-p}$ with the canonical value for interstellar grains of $p = 3.5$
 \citep{1977ApJ...217..425M}. The dust composition consists of a mixture of $60 \%$ amorphous laboratory silicate, $15 \% $ amorphous carbon, and $25 \%$ vacuum. The vacuum represents the porosity of the dust grains. We use a distribution of hollow spheres with a maximum hollow volume ratio $V_{\mathrm{hollow}}^{\mathrm{max}} = 0.8$. All the relevant dust properties are listed in Table \ref{table:1} and are taken from \cite{2016A&A...586A.103W}.
 
In this work, we use dust settling using the method of \cite{1995Icar..114..237D}. The dust composition and dust size distribution are constant throughout the entire disk. Furthermore, the dust composition, its maximum and minimum sizes, and the slope of the dust size distribution are assumed to be identical for all considered metallicities. This may be an oversimplification, but the process of dust growth in low-metallicity disks is still poorly known. 

\begin{table}
\caption{Main parameters model Z1.}
    \label{table:1}
    \centering
    \begin{tabular}{l|c|c} 
    \hline
    \hline
    Quantity & Symbol & Value \\
    \hline
    stellar mass & $M_{*}$ & 0.7 $M_{\sun}$ \\ 
    stellar effective temp. & $T_{*}$ & 4000 K \\ 
    stellar luminosity & $L_{*}$ & 1.0 $L_{\sun}$ \\
    X-ray luminosity & $L_{X}$ & 2.6$\times10^{-4} L_{\sun}$ \\
    \hline
    disk gas mass & $M_{\mathrm{disk}}$ & 0.03 $M_{\sun}$ \\
    disk inner radius & $R_{\mathrm{in}}$ & 0.07 au \\
    disk tapering-off radius & $R_{\mathrm{tap}}$ & 100 au \\
    column dens. pow. ind.  & \pmb{ $\lambda$} & 1.0 \\
    reference scale height & $H_{100}$ & 10 au \\
    flaring power index & $\beta$ & 1.15 \\
    \hline
    outer radius & $R_{\mathrm{out}}$ & 600 au \\
    \hline 
    dust-to-gas mass ratio & $\delta$ & 0.01 \\
    min. dust particle radius & $a_{\mathrm{min}}$ & 0.05 $\mu$m \\
    max. dust particle radius & $a_{\mathrm{max}}$ & 3000 $\mu$m \\
    dust size dist. power ind. & $a_{\mathrm{pow}}$ & 3.5 \\
    Max. hollow volume ratio   & $V_{\mathrm{hollow}}^{\mathrm{max}}$ & $80\%$ \\
    dust composition$\ ^{a}$  & $\mathrm{Mg_{0.7} Fe_{0.3} SiO_{3}}$ & $60\%$ \\
    (volume fractions) & amorph.carbon &  $15\%$\\
                     &  vacuum &  $25\%$\\
    \hline
    cosmic ray H ion. rate & $\zeta_{\mathrm{CR}}$ & $1.7\times10^{-17}\mathrm{s^{-1}}$ \\
    strength of interst. FUV & $\chi^{\mathrm{ISM}}$ & $1^{b}$ \\
    \hline
    distance & d & 140 pc \\
    \hline
    \end{tabular}
    \tablefoot{ $^{(a)}$ Optical constants are from ~\cite{1995A&A...300..503D}. and ~\cite{1996MNRAS.282.1321Z}, BE-sample). $^{(b)} \chi^{ISM}$ is given in units of the Draine field (~\cite{1996ApJ...468..269D}; ~\cite{2009A&A...501..383W}).}
\end{table}

\subsubsection{Chemical model}\label{ssec:chemical model}

After modelling the radial and vertical physical structure of the disk, \prodimo calculates a detailed continuum and line radiative transfer as well as the heating and cooling balance and the gas and surface chemistry of the disk. The net formation rate of a chemical species $i$ is calculated with the following equation:
\begin{equation}\label{eq:chem}
\begin{split}
   \frac{\mathrm{d}n_{i}}{\mathrm{d}t} = \sum_{jkl}R_{jk\to il}(T_{g})\ n_{j}n_{k} + \sum_{jl}  \left( R^{\rm ph}_{j\to il}+R^{\rm cr}_{j\to il}\right)n_{j} + ... \\
    - n_{i}\left( \sum_{jkl}R_{il\to jk}\ n_{l} + \sum_{jk}  \left( R^{\rm ph}_{i\to jk}+R^{\rm cr}_{i\to jk}\right) + ...\right)  
\end{split}
\end{equation}
Here, the first row represents all the formation reactions that produce species $i$ and the second row all the destruction reactions that destroy species $i$. $R_{jk\to il}$ indicates the two-body gas-phase reaction rate between two reactants ($j$ and $k$), forming two products ($i$ and $l$). $R^{\mathrm{ph}}_{j\to il}$ stands for the photo-reaction rate. This rate depends on the local strength of the UV radiation field. $R^{\mathrm{cr}}_{j\to il}$ indicates a cosmic ray-induced reaction rate. We only show three kinds of reactions in Eq.\eqref{eq:chem}. The dots indicate that there are other kinds of reactions involved (freeze-out, X-ray chemistry, etc.). The technical details regarding the chemistry are fully described in \cite{2009A&A...501..383W}.
We assume kinetic chemical equilibrium (steady-state chemistry) for this work. This means we have $\frac{\mathrm{d}n_{i}}{\mathrm{d}t} = 0 $ for the left side of Eq.~\eqref{eq:chem} for the steady-state model. We obtain i = 1...N$_{\mathrm{sp}}$ non-linear equations with j = 1...N$_{\mathrm{sp}}$ unknown particle densities $n_{j}$, where $N_{\rm sp}$ is the number of considered species. The chemistry of the disk is computed by solving Eq.~\eqref{eq:chem} using a globally convergent Newton-Raphson method as a steady-state solver \cite{2009A&A...501..383W} The steady-state chemistry approach has also been used and compared with time-dependent chemistry in \cite{2016A&A...586A.103W}.

The chemical network used in this work is based  on the one used in \cite{2017A&A...607A..41K}. We use additionally several surface chemistry reactions described in \cite{2020A&A...635A..16T} and Thi et al (in prep). The surface chemistry network~\cite[]{2020A&A...635A..16T} is based in \cite{1993MNRAS.261...83H}. 
The total amount of species is 250 and 5790 chemical reactions, respectively.
The modifications to the element abundances done for this work will be explained in detail in the next section.

\begin{figure}[!ht]
  \centering
  \begin{subfigure}{0.99\hsize}
    \includegraphics[width=\hsize]{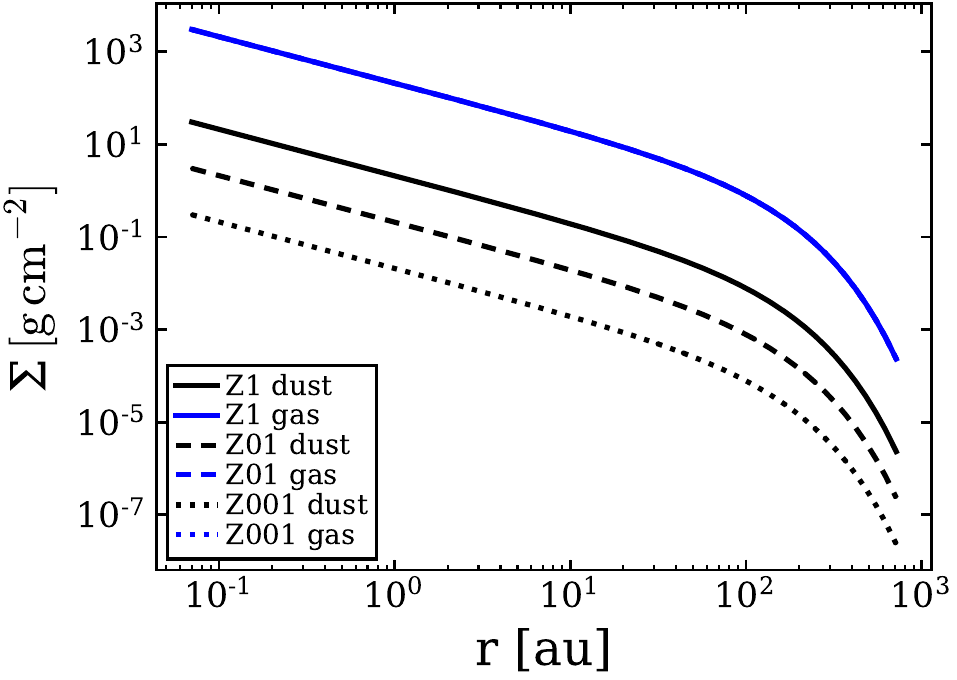}
  \end{subfigure}
  \caption{Surface density in the midplane for dust (black) and gas (blue) as a function of radius. The models Z1 (solid line), Z01 (dashed line), and Z001 (dotted line) are shown.}
  \label{fig:sdd}
\end{figure}

\subsection{Surface density}
Fig.~\ref{fig:sdd} shows the resulting gas and dust surface density after the mentioned changes in the models have been applied. It shows a decrease in the dust surface density for lower metallicities. Note that we keep the gas surface density and disk mass constant in all our models.

\section{Results}\label{Results}

\begin{figure}[h!]
  \centering
  \begin{subfigure}{0.99\hsize} 
    \includegraphics[width=\hsize]{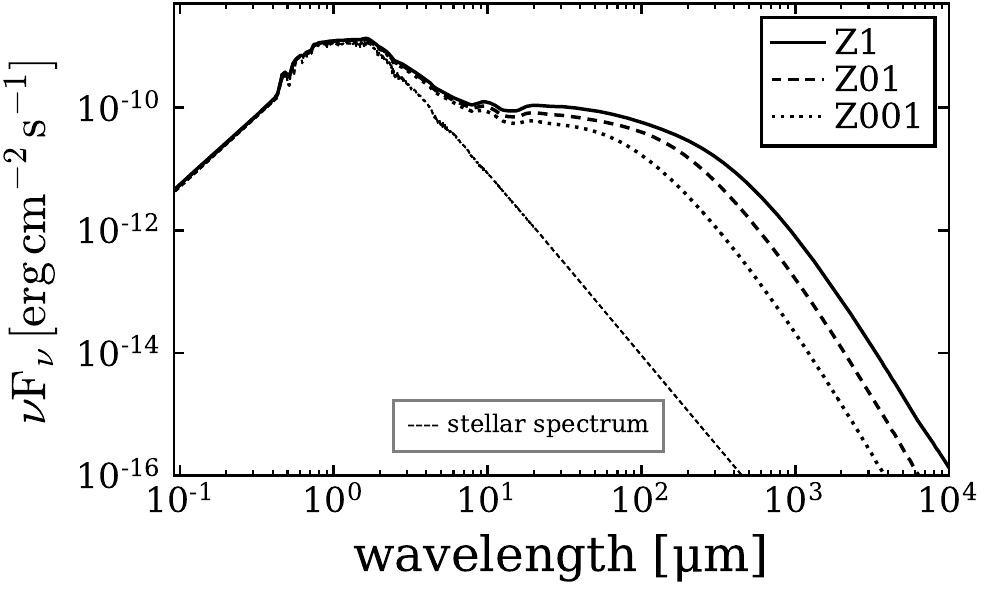}
   \end{subfigure}
  \caption{Spectral energy distribution of the different disk models and stellar spectrum. The models Z1 (solid line), Z01 (dashed line), and Z001 (dotted line) are shown. A lower metallicity leads to a decrease in the flux density for the SED.}
  \label{fig:sed}
\end{figure}

\begin{figure*}[!ht]
\centering
\begin{subfigure}{1.0\textwidth}
\includegraphics[width=\textwidth]{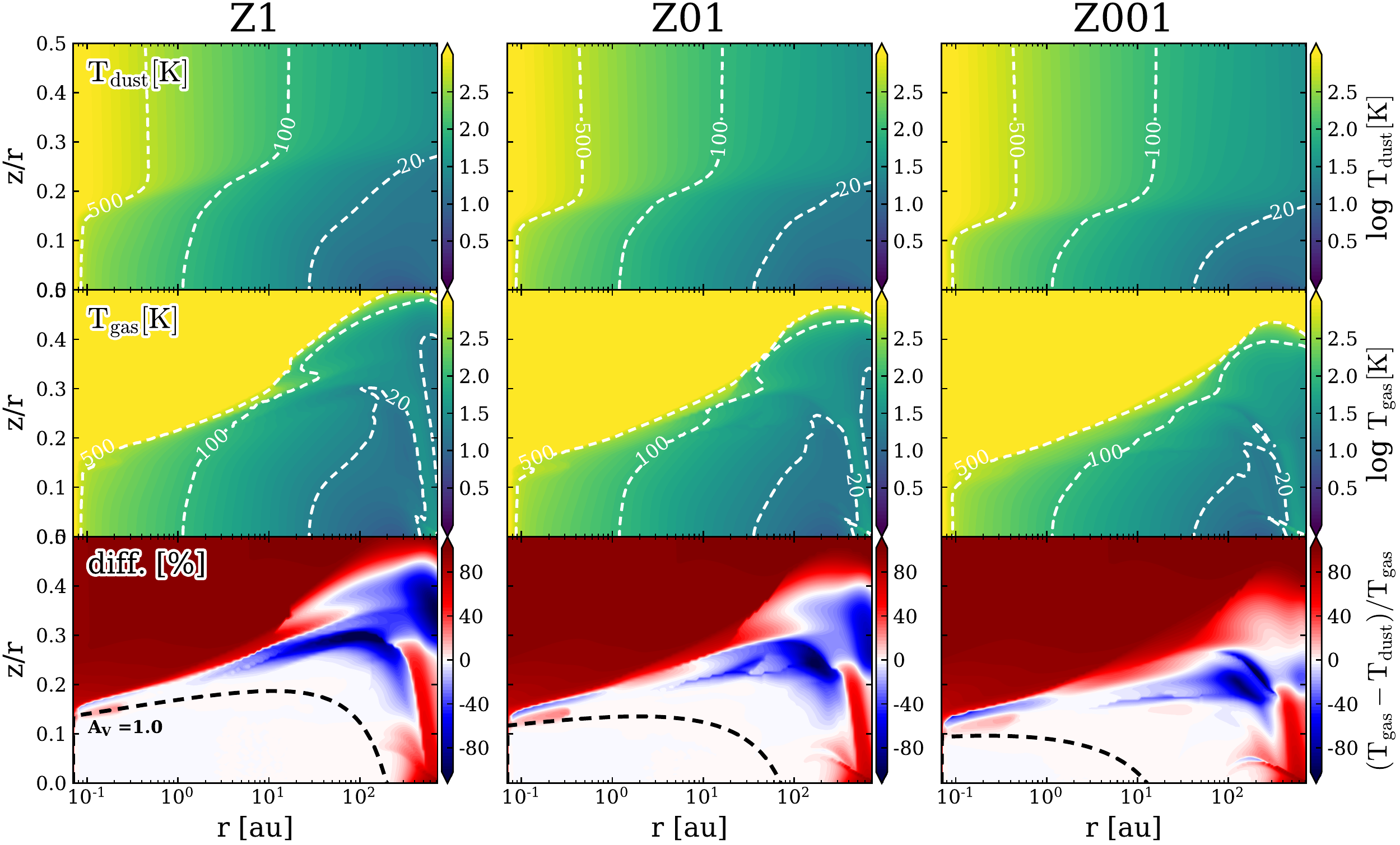}
\end{subfigure}
\caption{
Contour plots of the models with different metallicities. The left, middle and right columns of the plot represent the models Z1, Z01, and Z001. Z01 and Z001 with 1/10  and 1/100 of the reference metallicity, respectively. The top, middle, and bottom rows correspond to the dust temperature, gas temperature, and the difference between the gas and dust temperature relative to the gas temperature. The decreasing metallicity leads to an increase in both dust and gas temperature but with the gas temperature increasing stronger (bottom row). The black dashed line represents the surface corresponding to a visual extinction of A$_{v} = 1$.} 
\label{fig:cont_t}
\end{figure*}

\subsection{Spectral energy distribution (SED)}
Fig.~\ref{fig:sed} shows the impact of the metallicity in the models on the Spectral Energy Distribution (SED).  For the calculation of the SED, a distance to the source of 140 pc is assumed.
It is clear that for reduced dust-to-gas ratio the flux density of the disk also decreases. This is a consequence of the lesser amount of dust that a model with reduced metallicity has and is shown clearly in Fig. \ref{fig:sdd} where the dust surface density is displayed.
At shorter wavelengths, the emission of the disk comes from the part where it is optically thick. Therefore it does not vary significantly if a fraction of dust is removed from it. At $ \lambda < 1\ \mu m$ the SED flux density is dominated by the stellar emission, which remains unchanged for the three models as the star parameters are not being changed. This suggests that low metallicity disks are more difficult to observe at longer wavelengths. At shorter wavelengths (e.g.near-infrared), they are as easy to observe as disks with higher metallicity. The impact that lower metallicity has on the stellar parameters is a subject that will be covered in future studies.

\subsection{Dust and gas temperature}

\begin{figure}[!h]
  \centering
  \begin{subfigure}[b]{0.49\textwidth}
    \includegraphics[width=\textwidth]{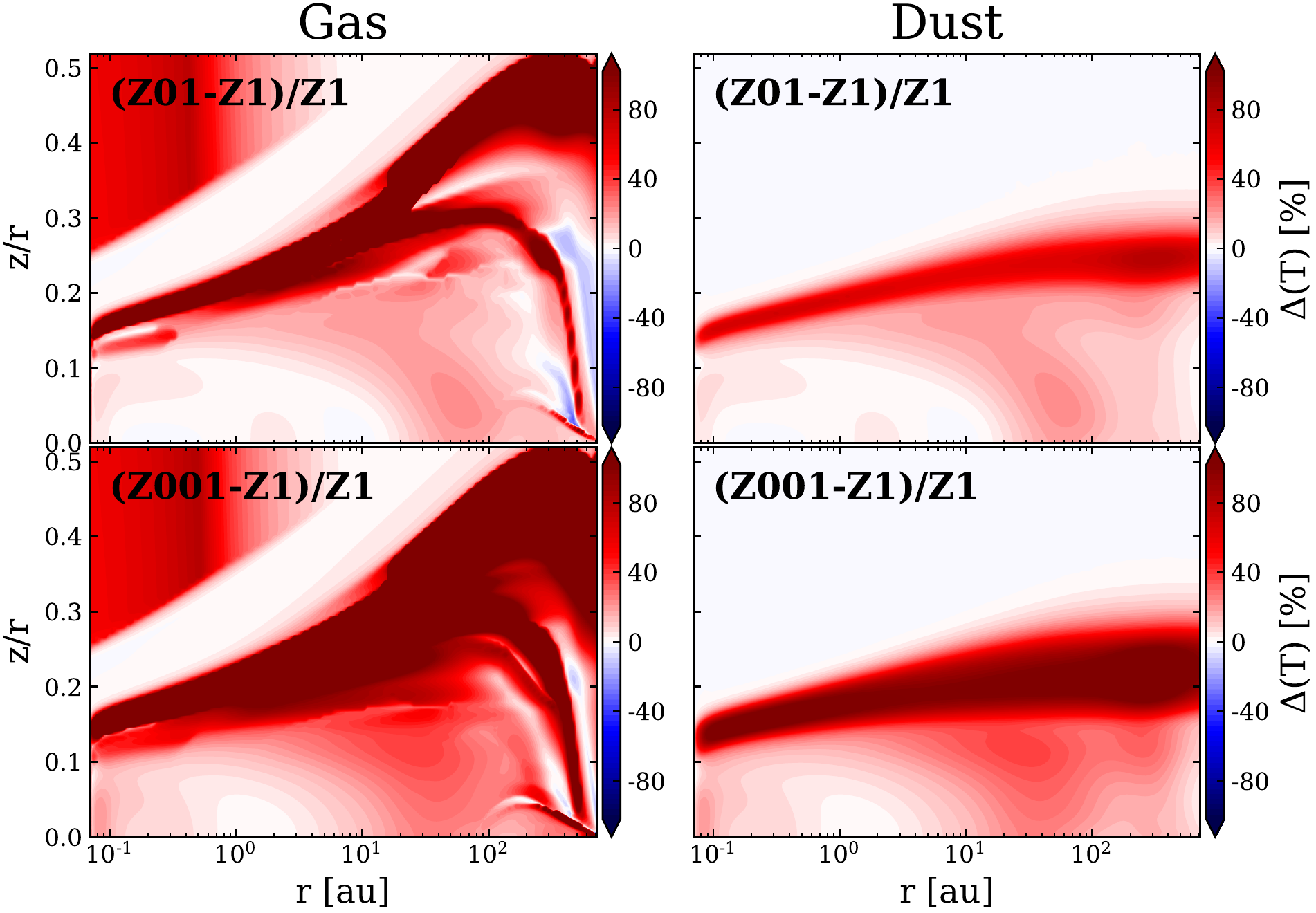}
  \end{subfigure}
   \caption{Contour plots that show the difference between the temperatures of the model Z1 and the Z01 model (top panel) and the Z001 model (bottom panel). The r.h.s column shows the difference in dust temperature and the l.h.s column the difference in the gas temperature. The red color represents the regions where the model with lower metallicity has a higher temperature. The white and blue colors represent the regions where both temperatures are similar and where the Z1 model has a higher temperature respectively.}
  \label{fig:diff_temps}
\end{figure}

\begin{figure*}
    \centering
    \begin{subfigure}[b]{0.99\textwidth}
    \includegraphics[width=\textwidth]{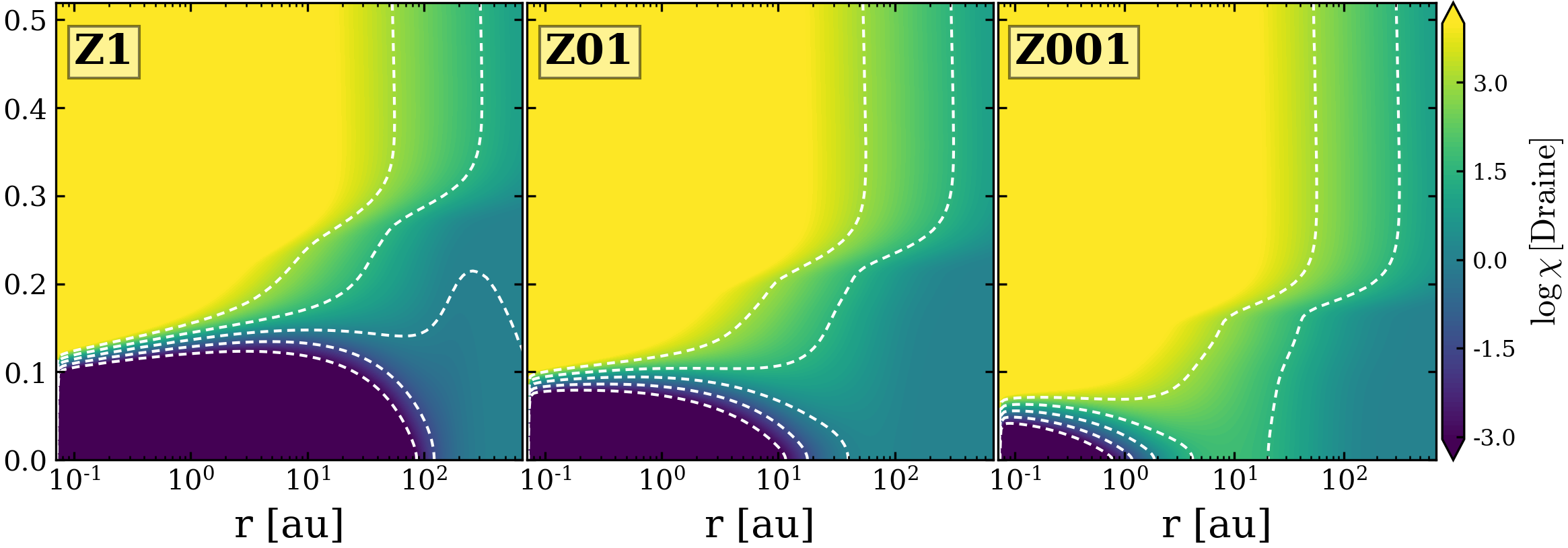}
  \end{subfigure}
  \caption{UV radiation field for the Z1 model (left panel), model Z01 (middle panel), and Z001 (right panel), in units of the Draine field. The dashed contour lines correspond to the tick values in the color bar. The decreasing metallicity enables the UV radiation to get to the midplane at a smaller radius.}
  \label{fig:diff_rad}
\end{figure*}

The first result we analyzed in the produced models was the influence of a lower metallicity on the gas and dust temperatures. {For a description of the properties of model Z1 we separate the structure of the disk into three radial zones and three vertical layers following \cite{2013ChRv..113.9016H}. The radial structure consists of three zones:
\begin{itemize}
    \item \textit{The inner zone}: it is located closest to the star and extends up to $\approx 1$  au and has a temperature above 100 K.
    \item \textit{The middle zone}: it is located between 1 and 30 au. The temperatures in this zone are in the range of 100 (middle and upper parts of the disk) $\geq T \geq 20$ K (midplane).
    \item \textit{The outer zone}: it starts at 30 au and extends to the edge of the disk. The temperature in this zone is below 20 K.
\end{itemize}
The vertical structure from bottom to top is as follows:
\begin{itemize}
    \item \textit{The midplane}: The coldest vertical layer of the disk. Here the chemical reaction is predominantly gas grain interaction and dust surface chemistry. 
    \item \textit{The rich molecular layer}: The temperature in this layer lies between 20 and 100 K. It is warm enough to keep some species in the gas phase (e.g. CO). It is also deep enough that the molecules are mostly shielded from photodissociation. The conditions in this layer allow rich molecular chemistry to take place. 
    \item \textit{The photon-dominated layer}: The hottest layer is characterized by temperatures between 100 and 5000 K. X-rays and UV radiation are the main drivers of the chemistry. Most of the species here are in atomic form and very often ionized.
\end{itemize}}
Fig.~\ref{fig:cont_t} displays dust and gas temperatures and their difference in all the models. The figure shows that the decrease in metallicity produces higher gas and dust temperatures.
{As already mentioned, the top layer of the disk (photon-dominated layer) is characterised by gas temperatures of $\rm{T}_{\rm{gas}} > 1000\,\rm{K}$.}
Both gas and dust increase in temperature as the metallicity decreases, mostly because the radiation is able to penetrate the disk more deeply with less shielding effect by the dust mass in the environment. This also affects the thermal accommodation process between dust and gas and hinders the dust to reach the same temperatures as the gas.

The middle layers (rich molecular layers) show a greater temperature for the gas for a small radius but a greater dust temperature for radii greater than 100 au. In this part of the disk, the collisional exchange of energy is not enough to make both dust and gas temperatures equal. Gas and dust are thermally decoupled and depending on the position in this layer,the dust or gas will be warmer. For a radius smaller than 100 au this region is being directly hit by the stellar high energy radiation and the dominant heating processes are chemical heating followed by heating by the formation of H$_{2}$ on dust and heating by collisional de-excitation of H$_{2}$. These UV-driven heating processes elevate the gas temperature. For radii beyond 100 au, we have regions where the dust is warmer than the gas. These regions of the disk are shielded more effectively from the stellar radiation so that the heating of the gas by direct radiation (mainly X-rays) is not efficient. Additionally, OI and other line cooling processes of the gas dominate this region which leads to dust being warmer than gas.  As the metallicity decreases, the region where dust is warmer than gas shrinks because the radiation is not being shielded as strongly and the gas heats up again. 

Finally, for all three models with different metallicity, we observe that the gas and dust temperatures are in equilibrium in the midplane of all models. This section of the disk has a relatively low temperature and high density. This means that the energy exchange rate by collisions between dust and gas particles is frequent enough to reach thermalization. Thus, the gas and dust temperature become the same through a process called thermal accommodation \cite[]{2009A&A...501..383W}.  As metallicity decreases, the dust density of these deep layers also decreases and the thermal accommodation becomes less efficient. This leads to a shrinking of the region (white area) where dust and gas have the same temperature.
  
We note that a general trend of increasing temperature with decreasing metallicity is true for the type of models considered in this work, which are often referred to as passive disks. For passive disks, the only heating source is stellar radiation and cosmic rays. In the case of active disks, the hydrodynamical processes provide additional heating via viscous and compressional heating, which operate predominantly in the disk's midplane. The active disk tends to decrease its temperature with a lower metallicity as the heat in a more diluted disk (characterized by lower opacity) can escape more easily and thus cool the disk (\cite{2020A&A...641A..72V}).

In order to have a closer look at the impact of the different metallicity on the gas and dust temperature of the disk, we compare the temperature of the lower metallicity models with model Z1.
Fig.~\ref{fig:diff_temps} shows the influence of a lower metallicity over the gas and dust temperature by displaying 2D plots of the relative temperature difference between the model Z1 and the ones with the reduced metallicity. A decreased metallicity leads to an increase in the gas and dust temperature in the disk. This effect is stronger in the middle layers of the disk but it also takes place in the midplane beyond a radius of 20 au for the Z01 model and a radius of 4 au for the Z001 model.

\subsection{Radiation field }

\begin{figure*}[!h]

  \centering
  \begin{subfigure}[b]{0.49\textwidth}
    \includegraphics[width=\textwidth]{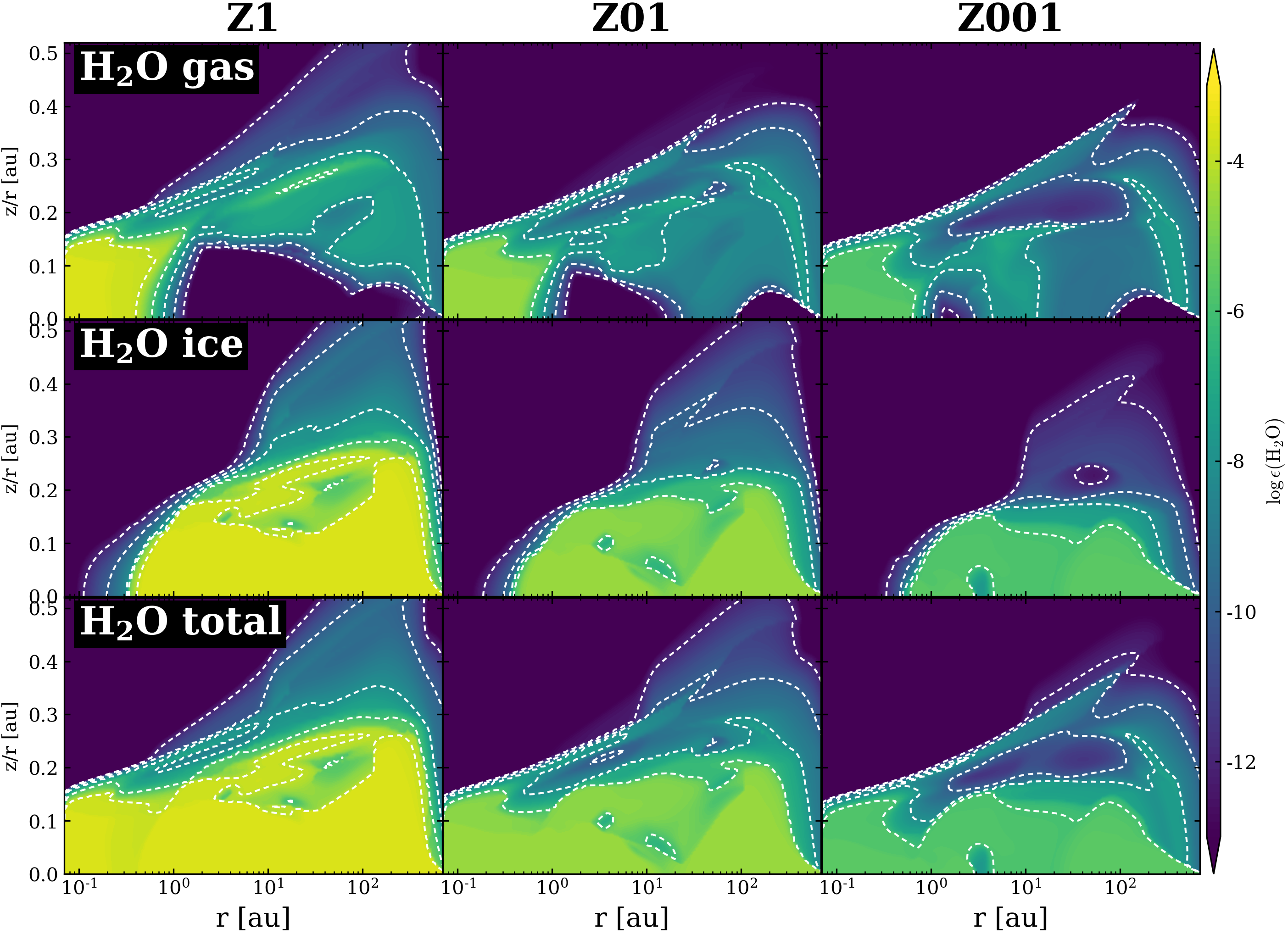}
    \end{subfigure}
\begin{subfigure}[b]{0.49\textwidth}
    \includegraphics[width=\textwidth]{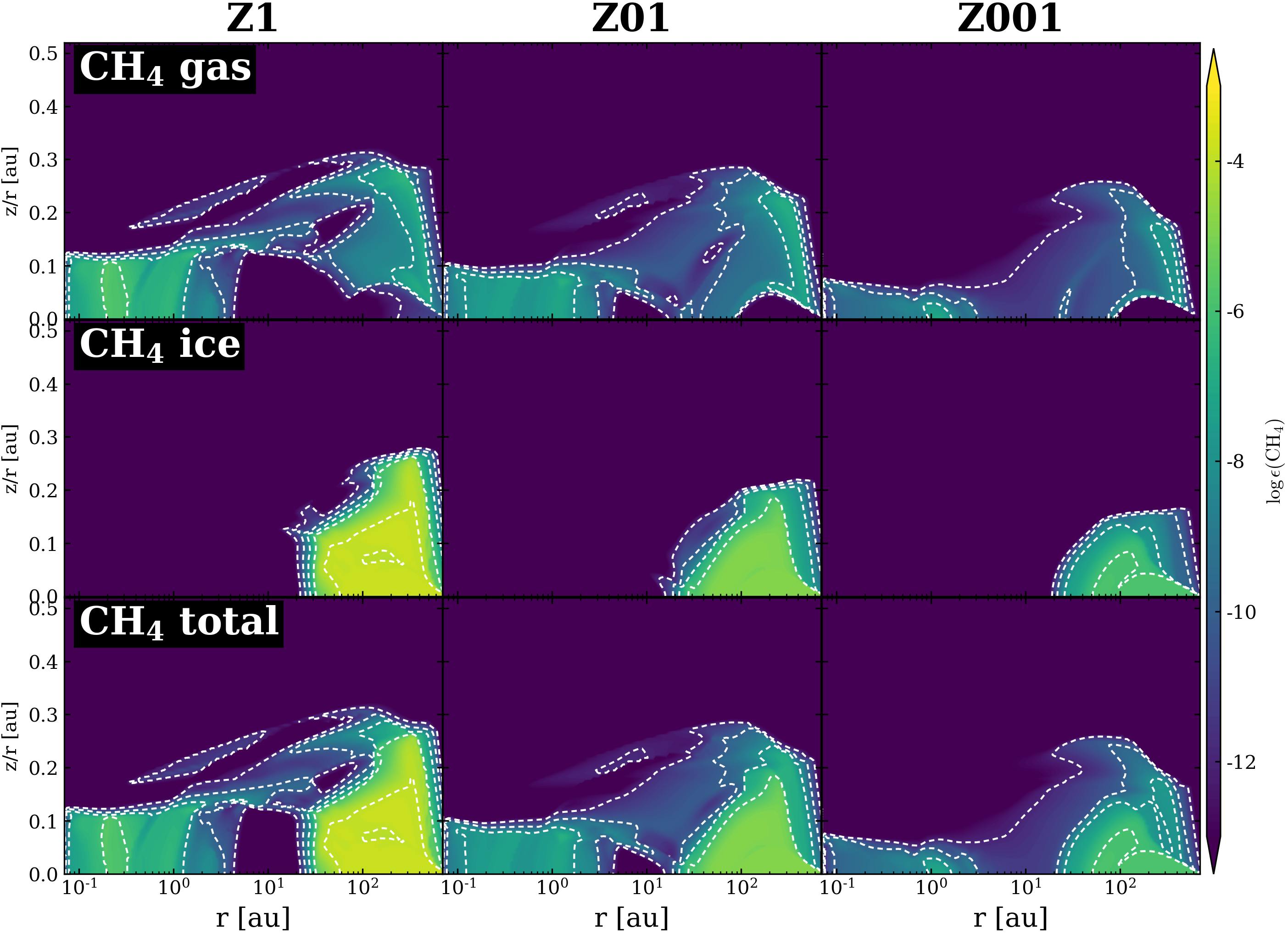}
  \end{subfigure}
\begin{subfigure}[b]{0.49\textwidth}
    \includegraphics[width=\textwidth]{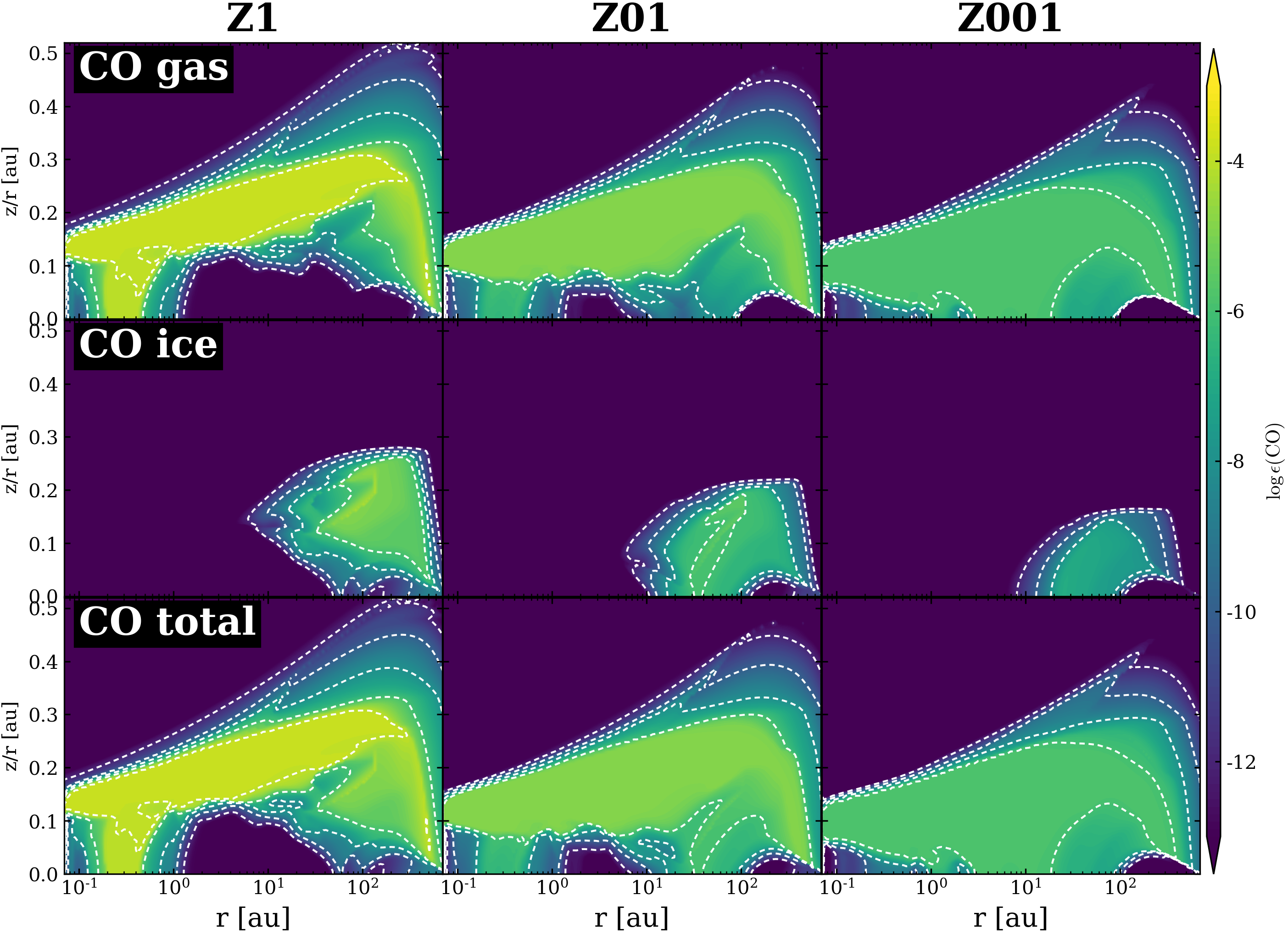}
  \end{subfigure}
\begin{subfigure}[b]{0.49\textwidth}
    \includegraphics[width=\textwidth]{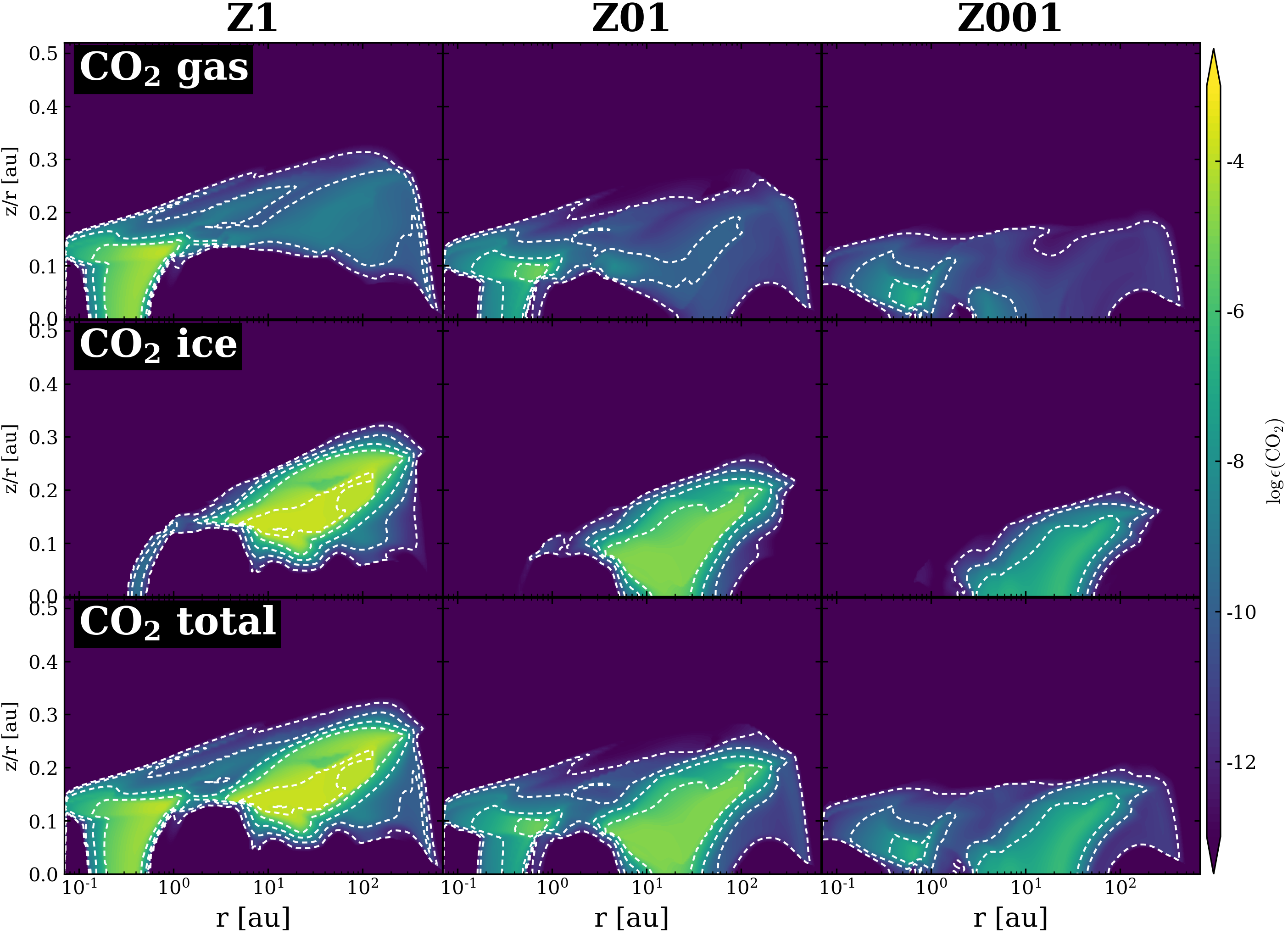}
  \end{subfigure}
  \begin{subfigure}[b]{0.49\textwidth}
    \includegraphics[width=\textwidth]{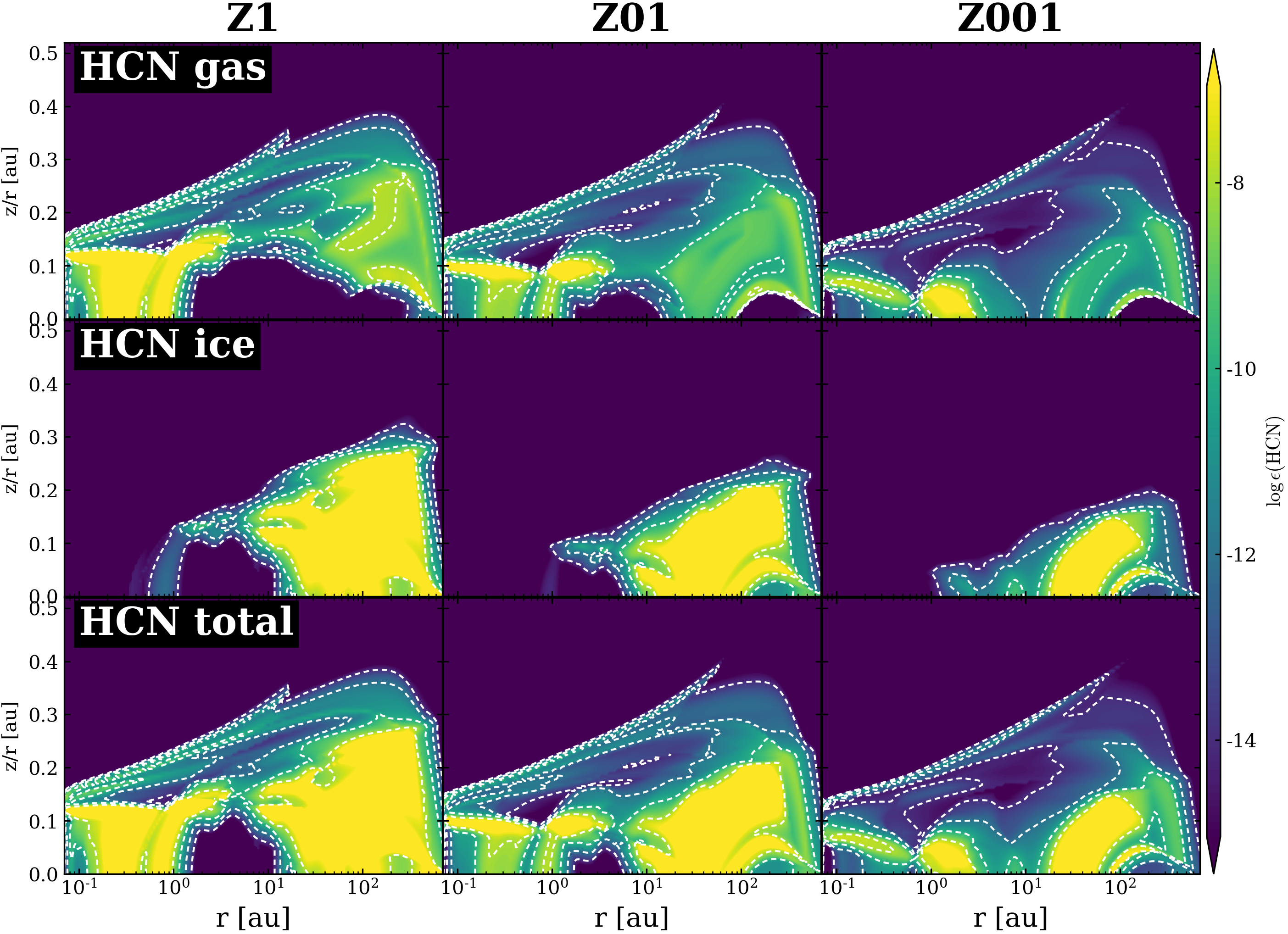}
  \end{subfigure}
  \begin{subfigure}[b]{0.49\textwidth}
    \includegraphics[width=\textwidth]{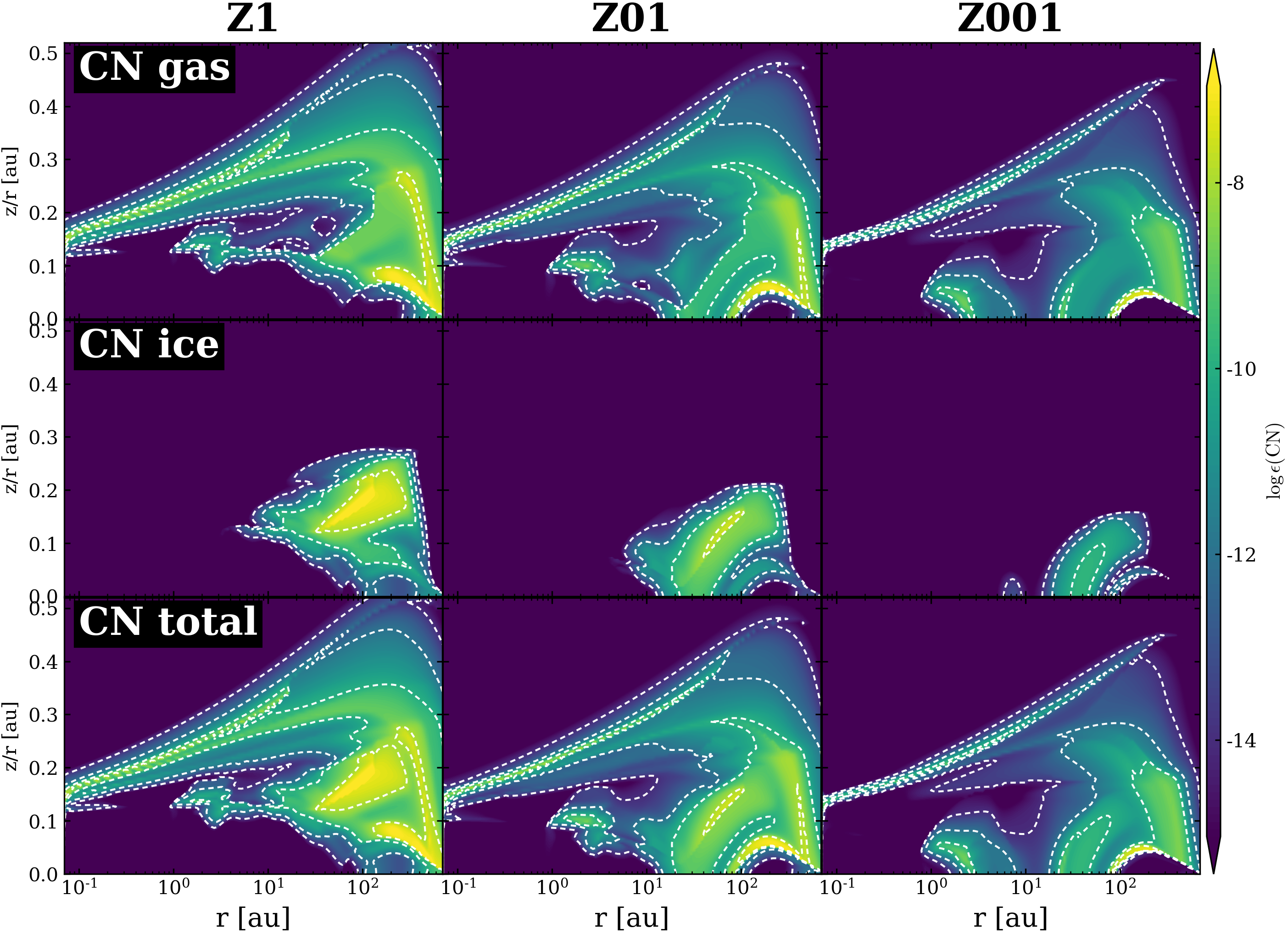}
  \end{subfigure}
\caption{Contour plots of the abundances. From left to right and top to bottom: H$_{2}$O, CH$_{4}$, CO, CO$_{2}$, HCN, and CN are represented relative to the total hydrogen nuclei number density. The dashed contour lines describe the tick values on the colorbar. The left column corresponds to model Z1. The middle column corresponds to the model Z01 and the left column to the model Z001. The upper row represents the gas species, the lower row the ice species, and the bottom row the total abundances. With decreasing metallicity the cavity in the midplane for all gas-phase species shown here shrinks (Z01) and vanishes (Z001). In the case of CH$_{4}, $ a second cavity in the upper layers of the disk grows as metallicity decreases.}
  \label{fig:abuncont}
\end{figure*}
In Fig.~\ref{fig:diff_rad} the UV radiation ($91.2\ \rm{nm} <\lambda < 250\ \rm{nm}$) field in Draine units is shown.
The contour lines in the plots show that for the reduced metallicity models the radiation has higher values along the midplane of the disk. For the Z1 model, the radiation value of $10^{-1.5}$  reaches the midplane at a radius greater than 100 au. For the Z01 model, the same radiation value is reached at a radius around 10 and 20 au and for the Z001 model, this value is reached at a radius around 1 au. 
This confirms that models with lower metallicity allow the radiation of the star to penetrate deeper regions of the disk. In this case, this is simply the effect of reducing the dust mass. The same effect would be achieved by using unchanged elemental abundances. The effect of a stronger radiation field in the disk on the respective  disk chemistry is presented in sections \ref{ssec:Abundances},  \ref{ssec:Detection of the snowline} and \ref{ssec:Vertical column densities of chemical species}.

\subsection{Spatial distribution of abundances} \label{ssec:Abundances}
 
\begin{figure}[!h]

  \centering
  \begin{subfigure}[b]{0.49\textwidth}
    \includegraphics[width=\textwidth]{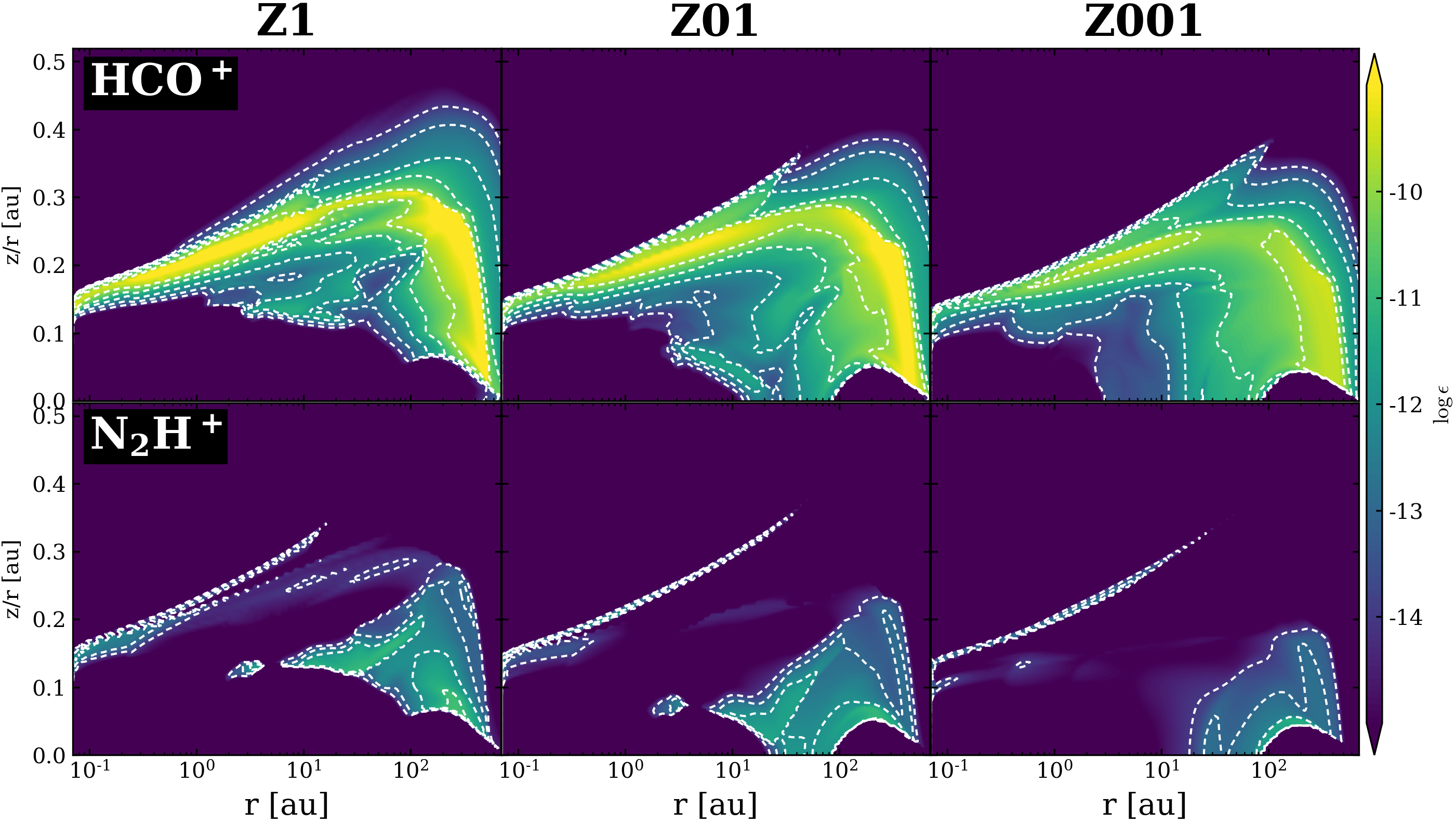}
  \end{subfigure}
  \caption{{Same as Fig. \ref{fig:abuncont} but for HCO$^+$ and N$_2$H$^+$. We only show the gaseous abundances because HCO$^+$ and N$_2$H$^+$ do not freeze-out in our models.}}
  \label{fig:abuncont_2}
\end{figure}

In this section, we present the impact of metallicity on the abundances of various molecules relative to hydrogen. The ice and gas abundances of H$_2$O, CH$_4$, CO, CO$_2$, HCN, CN, HCO$^+$ and N$_2$H$^+$ are displayed as a function of the radius and the height of the disk in Fig.~\ref{fig:abuncont} and Fig. \ref{fig:abuncont_2}. These species were selected because they are a product of reactions involving the two most abundant molecules in the gas phase (H$_2$ and CO) and because their molecular lines are known tracers of physical and chemical disk properties. For instance, lines of H$_{2}$O, CO in the IR regime can give us information about the temperature in the disk \cite[]{2013ChRv..113.9016H}. CN, HCN lines in the sub-mm and mm regime serve as photochemistry tracers {\cite[]{2013ChRv..113.9016H}}. HCO$^+$ and N$_{2}$H$^+$ lines in the same regime are used to trace the ionization \cite[]{2013ChRv..113.9016H}.
{In the study performed by \cite{2021arXiv211204930H} one of the main findings was that the abundances of CH$_4$ on cooler hot Jupiters can be used to link the composition of a planet's atmosphere to its formation location. 
Additionally, even though CO is still considered to be the best gas mass tracer, CO$_{2}$ and H$_2$O can potentially be used as alternative mass tracers as well \cite[]{2017ApJ...849..130M}}. We chose to take a deep look into these specific molecules also because they play an important role as the building blocks for complex carbon-based molecules which are also the building blocks for organic life. 

The impact of metallicity on the abundances can be described as follows. On the one hand, as the metallicity decreases the relative abundances of ice and gas species decrease the entire disk. This general decrease is the result of the reduction of the elemental abundances mentioned in section \ref{lower metallicity}. On the other hand, the dust reduction applied to obtain lower metallicities affects specific regions of the disk.
We focus on the impact on specific regions of the disk:  

\begin{enumerate}
    \item The midplane shows a cavity for the gaseous species that shrinks with decreasing metallicity. The ice species are also affected by this and exhibit a shrinking of their spatial distribution. This is a consequence of the increase of the input energy by a stronger radiation field in the disk as the metallicity decreases. The stronger radiation field desorbs or dissociates ice-phase species and enhances  the formation reaction of some gas-phase species. This leads to a depletion of icy species in the location of the midplane cavity and an increase in the abundance of the respective gaseous species.
    
    \item The total abundances displayed in the bottom row in the panes of Fig. \ref{fig:abuncont} also exhibit a cavity in the midplane that shrinks with decreasing metallicity. In the case of H$_2$O, the total abundance does not show a midplane cavity for any metallicity but a general decrease in all the disks is noticeable. Apart from H$_2$O, the behavior of the midplane cavity for the total abundances mirrors the replenishment of gaseous species as the metallicity decreases.
\end{enumerate}

We decided to make a simple comparison of the Z01 and Z001 models with models where we artificially reduced the chemical abundances of the Z1 model by a factor of 10 and 100, respectively. We do this comparison to show the non-linear behavior of the chemistry when reducing the metallicity. We named the models with the artificially reduced column density as scaled-down models.

\subsection{Vertical column densities}\label{ssec:Vertical column densities of chemical species}
In this section a detailed analysis of the vertical column density of the above-mentioned species is provided along with the chemical reactions responsible for the formation and destruction of the molecules studied in this work. We analyze the behavior of the total vertical column density of the already mentioned species. The ice species are followed by a $\#$ symbol. 

Figs. \ref{fig:column_H2O} - \ref{fig:column_HCO+_N2H+} show the total vertical column density of the mentioned chemical species. We again made a simple comparison of the Z01 and Z001 models with models where we artificially reduced the vertical column density of the Z1 model by a factor of 10 and 100, respectively. The red shaded area in the figures highlights the cases where the column density is higher than the simple scaled-down of the Z1 model abundances and the green shaded area where it is lower.
For each molecule, we describe the general trends of the column density. We also give in Table \ref{table:form_destr} the respective changes in the formation and destruction reactions for the molecules for different metallicities. A more detailed display of the behavior of the molecules in a frozen and gaseous state is given in Appendix \ref{appendix:a}. The relevant reactions are given in a detailed form in Table \ref{tab:react_eq} in the Appendix as well.
For completeness, we also show the impact of the dust reduction and the elemental reduction of the metals separately in Appendix \ref{appendix:b}.

\subsubsection{H$_{2}$O}\label{cd_h2o}
\begin{figure}[!h]
    \centering
    \includegraphics[width=0.99\hsize]{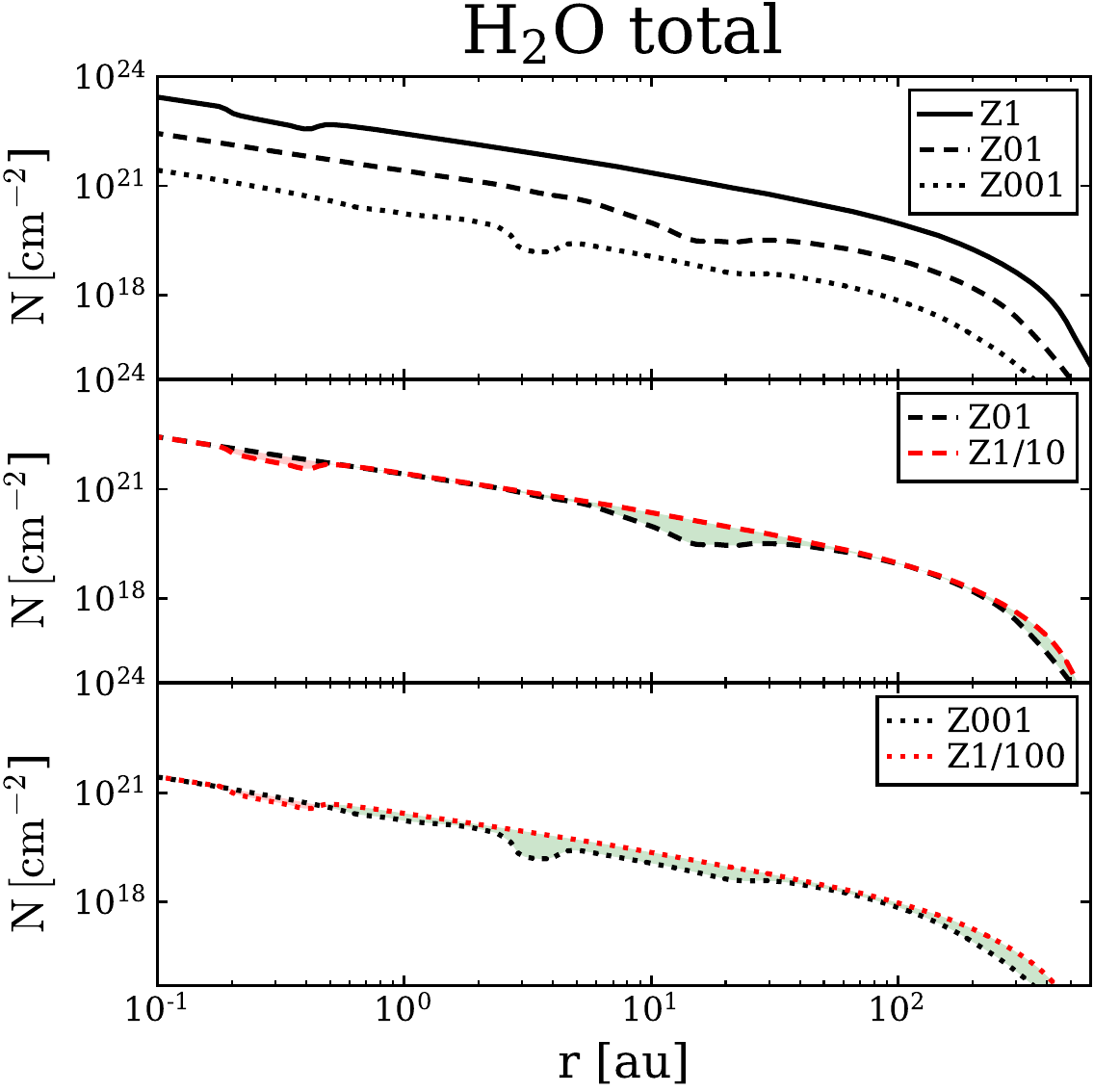}
    \caption{Total vertical column density of H$_{2}$O for the models. The top row shows models Z1, Z01, and Z001 (solid, dashed, and dotted lines respectively). The middle row shows the difference between the Z01 model (black dashed line) and the values by reducing the vertical column density of Z1 manually by a factor of 10 (red dashed line). The red and green areas show the cases where the vertical column density is higher or lower than the simple scaled-down values. The bottom row shows the same plots as the middle row but for the Z001 model and the Z1 model reduced by a factor of 100.}
    \label{fig:column_H2O}
    
\end{figure}

The top panel of Fig. \ref{fig:column_H2O} shows a total column density decrease with radius for all three models. However, there are small local drops that are located in different radii for each model. The total column densities (gas + ice) shown in the middle and bottom panels have equal or slightly lower values than the scaled-down values. The reason for the lower than {the scaled-down values} lies in the stronger radiation field and efficient photodissociation of $\mathrm{H_2O\#}$ (eq. \ref{eq:d_H2O_Z01_Z001}) in the deeper parts of the disk. It is the consequence of the radiation of the star being able to enter deeper regions of the disk as the metallicity of the model decreases. For Z1, the main destruction reaction for $\mathrm{H_2O\#}$ in the disk region of underabundance is thermal desorption, which does not affect the total column density. Table~\ref{table:form_destr} shows that for lower metallicities (Z01 and Z001) photodissociation takes over as the main destruction reaction because the radiation field is stronger in that part of the disk. $\mathrm{H_2O\#}$ is therefore not only being sublimated by the rising temperatures but is also being dissociated by photons that are able to reach the disk midplane (\cite{2006JChPh.124f4715A}). Particularly for Z001 the small drop beyond 2 au starts at the same radius where a great part of the UV radiation is not being strongly shielded and reaches the midplane for that model (see Fig. \ref{fig:diff_rad}). The resulting effect is a more efficient depletion of all H$_2$O as the metallicity decreases. This is shown clearly in Fig. \ref{fig:integ_ratio} where the total ratios $\mathcal{R}_{\mathrm{H_2O}} < 1$. 

\subsubsection{CH$_4$}\label{cd_ch4}

\begin{figure}[!h]
\centering
\includegraphics[width=0.99\hsize]{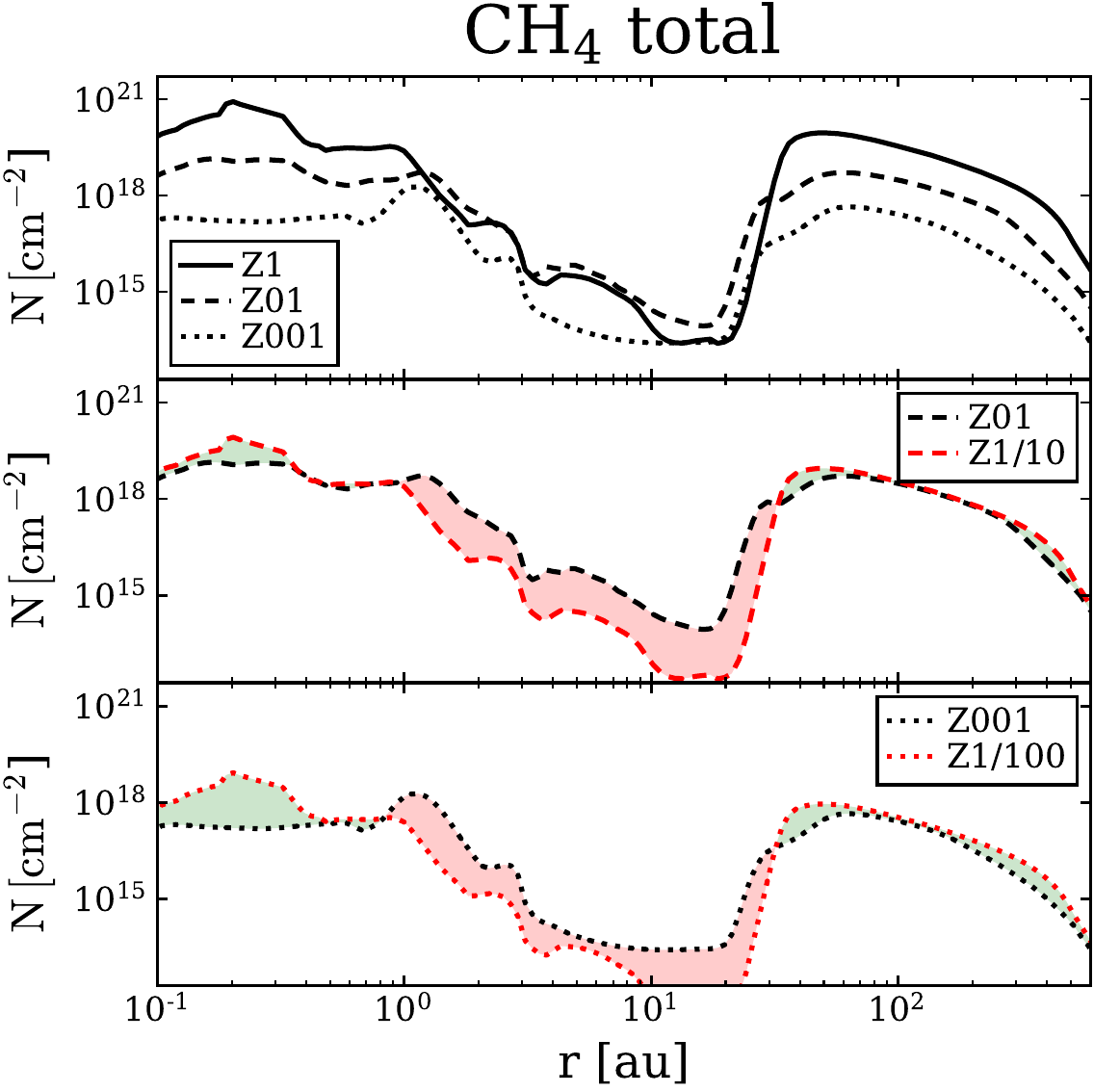}
\caption{Same as Fig. \ref{fig:column_H2O} but for CH$_{4}$.}
\label{fig:column_CH4}
\end{figure}

The three models show a decrease that starts at 1 au and continue until a radius of 20 au, where a strong {increase follows (top row of Fig. \ref{fig:column_CH4}). The decrease from 1 au to 20 au is due to the cavity for the total CH$_4$ that is visible in Fig. \ref{fig:abuncont}}. The middle and bottom rows show an underabundance inside $0.4$~au for the lower metallicity models. This is mainly due to the enhanced depletion by photodissociation of gaseous CH$_{4}$ caused by a stronger radiation field.  
It is also shown that roughly between 0.8 and 8 au the lower metallicity models have higher values than scaled-down values. The overabundance of Z01 and Z001 between 0.8 and 8 au is the effect of the larger formation rates that the lower metallicity models have near the midplane and around a radius of 1.5 au. For instance, the CH$_4$ formation rates in that region for the Z1, Z01 and Z001 model are 3.44 x$10^{-13}$, 3.84 x$10^{-11}$ and 1.21 x$10^{-7}$ cm$^{-3}$s$^{-1}$ respectively. The main destruction rates of CH$_{4}$ also increase with decreasing metallicity but not enough to counteract the enhanced formation of CH$_{4}$ (see Table~\ref{table:form_destr}).
Beyond 20 au the three models follow the scaling-down by one and two orders of magnitude very closely and have either equal or slightly lower values than the scaled-down values.
The {total molecular amount} (Fig. \ref{fig:integ_ratio}) shows that the Z01 and Z001 models have a lower than 1 ratio in comparison to the scaled-down values. Thus, the overabundance is present in regions of the disk where the total abundance of CH$_4$ is very low and it does not compensate for the underabundance in the parts of the disk with a radius greater than 25 au.

\subsubsection{CO}\label{cd_co}

\begin{figure*}[h]
    \centering
    \begin{subfigure}{0.48\textwidth}
    \includegraphics[width=\textwidth]{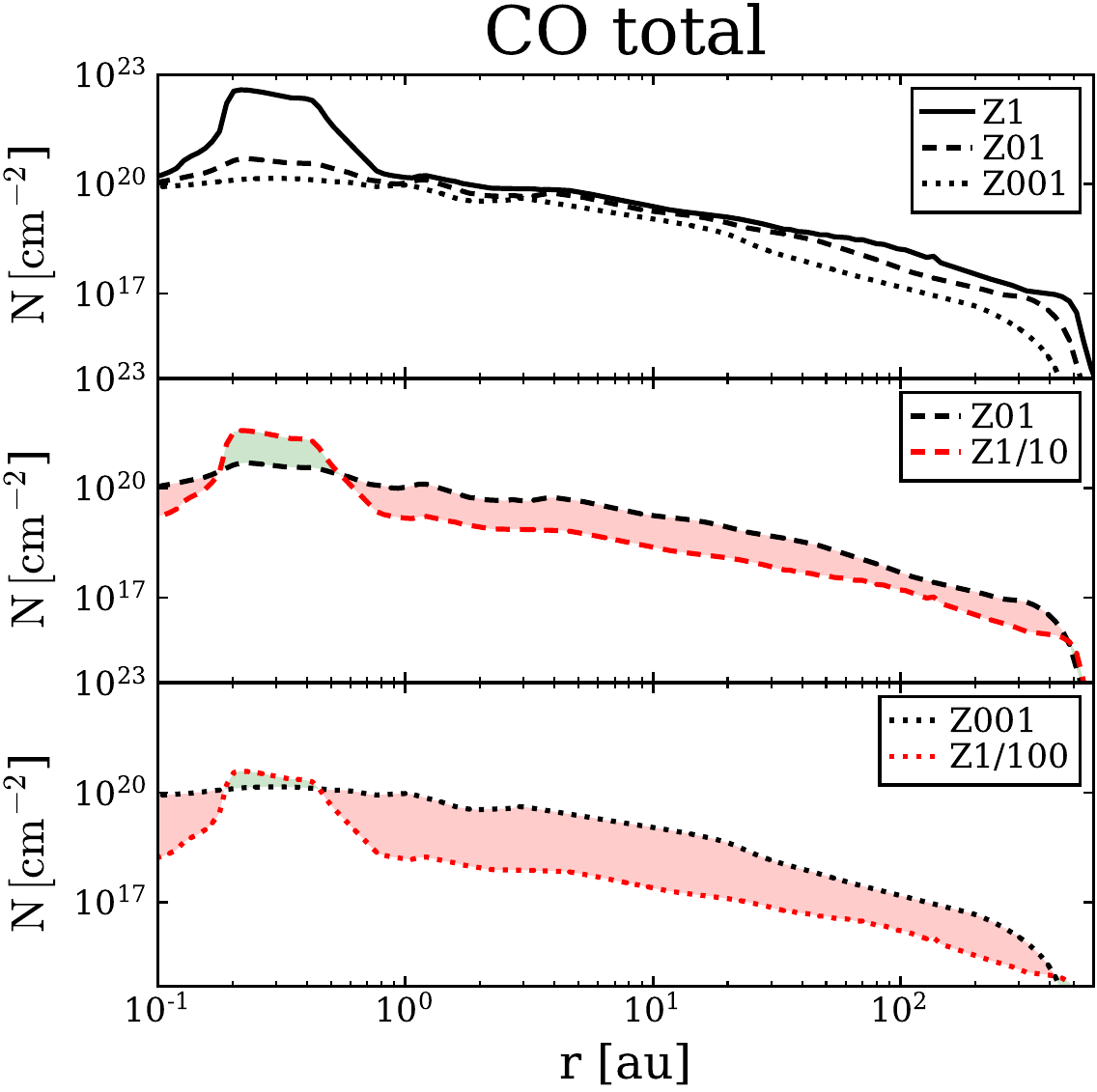}
    \end{subfigure}\hspace{0.02\textwidth}
    \begin{subfigure}{0.48\textwidth}
    \includegraphics[width=\textwidth]{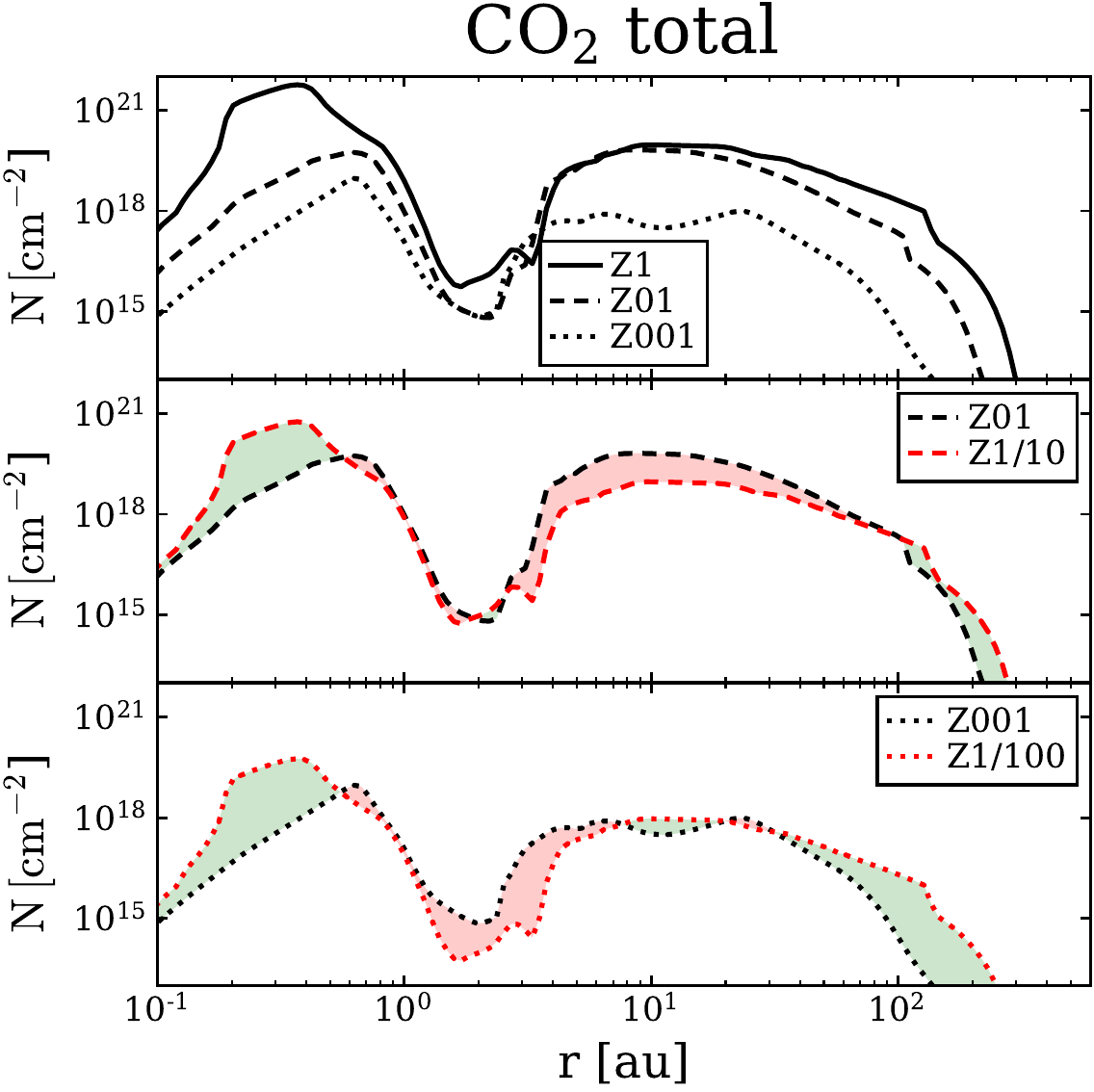}
    \end{subfigure}
    \par\bigskip
    \begin{subfigure}{0.48\textwidth}
    \includegraphics[width=\textwidth]{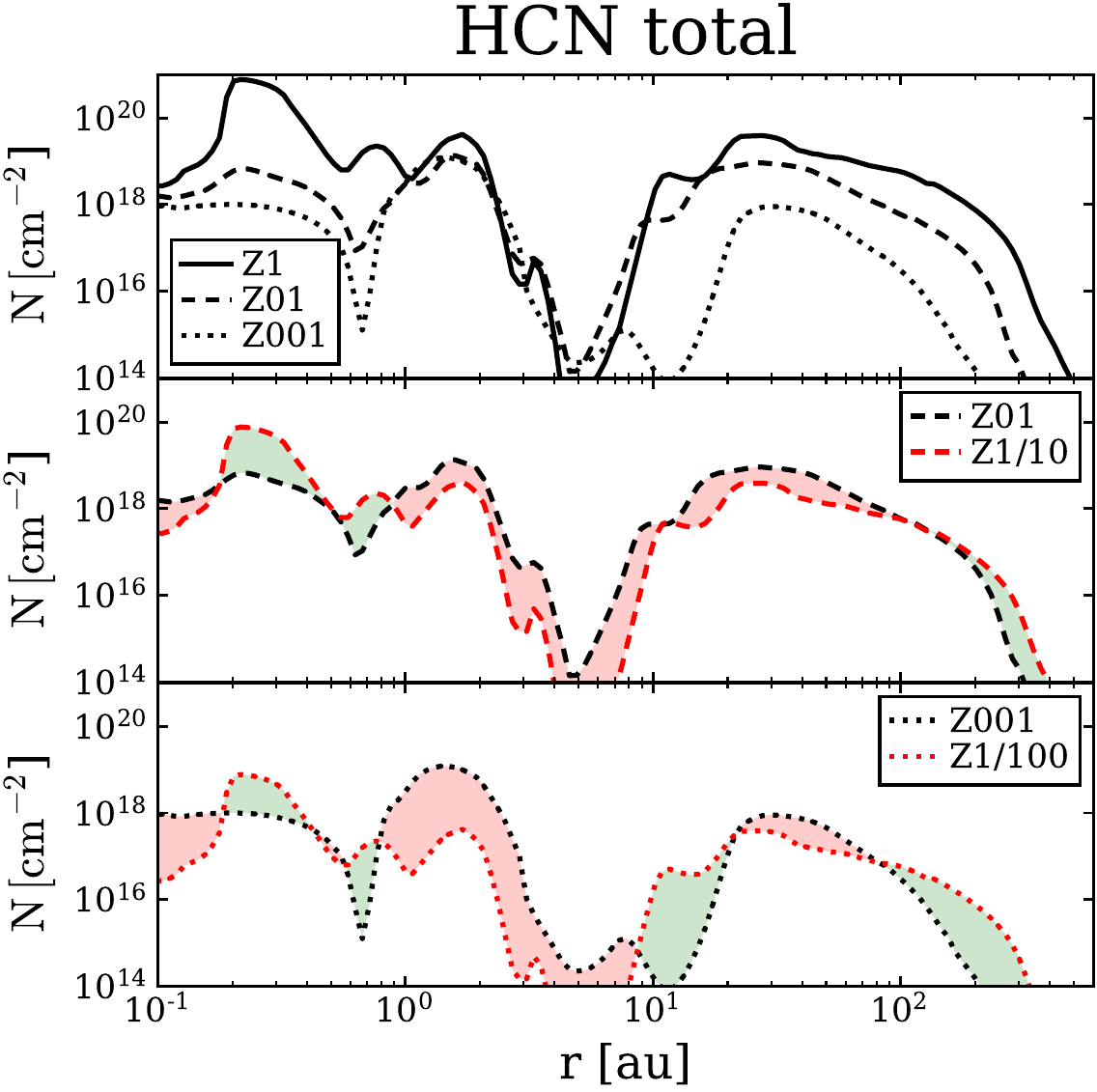}
    \end{subfigure}\hspace{0.02\textwidth}
    \begin{subfigure}{0.48\textwidth}
    \includegraphics[width=\textwidth]{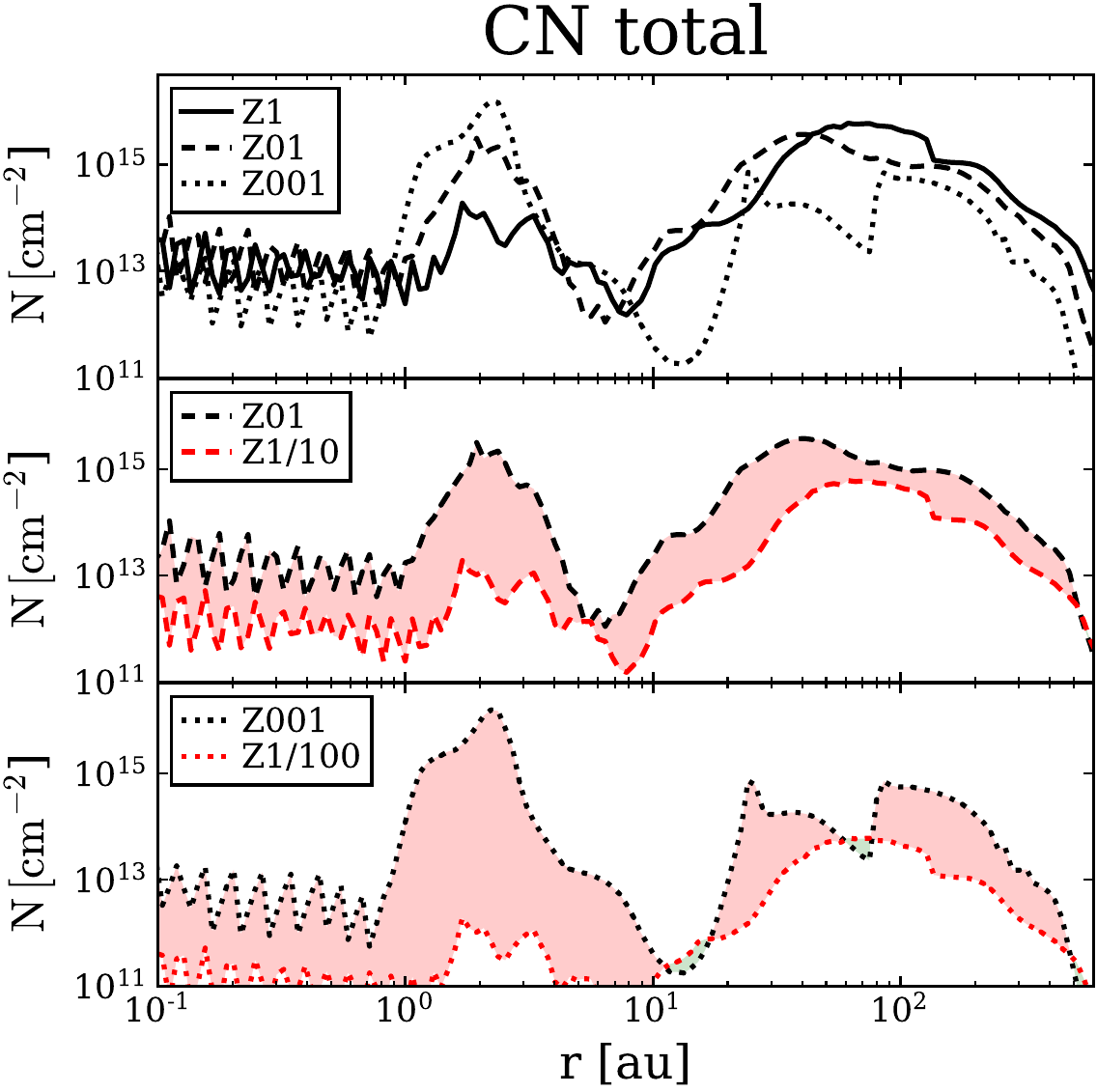}
    \end{subfigure}
    \caption{Same as Fig, \ref{fig:column_H2O} but for {CO}, CO$_{2}$, HCN and CN.}
    \label{fig:column_CO_CO_CH4_HCN}
\end{figure*}

CO shows a very similar trend to H$_2$O with the very important exception that the magnitudes of the differences to the scaled-down values are greater than for H$_2$O. For all three models, the total column density decreases with increasing radius. Z1 model has a strong bump for a radius between 0.1 and 1 au. This feature is absent or much less pronounced in the case of lower metallicity models. The reason for the existence of the mentioned bump is the different formation reactions of CO for the three different models. In that part of the disk, the reactant for the most important formation reaction of CO is gaseous H$_2$O. This reactant has an abundance that is orders of magnitude lower for models Z01 and Z001, so it is to be expected that CO will be much less abundant in that region for models {Z01} and Z001. This will be explained in more detail below in section \ref{det_bump}. As a result, CO is less abundant in the lower metallicity models than the scaled-down values in the vicinity of the bump.  
Apart from that bump, models Z01 and Z001 show a CO overabundance along the complete disk. This overabundance increases as the metallicity decreases. Table \ref{table:form_destr} shows that the formation reaction for CO in that region changes with decreasing metallicity and also that the respective reaction rate increases. For example, the most important formation reaction of CO in the Z1 model is dissociative recombination of H$_2$CO$^+$ while for the Z01 and Z001 models the formation reaction is the neutral-neutral reaction of H and HCO and photodissociation of CO$_2$, respectively. We can see in Fig. \ref{fig:integ_ratio} that CO is overabundant relative to the scaled-down values as a result of the enhanced formation reactions. Table \ref{table:integ_ratios} confirms this overabundance with ratios for the gaseous state of 4.78 and 16.92 for models Z01 and Z001 respectively. 
This is an important finding as CO is often used for observations. The resulting column density seen in Fig. \ref{fig:column_CO_CO_CH4_HCN} for models Z01 and Z001 does not show differences of order of magnitude compared to model Z1. Thus, if observations were performed on a sub-solar metallicity region, it would be incorrect to expect lower CO abundances than in the solar metallicity case. This assumption would then result in an underestimation of CO column density by an order of magnitude.

\subsubsection{CO$_2$}\label{cd_co2}
A similar bump to the CO case for small radii is also present here for model Z1. This bump is a consequence of the bump in the CO case. The origin of this bump will also be explained in more detail in Section \ref{det_bump}. Apart from that, all three models show a drop that reaches its minimum value at around 2 au. This drop is consistent with the location of the midplane cavity for this molecule. From 2 au on all models show an increase of the column density with radius until it starts to drop again towards the outer disk radius. All models show a similar increase that drops again at the end of the disk's radius. 
This decrease and rise of the total column density around 2 au follow the depletion of gas phase CO$_2$ due to ion-neutral reactions (model Z1 and Z01) and photodissociation (model Z001). On the other hand, CO$_2\#$ is also being depleted in that part of the disk due to photodissociation (model Z1) and photodesorption (models Z01 and Z001). It is only beyond that radius where CO$_2\#$ builds up and is responsible for the increase of the total column density.  
The comparison between the Z01 and Z001 case (middle and bottom panel for CO$_{2}$ in Fig. \ref{fig:column_CO_CO_CH4_HCN} shows an underabundance in the region where the bump is placed.
Between the already mentioned bump and the outer parts of the disk, CO$_{2}$ shows an overabundance. This overabundance present in the Z01 model decreases in size in model Z001. In the part of the disk with overabundance, we see again a change in the formation reaction type and in the reaction rate as metallicity decreases. For instance, the formation reaction changes from thermal desorption to photon desorption to the neutral-neutral reaction of OH and CO for models Z1, Z01, and Z001 respectively.  
There is an underabundance in the outer parts of the disk that becomes more noticeable and extends into the disk with decreasing metallicity. This increase in size and magnitude in the underabundance as the metallicity decreases leads to the resulting ratios in Fig. \ref{fig:integ_ratio} for CO$_2$ in the sense that while the Z01 model shows a net overabundance, the Z001 model has a net underabundance.

\subsubsection{HCN}\label{cd_hcn}
The trend shown in the top panel for HCN {in Fig. \ref{fig:column_CO_CO_CH4_HCN}} exhibits a bump in the Z1 model for small radii. Apart from that, all models have a depression that reaches a minimum value at around 6 au and then show an increase beyond that radius. The values for a radius smaller than 2 au are caused by the amount of gaseous HCN while the values beyond 9 au by the amount of HCN\#.
In the middle and bottom panel, models Z01 and Z001 show an overabundance for different parts of the disk (for very small radii, for a radius between 0.9 and 9, and for a region around 35 au). The region with the most overabundance is located around 1.8 au and Table \ref{table:form_destr} confirms the rising of the respective reaction rates as the metallicity is reduced. The formation reaction also changes. For models Z1 and Z01, it is the dissociative recombination of HCNH$^+$ and for model Z001 it is the neutral-neutral reaction of CN and C$_2$H$_2$. The outer regions of the disk show an underabundance that increases for the Z001 model. Then Z001 model also shows an underabundance between 9 and 20 au. This underabundance is the result of the depletion of HCN\# due to the greater values for the respective destruction reaction as metallicity decreases (green shaded rows for HCN\# in Table \ref{table:form_destr}. Although the underabundance seems to increase as the metallicity is reduced, we see in Fig. \ref{fig:integ_ratio} that the ratios to the scaled-down values are still greater than 1 for the complete disk.     

\begin{table*}[!ht]

\centering
\caption{Formation and destruction rates for the different species at different locations in the midplane of the disk for each model. The ice species are followed by a $\#$. The red and green shaded cells represent the reactions responsible for a relative overabundance (red) or underabundance (green).
The reaction rate is the reaction coefficient multiplied by the {number} densities of the reactants. The reaction type is given in the brackets under the reaction rate with the corresponding reaction equation number. The equations are given in appendix \ref{appendix:a}. The location was determined by choosing the point in the midplane where the differences between the column density of models Z01 and Z001 and the respective scaled-down column densities are the largest.} 
\begin{adjustbox}{width=1\textwidth}
\begin{tabular}{c|c|c|c|c|c|c|c}
\hline
\hline
species & location (x, z) {[}au{]} &
  \multicolumn{3}{c|}{main formation reaction rate [cm$^{-3}$ s$^{-1}$]} & \multicolumn{3}{c}{main destruction reaction rate [cm$^{-3}$ s$^{-1}$]} \\  & &Z1 & Z01 & Z001 & Z1 & Z01 & Z001   \\ \hline
 
H2O\# & (2.70, 0)&  \begin{tabular}[c]{@{}c@{}}4.35 x$10^{-14}$\\ (freeze out)\end{tabular} & \begin{tabular}[c]{@{}c@{}}5.19 x$10^{-14}$\\ (surface reaction)\end{tabular} &
    \begin{tabular}[c]{@{}c@{}}4.9 x$10^{-3}$\\ (surface reaction)\end{tabular} &
  \cellcolor[rgb]{0.592,1,0.592} \begin{tabular}[c]{@{}c@{}}4.37 x$10^{-15}$\\ (thermal desorption)\end{tabular} & \cellcolor[rgb]{0.592,1,0.592} \begin{tabular}[c]{@{}c@{}}4.79 x$10^{-14}$\\ (thermal desorption)\end{tabular} &
   \cellcolor[rgb]{0.592,1,0.592} \begin{tabular}[c]{@{}c@{}}4.92 x$10^{-3}$\\ (photodiss.,Eq.\ref{eq:d_H2O_Z01_Z001})\end{tabular} 
   \\ \hline

CH$_{4}$&  (1.54, 0)  & \cellcolor[rgb]{1,0.675,0.675} \begin{tabular}[c]{@{}c@{}}3.44 x$10^{-13}$\\ (diss. recombination, Eq.\ref{eq:f_CH4_Z01})\end{tabular} &
\cellcolor[rgb]{1,0.675,0.675} \begin{tabular}[c]{@{}c@{}}3.84 x$10^{-11}$\\ (diss. recombination, Eq.\ref{eq:f_CH4_Z01})\end{tabular} & 
\cellcolor[rgb]{1,0.675,0.675} \begin{tabular}[c]{@{}c@{}}1.21 x$10^{-7}$\\ (ion-neutral, Eq.\ref{eq:f_CH4_Z001})\end{tabular} & \begin{tabular}[c]{@{}c@{}}2.43 x$10^{-13}$\\ (ion-neutral)\end{tabular} & \begin{tabular}[c]{@{}c@{}}2.44 x$10^{-11}$\\ (ion-neutral)\end{tabular}  & \begin{tabular}[c]{@{}c@{}}9.98 x$10^{-8}$\\ (photodiss.)\end{tabular} \\ \hline
   
CO  &  (6.4, 0)  & \cellcolor[rgb]{1,0.675,0.675} \begin{tabular}[c]{@{}c@{}}7.41 x$10^{-25}$\\ (diss. recombination, Eq.\ref{eq:f_CO_Z1})\end{tabular} &
\cellcolor[rgb]{1,0.675,0.675} \begin{tabular}[c]{@{}c@{}}3.74 x$10^{-14}$\\ (neutral-neutral, Eq.\ref{eq:f_CO_Z01})\end{tabular}& 
\cellcolor[rgb]{1,0.675,0.675} \begin{tabular}[c]{@{}c@{}}1.31 x$10^{-5}$\\ (photodiss, Eq.\ref{eq:f_CO_Z001})\end{tabular}  & \begin{tabular}[c]{@{}c@{}}9.47 x$10^{-24}$\\ (ion-neutral)\end{tabular}  & \begin{tabular}[c]{@{}c@{}}8.31 x$10^{-13}$\\ (ion-neutral)\end{tabular}   & \begin{tabular}[c]{@{}c@{}}1.37 x$10^{-5}$\\ (neutral-neutral)\end{tabular}        \\  \hline

CO\#  & (27.6, 0)& \cellcolor[rgb]{1,0.675,0.675} \begin{tabular}[c]{@{}c@{}}4.36 x$10^{-21}$\\ (surface reaction, Eq.\ref{eq:f_COi_Z1} )\end{tabular} &
\cellcolor[rgb]{1,0.675,0.675} \begin{tabular}[c]{@{}c@{}}9.56 x$10^{-9}$\\ (freeze out)\end{tabular} &
\cellcolor[rgb]{1,0.675,0.675} \begin{tabular}[c]{@{}c@{}}3.76 x$10^{-7}$\\ (freeze out)\end{tabular} & \begin{tabular}[c]{@{}c@{}}4.40 x$10^{-21}$\\ (surface reaction)\end{tabular}    & \begin{tabular}[c]{@{}c@{}}1.82 x$10^{-8}$\\ (photodiss.)\end{tabular}      & \begin{tabular}[c]{@{}c@{}}3.90 x$10^{-7}$\\ (photodiss.)\end{tabular}       \\ \hline 

CO2   & (9.55, 0)& \cellcolor[rgb]{1,0.675,0.675} \begin{tabular}[c]{@{}c@{}}3.27 x$10^{-30}$\\ (thermal desorption)\end{tabular} &
\cellcolor[rgb]{1,0.675,0.675} \begin{tabular}[c]{@{}c@{}}2.74 x$10^{-14}$\\ (photon desorption)\end{tabular} &
\cellcolor[rgb]{1,0.675,0.675} \begin{tabular}[c]{@{}c@{}}1.32 x$10^{-6}$\\(neutral-neutral,Eq.\ref{eq:f_CO2_Z001})\end{tabular} &
\begin{tabular}[c]{@{}c@{}}2.72 x$10^{-30}$\\ (ion-neutral)\end{tabular}     &
\begin{tabular}[c]{@{}c@{}}2.77 x$10^{-14}$\\ (ion-neutral)\end{tabular}      & 
\begin{tabular}[c]{@{}c@{}}1.36 x$10^{-6}$\\ (photodiss.)\end{tabular}        \\ \hline

CO2\#    & (9.55, 0) & \cellcolor[rgb]{1,0.675,0.675} \begin{tabular}[c]{@{}c@{}}6.79 x$10^{-31}$\\ (freeze out)\end{tabular} &
\cellcolor[rgb]{1,0.675,0.675} \begin{tabular}[c]{@{}c@{}}7.20 x$10^{-16}$\\ (freeze out)\end{tabular} &
\cellcolor[rgb]{1,0.675,0.675} \begin{tabular}[c]{@{}c@{}}2.86 x$10^{-8}$\\ (freeze out)\end{tabular} & \begin{tabular}[c]{@{}c@{}}2.51 x$10^{-29}$\\ (photodiss.)\end{tabular}     &
\begin{tabular}[c]{@{}c@{}}2.74 x$10^{-14}$\\ (photodesorption)\end{tabular}      &
\begin{tabular}[c]{@{}c@{}}3.10 x$10^{-8}$\\ (photodesorption)\end{tabular}       \\ \hline

HCN   &    (1.81, 0) & \cellcolor[rgb]{1,0.675,0.675} \begin{tabular}[c]{@{}c@{}}1.89 x$10^{-12}$\\ (diss. recombination,Eq.\ref{eq:f_HCN_Z1_Z01})\end{tabular} &
\cellcolor[rgb]{1,0.675,0.675} \begin{tabular}[c]{@{}c@{}}3.47 x$10^{-11}$\\ (diss. recombination,Eq.\ref{eq:f_HCN_Z1_Z01})\end{tabular} &
\cellcolor[rgb]{1,0.675,0.675} \begin{tabular}[c]{@{}c@{}}2.49 x$10^{-4}$\\ (neutral-neutral,Eq.\ref{eq:f_HCN_Z001})\end{tabular} &
\begin{tabular}[c]{@{}c@{}}2.64 x$10^{-12}$\\ (ion-neutral)\end{tabular}    & 
\begin{tabular}[c]{@{}c@{}}4.93 x$10^{-11}$\\ (ion-neutral)\end{tabular}      &
\begin{tabular}[c]{@{}c@{}}1.89 x$10^{-4}$\\ (photodiss.)\end{tabular}   \\ \hline

HCN\# & (10.20, 0) &  \begin{tabular}[c]{@{}c@{}}1.94 x$10^{-22}$\\ (surface reaction)\end{tabular} &
\begin{tabular}[c]{@{}c@{}}6.77 x$10^{-12}$\\ (surface reaction)\end{tabular} &
\begin{tabular}[c]{@{}c@{}}9.34 x$10^{-8}$\\ (surface reaction)\end{tabular} &
\cellcolor[rgb]{0.592,1,0.592}  \begin{tabular}[c]{@{}c@{}}2.41 x$10^{-22}$\\ (thermal desorption)\end{tabular} & \cellcolor[rgb]{0.592,1,0.592} \begin{tabular}[c]{@{}c@{}}6.76 x$10^{-12}$\\ (photodiss., Eq.\ref{eq:d_HCN_Z001})\end{tabular} &
\cellcolor[rgb]{0.592,1,0.592} \begin{tabular}[c]{@{}c@{}}9.41 x$10^{-8}$\\ (photodiss., Eq.\ref{eq:d_HCN_Z001})\end{tabular} \\ \hline

CN  & (2.53, 0)& \cellcolor[rgb]{1,0.675,0.675} \begin{tabular}[c]{@{}c@{}}3.46 x$10^{-20}$\\ (diss. recombination, Eq.\ref{eq:f_CN_Z1_Z01})\end{tabular} &
\cellcolor[rgb]{1,0.675,0.675} \begin{tabular}[c]{@{}c@{}}1.45 x$10^{-12}$\\ (diss. recombination, Eq.\ref{eq:f_CN_Z1_Z01})\end{tabular} &
\cellcolor[rgb]{1,0.675,0.675} \begin{tabular}[c]{@{}c@{}}1.16 x$10^{-3}$\\ (photodiss., Eq.\ref{eq:f_CN_Z001})\end{tabular} & \begin{tabular}[c]{@{}c@{}}2.61 x$10^{-20}$\\ (neutral-neutral)\end{tabular} &
\begin{tabular}[c]{@{}c@{}}1.10 x$10^{-12}$\\ (neutral-neutral)\end{tabular} &
\begin{tabular}[c]{@{}c@{}}8.28 x$10^{-4}$\\ (neutral-neutraL)\end{tabular}    \\ \hline

CN\#  &  (95.36, 0)& \cellcolor[rgb]{0.592,1,0.592} \begin{tabular}[c]{@{}c@{}}9.77 x$10^{-14}$\\ (photodiss., Eq.\ref{eq:d_HCN_Z001})\end{tabular} &
\cellcolor[rgb]{0.592,1,0.592}  \begin{tabular}[c]{@{}c@{}}1.13 x$10^{-17}$\\ (photodiss., Eq.\ref{eq:d_HCN_Z001})\end{tabular} &
  \cellcolor[rgb]{0.592,1,0.592} \begin{tabular}[c]{@{}c@{}}7.80 x$10^{-21}$\\ (photodiss., Eq.\ref{eq:d_HCN_Z001})\end{tabular} &
  \begin{tabular}[c]{@{}c@{}}9.31 x$10^{-14}$\\ (surface reaction)\end{tabular} &
  \begin{tabular}[c]{@{}c@{}}7.48 x$10^{-18}$\\ (surface reaction)\end{tabular} &
  \begin{tabular}[c]{@{}c@{}}4.64 x$10^{-21}$\\ (surface reaction)\end{tabular} \\ \hline
  
HCO+ & (50.23, 0)& \cellcolor[rgb]{1,0.675,0.675} \begin{tabular}[c]{@{}c@{}}5.10 x$10^{-19}$\\ (ion-neutral, Eq.\ref{eq:f_HCO+_Z1})\end{tabular} &
\cellcolor[rgb]{1,0.675,0.675} \begin{tabular}[c]{@{}c@{}}1.70 x$10^{-8}$\\ (ion-neutral, Eq.\ref{eq:f_HCO+_Z01_Z001})\end{tabular} &
\cellcolor[rgb]{1,0.675,0.675} \begin{tabular}[c]{@{}c@{}}5.32 x$10^{-8}$\\ (ion-neutral, Eq.\ref{eq:f_HCO+_Z01_Z001})\end{tabular} &
\begin{tabular}[c]{@{}c@{}}1.47 x$10^{-18}$\\ (diss. recombination)\end{tabular} &
\begin{tabular}[c]{@{}c@{}}9.21 x$10^{-9}$\\ (diss. recombination)\end{tabular} &
\begin{tabular}[c]{@{}c@{}}4.66 x$10^{-8}$\\ (diss. recombination)\end{tabular}   \\ \hline

N2H+   & (10.36, 0)& \cellcolor[rgb]{1,0.675,0.675} \begin{tabular}[c]{@{}c@{}}5.21 x$10^{-47}$\\ (ion-neutral, Eq.\ref{eq:f_N2H+})\end{tabular} &
\cellcolor[rgb]{1,0.675,0.675} \begin{tabular}[c]{@{}c@{}}1.68 x$10^{-23}$\\ (ion-neutral, Eq.\ref{eq:f_N2H+})\end{tabular} & \cellcolor[rgb]{1,0.675,0.675} \begin{tabular}[c]{@{}c@{}}7.52 x$10^{-7}$\\ (ion-neutral, Eq.\ref{eq:f_N2H+})\end{tabular} &
\begin{tabular}[c]{@{}c@{}}4.85 x$10^{-47}$\\ (diss. recombination)\end{tabular}&
\begin{tabular}[c]{@{}c@{}}1.56 x$10^{-23}$\\ (diss. recombination)\end{tabular}&
\begin{tabular}[c]{@{}c@{}}6.93 x$10^{-7}$\\ (ion-neutral)\end{tabular}  \\ \hline
\end{tabular}
\end{adjustbox}
\label{table:form_destr}
\end{table*}

\subsubsection{CN}\label{cd_cn}
In the top panel that displays the total column density for CN {in Fig. \ref{fig:column_CO_CO_CH4_HCN} we see that none of the models have a bump for a small radius. However, an oscillating behavior for the three models (Z1, Z01, and Z001) is noticeable.} This is because most CN at a small radius is located in a very thin layer of the disk (see Fig.~\ref{fig:abuncont}). The disk model is not able to resolve this thin layer properly and the result is the oscillating behavior of the total column density. We note two important aspects regarding this oscillatory behavior: First, the values for the CN column density are orders of magnitude less than the other molecules presented in this study. Second, we produced models with higher resolution to check how this would influence the oscillations. We found out that the produced column densities of CN are almost the same as in Fig. \ref{fig:column_CO_CO_CH4_HCN}, but with weaker oscillations.   
Beyond 1 au CN behaves in a unique way compared to the other molecules. For a radius between 1 and 5 au the total column density increases by a couple of orders of magnitude as the metallicity decreases. This is the opposite impact that metallicity has on the total column density of the other molecules. It mirrors the clear increase of the CN abundance in the midplane (between a radius of 1 and 5 au) as metallicity decreases (see bottom row of the CN panels in Fig. \ref{fig:abuncont}). For the rest of the disk models, Z1 and Z01 have similar values, and model Z001 has lower values with a clear drop at 90 au.  
Apart from the drop at 90 au for the Z001 model, both models (Z01 and Z001) exhibit an overabundance throughout the complete disk. The region around 2.5 au shows the greatest overabundance and the reaction rates in Table \ref{table:form_destr} show a clear increase as the metallicity decreases. Models Z1 and Z01 have dissociative recombination of HCNH$^+$ as the main formation reaction and for model Z001 the respective reaction is photodissociation of the HCN molecule (eq. \ref{eq:f_CN_Z001}). The clear dominant overabundance along the disk confirms the ratios for CN shown in Fig. \ref{fig:integ_ratio}.

\subsubsection{Bump at small radii for CO, CO$_2$, and HCN}\label{det_bump}
 
The already mentioned bump at small radii for CO has the following origin:
In the location of the bump, the rate of the most important formation reaction of CO for model Z1 is greater than the main formation reaction for the Z01 and Z001 models by more than three orders of magnitude. This is mainly the result of the differences in the abundances of gaseous H$_{2}$O between the different models. The main formation reaction of CO in that disk region is the ion-neutral reaction of H$_2$O and HCO$^+$ (Eq. \ref{eq:f_COb_Z1}). Gas-phase H$_{2}$O is the reactant of the formation reaction for CO and is much more abundant in the Z1 model in that particular region. Therefore, the formation of CO with the mentioned ion-neutral reaction is much more efficient for the Z1 model relative to models Z01 and Z001. Similarly, in the case of CO$_{2}$, the bump is the result of the greater CO abundance of the Z1 model. CO is the reactant for the main formation reaction of CO$_2$. Therefore a greater CO abundance leads to a greater CO$_2$ abundance.
For HCN all three models have the same main formation reaction. It is the ion-neutral reaction of NH$_{3}$ with HCNH$^{+}$ producing HCN and NH$_{4}^{+}$. Although all the models have the same main formation reaction, the bump is the result of the different reaction rates each model has.

\subsubsection{HCO$^{+}$ \emph{and} N$_{2}$H$^{+}$}\label{cd_hco+_n2h+}

\begin{figure}[!h]
       \centering
    \begin{subfigure}{0.24\textwidth}
    \includegraphics[width=\textwidth]{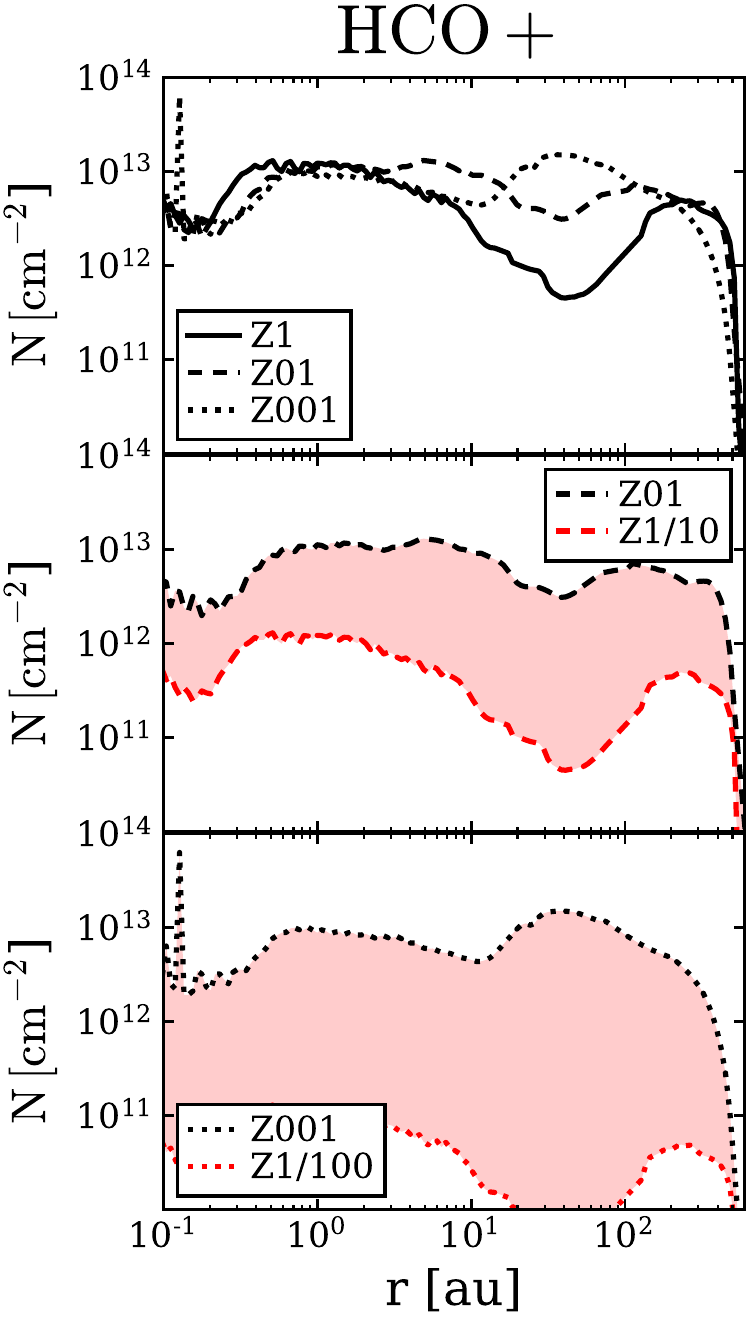}
    \end{subfigure}
    \begin{subfigure}{0.24\textwidth}
    \includegraphics[width=\textwidth]{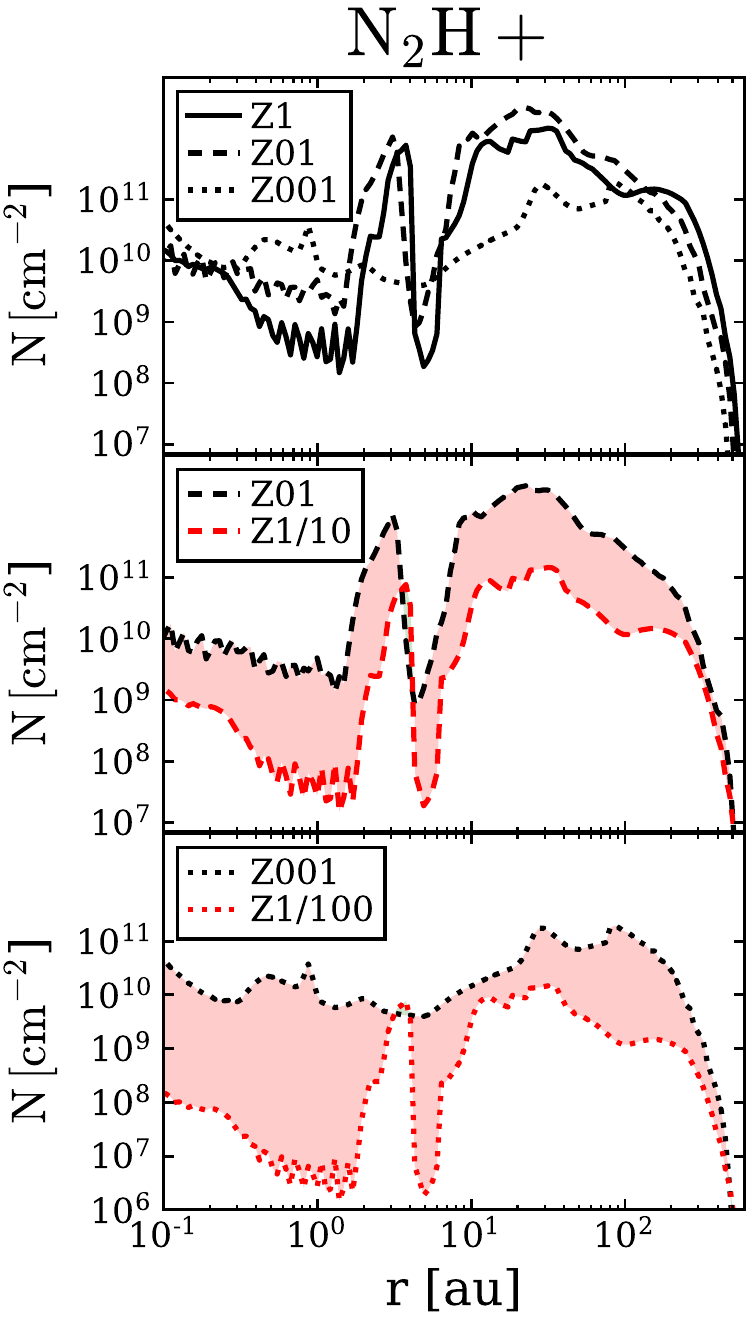}
    \end{subfigure}
    
    \caption{Same as Fig. \ref{fig:column_H2O} but for HCO$^{+}$ and N$ _{2}$H$^+$.}
    \label{fig:column_HCO+_N2H+}
\end{figure}

HCO$^{+}$ and N$_{2}$H$^{+}$ were put in this section together because they are electrically charged and are only present in the gas phase. The column density of the two ions N$_2$H$^+$ and HCO$^+$ is displayed in Fig.~\ref{fig:column_HCO+_N2H+}. The column density of N$_2$H$^+$ present an oscillating behavior similar to CN for a small radius. {As with CN, higher resolution models show that the oscillations become weaker as the resolution of the model increases.} This effect is also the result of a very thin layer in the inner parts of the disk which cannot be properly resolved in the model. In the top panel, HCO$^+$ shows similar values for the three models until a radius of 2 au is reached. Beyond that radius, models Z01 and Z001 have clearly higher total column densities than model Z1. This is expected because of the stronger radiation field that results in a more efficient ionization in the lower metallicity models. N$_{2}$H$^{+}$ shows a strong bump between 1 and 5 au and a following increase for models Z1 and Z01. Model Z001 does not have these strong features but it increases slightly as the radius increases. HCO$^{+}$ and N$_{2}$H$^{+}$ are more abundant than the scaled-down values for all radii. In fact, the overabundance is almost everywhere a couple of orders of magnitudes in both lower metallicity models (Z01 and Z001). Table \ref{table:form_destr} shows that for HCO$^{+}$ the respective ion-neutral formation reaction changes from H$^+$ + H$_2$CO for model Z1 to H$^+_3$ + CO for models Z01 and Z001. For N$_2$H$^+$ the ion-neutral formation reaction H$^+_3$ + N$_2$ does not change with metallicity but the rate of the reaction has a dramatic rise. The destruction reaction for the two ions is dissociative recombination (Table \ref{table:form_destr}). The electrons needed for this reaction usually come from the ionization of metals. However, as metallicity decreases the number of available metals is also reduced. The reduction of metals has a stronger impact on the destruction rate of HCO$^{+}$ and N$_{2}$H$^{+}$ than a stronger radiation field. {This is shown in Appendix \ref{appendix:b} where the reduction of the metals and the dust reduction are handled separately.} Eventually, this leads to an overabundance of the two ions. This general overabundance of both ions is displayed in a very clear way in the respective ratios shown in Fig. \ref{fig:integ_ratio}.
All the chemical reactions mentioned in this section are shown in detail in Table \ref{tab:react_eq} in Appendix \ref{appendix:a}. 

\subsubsection{General trend for column densities and quantitative analysis}

In most cases, the decreased metallicity leads to a decrease in the column density in comparison to the Z1 model. However, the actual reduction does not follow the simple reduction of the abundances by the factors 10 and 100.
It is important to reiterate that the over- or underabundances of the species are not only the result of different rates of adsorption and desorption of each species. The different trends of the column density for the lower metallicity models are also the result of surface reactions on the dust grains, photodissociation in the gaseous and ice phase, and recombination processes. The different values for the formation and destruction rates of the species are triggered by the higher temperatures and the stronger radiation field as the metallicity decreases.

{In Table \ref{table:integ_ratios} we summarize the quantitative analysis of this comparison for each of the considered species. {For each species, the vertical column density $N_{sp}$ is defined as   
\begin{equation}\label{vcoldens}
N_{sp}(r) =\int_{0}^{z_{max}(r)} n_{sp}(r,z) \,dz    
\end{equation}
with $z$ as the disk height, $z_{max}$ as the maximal disk height, $r$ as the radius, and $n_{sp}$ as the number density of a chemical species at each point of the disk.}   
For each species, we integrated the vertical column density $N_{sp}$ over the radius $r$ of models Z01 and Z001 and their scaled-down counterparts. 
We used a simple integration from the inner radius $R_{in}$ to the outer radius $R_{out}$ of the disk for this step (Eq. \ref{integ}).} 
{\begin{equation}\label{integ}
n_{tot}=2\pi \int_{R_{in}}^{R_{out}} N_{sp}r \,dr      
\end{equation}
\begin{equation}\label{ratio}
\mathcal{R}= n_{tot}/n_{tot(sd)}
\end{equation}}

\begin{figure}[!h]
       \centering
    \begin{subfigure}{0.49\textwidth}
    \includegraphics[width=\textwidth]{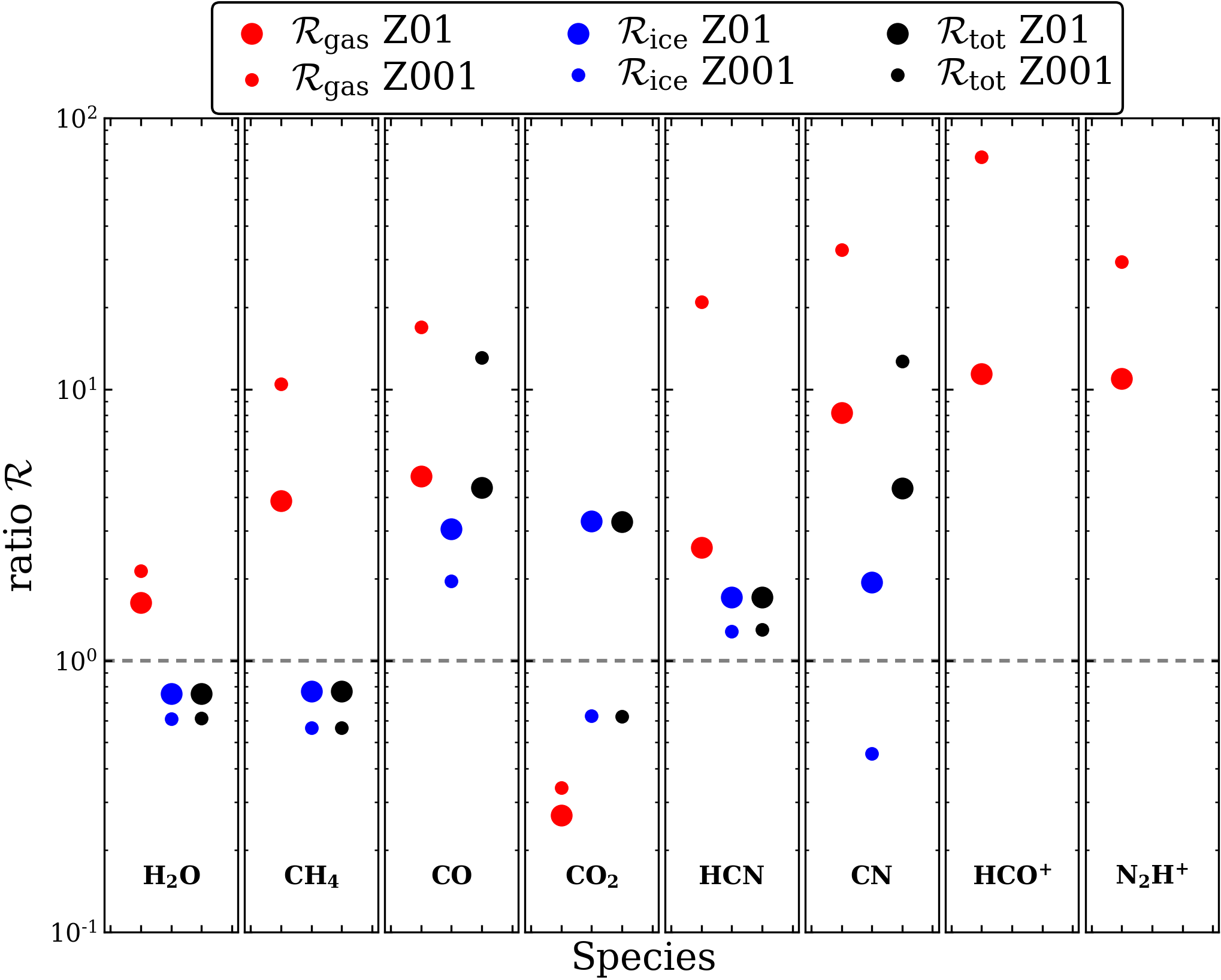}
    \end{subfigure}
    \caption{Ratio $\mathcal{R}$ (see Eq. \ref{ratio}) of the total molecular amount of the scaled-down Z1 and the low metallicity models Z01 and Z001 for each species. The red, blue, and black circles correspond to the ratios of the gaseous, ice, and total (gas + ice) species respectively. The dashed horizontal line shows the value for the ratio equal to 1.}
    \label{fig:integ_ratio}
\end{figure}

 Next, we calculated the ratio of {the total molecular amount} of models Z01 and Z001 (n$_{tot}$) relative to the scaled-down results (n$_{tot(sd)}$). The ratio is calculated with Eq. \ref{ratio}. These results are also presented in Fig.~\ref{fig:integ_ratio}.
The ratio is higher than 1 in almost all gaseous species. It is particularly high for the ion-species HCO$^{+}$ and N$_2$H$^{+}$. The exception of gaseous CO$_2$ is due to the fact that the underabundance in the disk zone closest to the star is also the zone where most part of CO$_2$ is stored. However, all gaseous species show that the value for the ratio has a positive trend as the metallicity decreases.

\begin{table}[!h]
\begin{center}
\caption{Comparison of the total particle number of the considered gaseous species (first column) for the model abundances and scaled-down abundances. The second and third column is the ratio relative to the scaled-down of the gaseous species ($\mathcal{R}_{\rm{gas}}$) for the Z01 and the Z001 model respectively. The fourth and fifth columns correspond to the ice species ($\mathcal{R}_{\rm{ice(\#)}}$) and the sixth and seventh columns represent the total (gas + ice) values ($\mathcal{R}_{\rm{tot}}$).}
\label{table:integ_ratios}
\begin{tabularx}{0.49\textwidth}{c|*{2}{Y}|*{2}{Y}|*{2}{Y}}
    \hline
   species &
  \multicolumn{2}{c|}{$\mathcal{R}$$_{\rm{gas}}$} &
  \multicolumn{2}{c|}{$\mathcal{R}$$_{\rm{ice}(\#)}$}  & 
  \multicolumn{2}{c}{$\mathcal{R}$$_{\rm{tot}}$}\\ 
      & Z01 & Z001 & Z01 & Z001 & Z01 & Z001  \\
    \hline
    \hline

H$_2$O & 1.63 & 2.14 & 0.75 & 0.61 & 0.76 & 0.61\\
CH$_4$ & 3.67 & 10.42 & 0.77 & 0.56 & 0.77 & 0.56\\
CO  & 4.78  & 16.92 & 3.05 & 1.96 & 4.34 & 13.06   \\
CO$_2$   & 0.27 & 0.34 & 3.25 & 0.62 & 3.24 & 0.62 \\
HCN & 2.61 & 20.97  & 1.71 & 1.28 & 1.71 & 1.30\\
CN  & 8.20    & 32.57 & 1.94  & 0.45 & 4.32 & 12.65 \\

\hline
HCO$^{+}$  & 11.39    & 71.83  & & \\
N$_2$H$^{+}$  & 10.92    & 29.51  & & \\
\hline
\end{tabularx}
\end{center}
\end{table}

The ice species show a smaller than 1 ratio for H$_2$O$\#$ and CH$_4\#$ which decreases as the metallicity decreases. The rest of the ice species show a higher than 1 ratio for the Z01 model which decreases for the lowest metallicity model Z001. CO$_2\#$ and CN$\#$ even exhibit a smaller than 1 ratio for the Z001 model. All ice species show a negative trend as the metallicity decreases. This indicates that although there is still chemical production of certain ice species, the destruction due to photodesorption or thermal desorption of these species becomes more dominant with a decreasing metallicity. The total (gas + ice) values show a smaller than 1 ratio for H$_2$O and CH$_4$ with a negative trend with decreasing metallicity. CO and CN show a greater than 1 ratio with about three times higher ratios for the lowest metallicity Z001. CO$_2$ and HCN have greater than 1 ratios for the Z01 model but exhibit a negative trend as the metallicity decreases. CO$_2$ even shows a smaller than 1 ratio for the Z001 model. In summary, the strongest deviation from the scaled-down values for gaseous species is found in the ions HCO$^+$ and N$_2$H$^+$, followed by CN, HCN, CO, and CH$_4$. H$_2$O shows a slight deviation and CO$_2$ even shows smaller particle numbers than the scaled-down values. For the frozen species CN, CO, and CO$_2$ exhibit the strongest deviation from scaled-down values. Regarding the total (gas + ice) numbers, CO and CN are the molecules that have the strongest deviation.

\begin{figure*}[h!]
  \centering
  \begin{subfigure}[b]{0.8\textwidth}
    \includegraphics[width=\textwidth]{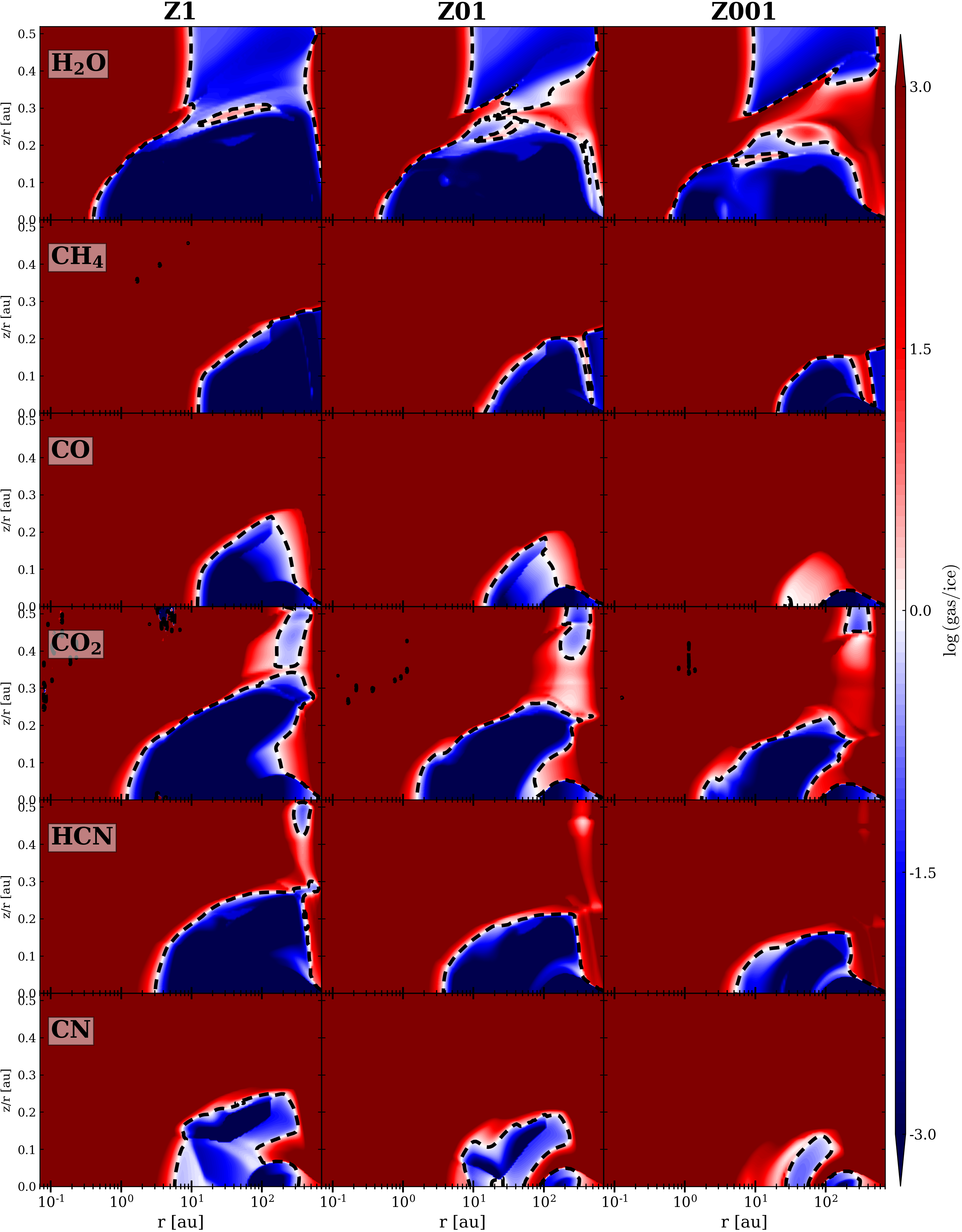}
  \end{subfigure}
  \caption{Gas-to-ice ratio of the molecules. The left, middle and right columns represent the Z1, Z01, and Z001 models respectively. Red and blue regions represent the regions where gas and ice are dominant respectively. The black dashed line shows where the ice and gas abundances are equal (snowline). The locations of the snowlines at the midplane are summarized in Fig.\ref{fig:snowline_loc}.}
  \label{fig:snowline}
\end{figure*}

\begin{figure}[!h]
       \centering
    \begin{subfigure}{0.49\textwidth}
    \includegraphics[width=\textwidth]{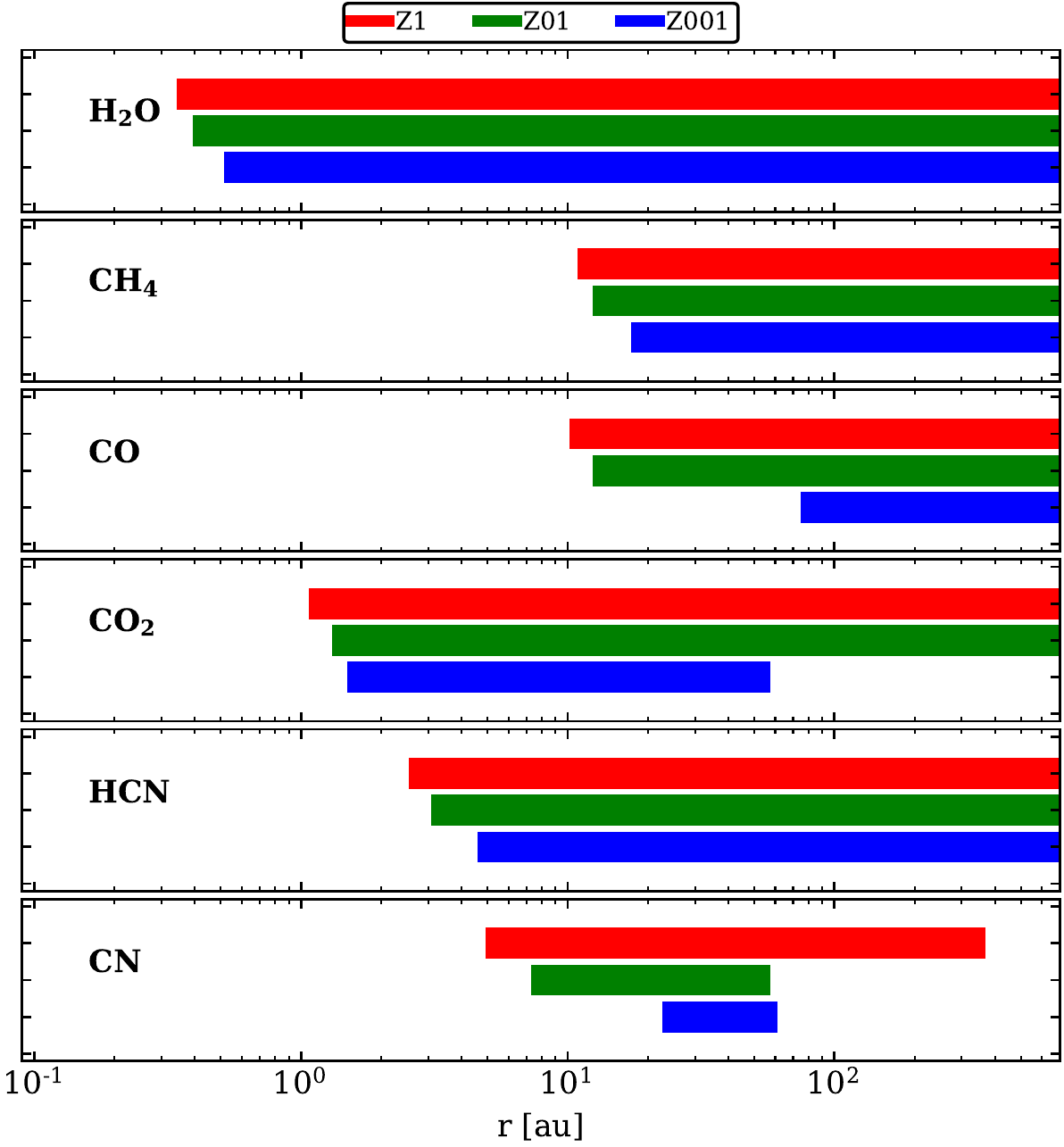}
    \end{subfigure}
    \caption{Snow regions at the midplane for the Z1(red), Z01(green), and Z001(blue) models for all neutral species. {The horizontal bars show the region in the midplane where the abundances of the ice species are larger than the abundances of the gaseous species. These regions are enclosed by the inner and outer snowline.}}
    \label{fig:snowline_loc}
\end{figure}

\subsection{Snowline for each molecule}\label{ssec:Detection of the snowline}

Snowlines are the regions in protoplanetary disks where the temperature reaches the sublimation point for a particular volatile species. The position of the snowline is of great importance to the formation and basic architecture of a planetary system. It plays an important role in the formation and chemical composition of planetesimals, and therefore in the formation and composition of planets, e.g. \citep{2011ApJ...743L..16O,2016ApJ...821...82O,2015ApJ...815L..15B}  
The temperature in a protoplanetary disk has a radial and a vertical gradient. Therefore the snowline is not a single radius but rather a two-dimensional surface. We show the location of this surface by displaying the regions in the disk where the chemical species are more abundant in the gas phase (red area) or the ice phase (blue area) (see Fig. \ref{fig:snowline}). The white regions represent a similar abundance of ice and gas and the black dashed line represents the location in the disk where gas and ice abundances are equal. Each row in the figure represents a different molecule and each column stands for a different metallicity model.
 As a result of the radial and vertical temperature gradient in the disk, a snow region is enclosed spatially by the two-dimensional surface where ice and gas abundances are equal. We further define the intersection of this two-dimensional surface with the midplane as the snowline. The innermost intersection is defined as the inner snowline and the next intersection as the outer snowline. For most of the molecules studied here the outer snowline is not present. However, there are some cases where the outer snowline lies inside the disk and we speak of an ice ring inside the disk.
 
 From Figure \ref{fig:snowline} it is clear that as the metallicity decreases, the snow region shrinks in its vertical and radial extension for all molecules, and in some cases, the snow region almost completely vanishes (e.g. CO$_2$ and CN). This is the result of the general increase in the disk's temperature caused by the stronger radiation field in the disk.
Fig.~\ref{fig:snowline_loc} summarizes the extension of the snow region and the location of the respective snowline at the midplane for each molecule for all models. The comparison between the models shows an influence of the metallicity on the location of the snowlines for all molecules. All the inner snowlines of each species are pushed outwards with decreasing metallicity. In some cases, the decrease in metallicity moves the outer snowline inside the disk so that an ice ring inside the disk is present. 
Two of the species (CO$_{2}$ and CN) used for this section exhibit the presence of ice rings inside the disk. Figs.~\ref{fig:snowline} and ~\ref{fig:snowline_loc} show that these ice rings shrink with decreasing metallicity (CO$_2$ and CN in particular).  

The effect of metallicity on the inner (outwards shift) and outer (inwards shift) snowlines is an indication that reducing the metallicity leads to a higher dust temperature (which is the temperature that determines if thermal desorption of frozen species from dust grains takes place) and an enhanced photo-desorption of ice species due to a stronger radiation field in the disk. Additionally, the fact that lower metallicity models have less amount of dust means that the gas phase molecules have less surface available to freeze out.
 All this favors the depletion of ice species and, consequently, causes the ice rings to shrink as the metallicity decreases.
 
 As a relevant example for planet formation, the snowline for H$_2$O (0.39, 0.45, and 0.59 au for models Z1, Z01, and Z001 respectively) does not change dramatically with metallicity. {Thus, the position of the H$_2$O snowline on planet formation would not be affected by metallicity changes.} Another snowline that does not exhibit a strong change with metallicity is the CO$_2$ snowline (1.22, 1.49, and 1.70 au). In the case of the CO snowline, the impact of metallicity is more noticeable (11.64, 14.21, and 85.45 au). In particular, for the Z001 model, the CO snowline is being pushed outwards to 85 au. The location of the snowlines for these three molecules has special significance in the formation and composition of planets, asteroids, and comets in a protoplanetary disk as these molecules are among the main carriers of C and O. Whether the atmosphere of a planet or a comet is carbon-rich or oxygen-rich, and whether it will be rich in more complex molecules, will depend on radial variations in disk midplane composition, and specifically the locations of the snowlines \citep{2016ApJ...823L..10W}.
 As CO is considered to be the starting point for the formation of many complex organic molecules (COMs), the metallicity will certainly impact the location in the disk where COMs are more likely to be found and therefore affect the chemical composition of the planets, asteroids or comets that are formed at those radii.

\section{Limitations of the model and outlook for future studies}\label{Limitations of model}

 In this section, we briefly discuss certain limitations of our model and their possible consequences.
In section \ref{ssec:gas disk structure} we introduce the physical structure of our model. We note that the column density power index \pmb{$\lambda$} of our models shown in Eq.(\ref{eq:sigma}) depends on the mass transport mechanisms operating in the disk, which in turn may depend on metallicity \citep{2013ApJ...775...68D}. This could lead to different timescales for accretion processes and the dissipation of the disk. In this study, we neglect these possible complications and assume that the disk structure is identical in the considered metallicity range. 
  
We use a fixed stellar spectrum for all the models. In order to provide a more realistic model, we should take into account that the stellar spectrum changes with a different metallicity. This limitation is due to the fact that the model uses stellar spectra with solar metallicity by default. A consequence of this simplification is that the effective temperature of the star is underestimated as stars with lower metallicity tend to be hotter.

For this work, we focused on the behavior of certain molecules. We chose this set of molecules because of their relevance in dust growth, planetary atmospheric composition, and importance for observations. The inclusion of more complex organic molecules in the analysis will be addressed in future papers.  

For this study, we assumed steady-state chemistry. Although this does not affect the consistency of our results, time-dependent chemistry becomes relevant when studying the impact of processes (e.g. accretion burst) that have a strong influence on the disk. {In a future study, we intend to include the impact of metallicity and accretion bursts in the chemistry of the disk.}

\begin{figure}
    \centering
    \begin{subfigure}{0.45\textwidth}
    \includegraphics[width=\textwidth]{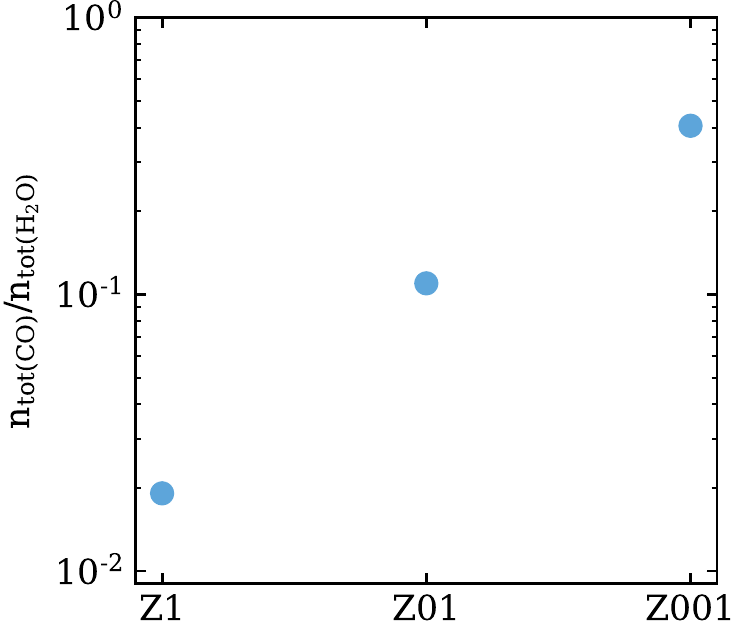}
    \end{subfigure}
    \caption{CO/H$_2$O ratio for models with different metallicity}
    \label{fig:co_h2o}
\end{figure}

We also note the particular case of CO. This molecule is very relevant because of its importance as a gas mass tracer (\cite{2017ApJ...849..130M}). Two very important conditions that allow the usage of CO as a gas mass tracer are that its emission has to be optically thin and it cannot be too strongly depleted in the gaseous state. {Our results suggest that changing the metallicity does not strongly affect the total abundance (gas + ice) of CO (relative to the response of other molecules and excluding the bump inside 1 au).} This would mean that CO-based observations would not be able to differentiate disks with different metallicity. This could be linked to the lower water abundance for lower metallicity disks and would suggest that low metallicity disks have a higher CO/H$_2$O ratio compared to "normal" disks (see Fig. \ref{fig:co_h2o}). To investigate the impact on observables requires proper modeling of line emission of the species in question, which is out of the scope of this paper but is planned in a future study.
 
  In future studies, we will be looking more deeply at different aspects of the impact of metallicity on protoplanetary disks. In particular, the effects of time-dependent chemistry in the context of an accretion burst, the implementation of a more realistic stellar spectrum, and dust size distribution in the disk. The behavior of more complex organic species and the modeling of their observation will be studied as well.
 
\section{Summary and conclusions}\label{Summary and conclusions}

We started this study to look into the impact of lower metallicities on certain chemical species in the protoplanetary disk. We used the radiation thermo-chemical code for protoplanetary disks \prodimo to create a model with a reference metallicity (Z1) and two models with lower metallicities by a factor of 10 (Z01) and 100 (Z001). We simulated the lower metallicities by decreasing the dust-to-gas ratio in the disk by one and two orders of magnitude and by decreasing the initial element abundances by the same factor in the initial setup of the models respectively. 
We analyzed the impact of metallicity on the chemical species H$_{2}$O, CO,  CO$_{2}$, CH$_{4}$, CN, HCN, HCO+, and N$_{2}$H+ because they are often used in observations as useful tracers of different disk properties. The findings in this work could therefore give useful information for future observations in low metallicity environments.  
One of the most important aspects of this study is the comparison of the lower metallicity model's vertical column density of chemical species with the reference metallicity vertical column density scaled-down by factors of 10 and 100.
We defined the notion of over-and underabundance of the models whenever the lower metallicity models show greater or smaller values for the vertical column density than the scaled-down vertical column density. We analyze the chemical formation and destruction reactions in the midplane wherever the over-or underabundance is strongest. 
Our main findings are: 

\begin{itemize}
    \item The effects shown in the lower metallicity models cannot be explained by a simple scaled-down of the abundances of the chemical species. The obtained values for the abundances of the Z01 and Z001 models differ from the scaled-down values almost everywhere in the disk.  This is because the lower metallicity models allow the radiation field to penetrate deeper regions of the disk. The temperature also increases in the regions where the radiation is stronger than in the Z1 model. This affects the formation and destruction reactions of the species.
     
    \item The main effect of a stronger radiation field appears to be the enhancement of the different processes responsible for the destruction of ice species and the formation of gas species and the shifts in the type of destruction and formation reactions. Thus, the stronger radiation field causes a substantial increase in the formation efficiencies of gas-phase species. This is clearly shown in Fig. \ref{fig:integ_ratio} where the ratios between the scaled-down values and the lower metallicity values of all gaseous species (red dots) and the ice species (blue dots) are displayed. Furthermore, the impact of metallicity on the chemical reactions is different for each individual species. In some cases, the reaction type changes with metallicity, and in other cases the type of reaction does not change with metallicity but the rate of the reaction is clearly affected. This is shown in detail in Table \ref{table:form_destr} where the different main reactions and the respective rates of the three models are compared.
    
    \item Our results show that CO shows a greater abundance than {the scaled-down values} for lower metallicities. This molecule shows a clear non-linear response to the change in metallicity. Beyond the bump at small radii, metallicity does not seem to have a strong influence on the column density of CO. This finding has very relevant implications for observations as CO would not be a useful indicator for the metallicity of the disk.
    
    \item The impact of metallicity on the position of the snowlines is also present. With decreasing metallicity the snowline of the molecules analyzed in this study is "pushed" outwards. Additionally, decreasing the metallicity leads to the creation of ice rings for some species (e.g. CO$_2$) and to the shrinking of already existing ice rings for other species (e.g. CN). 
\end{itemize}

The inclusion of other metallicity-dependent features will be performed in future studies (star spectrum and dust size distribution among others). The analysis of other scenarios (e.g. accretion bursts) will also help us gain more insight into the impact of metallicity in protoplanetary disks.

\begin{acknowledgements}
We would like to express our gratitude to the anonymous referee for the valuable comments that helped us improve the presentation of this work. The computational results presented have been achieved using the Vienna Scientific Cluster (VSC). This work was supported by the Austrian Science Fund (FWF) under research grant P31635-N27. Ch. Rab is grateful for support from the Max Planck Society and acknowledges funding by the Deutsche Forschungsgemeinschaft (DFG, German Research Foundation) - 325594231.

\end{acknowledgements}

%
%

\bibliographystyle{aa}
\bibliography{Lowmetpaper}

\appendix

\section{Detailed chemical reactions and display of the ice and gaseous column densities}
\label{appendix:a}
\begin{figure*}[!h]

  \centering
  \begin{subfigure}[b]{0.49\textwidth}
    \includegraphics[width=\textwidth]{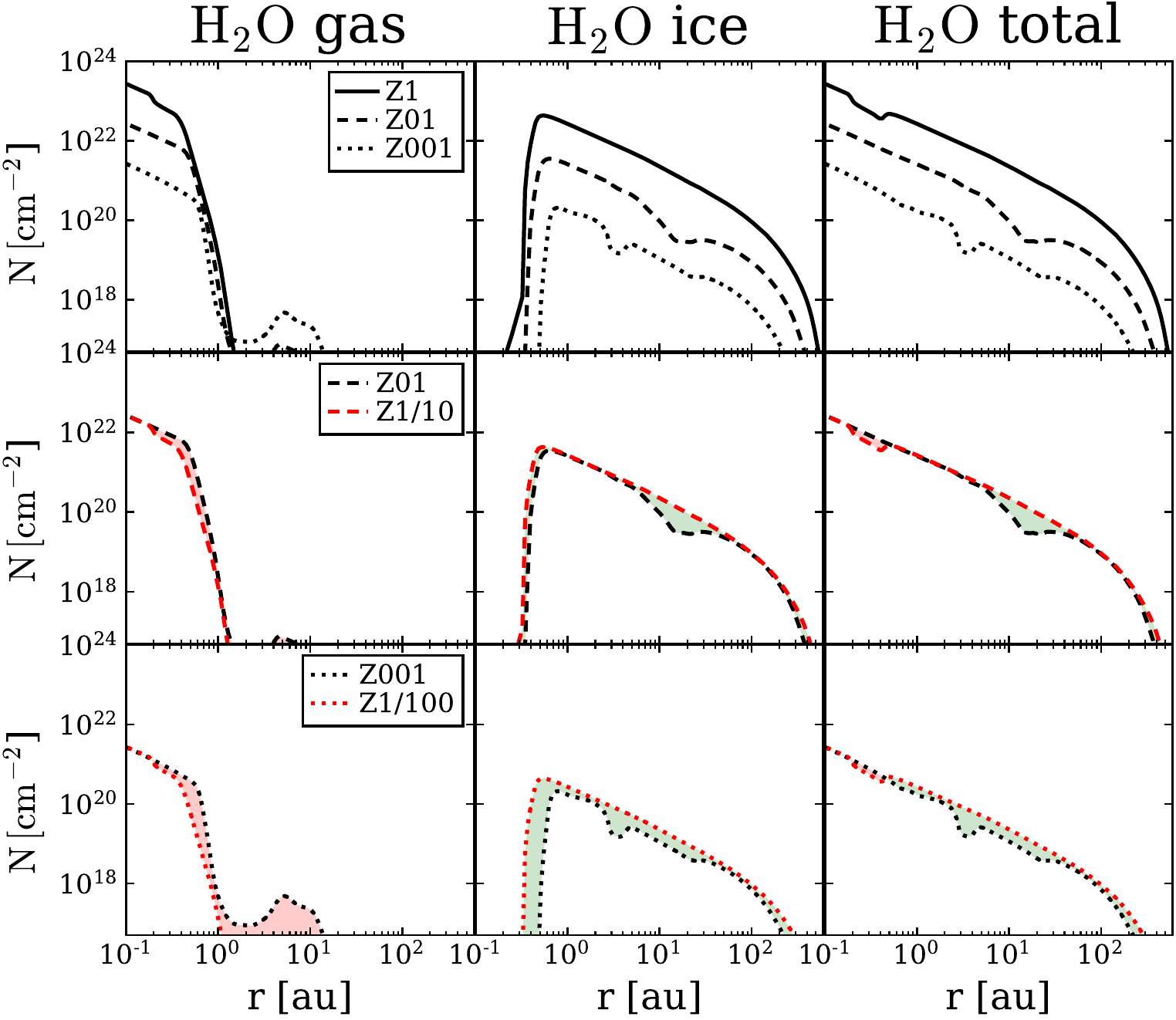}
    \end{subfigure}
\begin{subfigure}[b]{0.49\textwidth}
    \includegraphics[width=\textwidth]{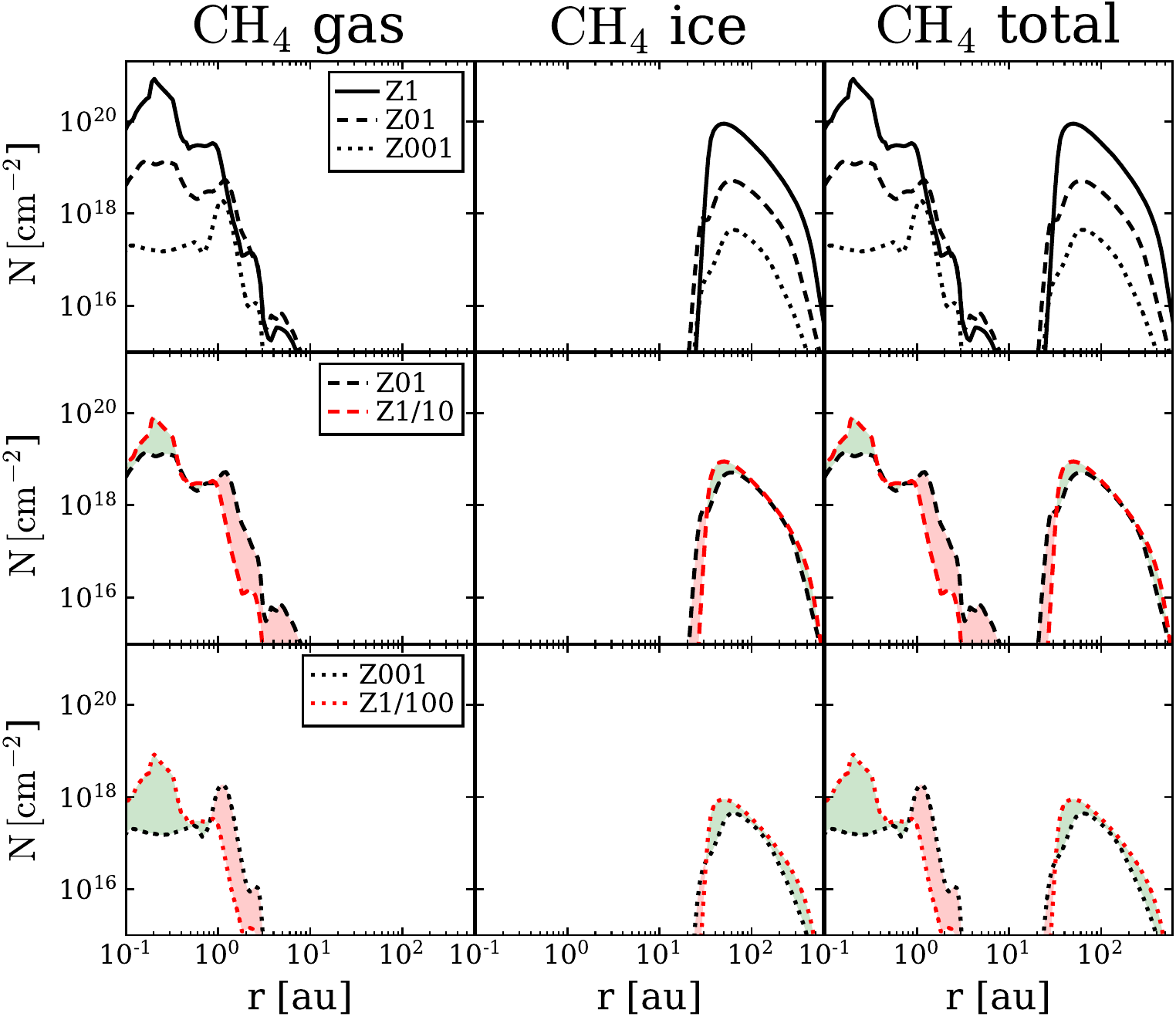}
  \end{subfigure}
\begin{subfigure}[b]{0.49\textwidth}
    \includegraphics[width=\textwidth]{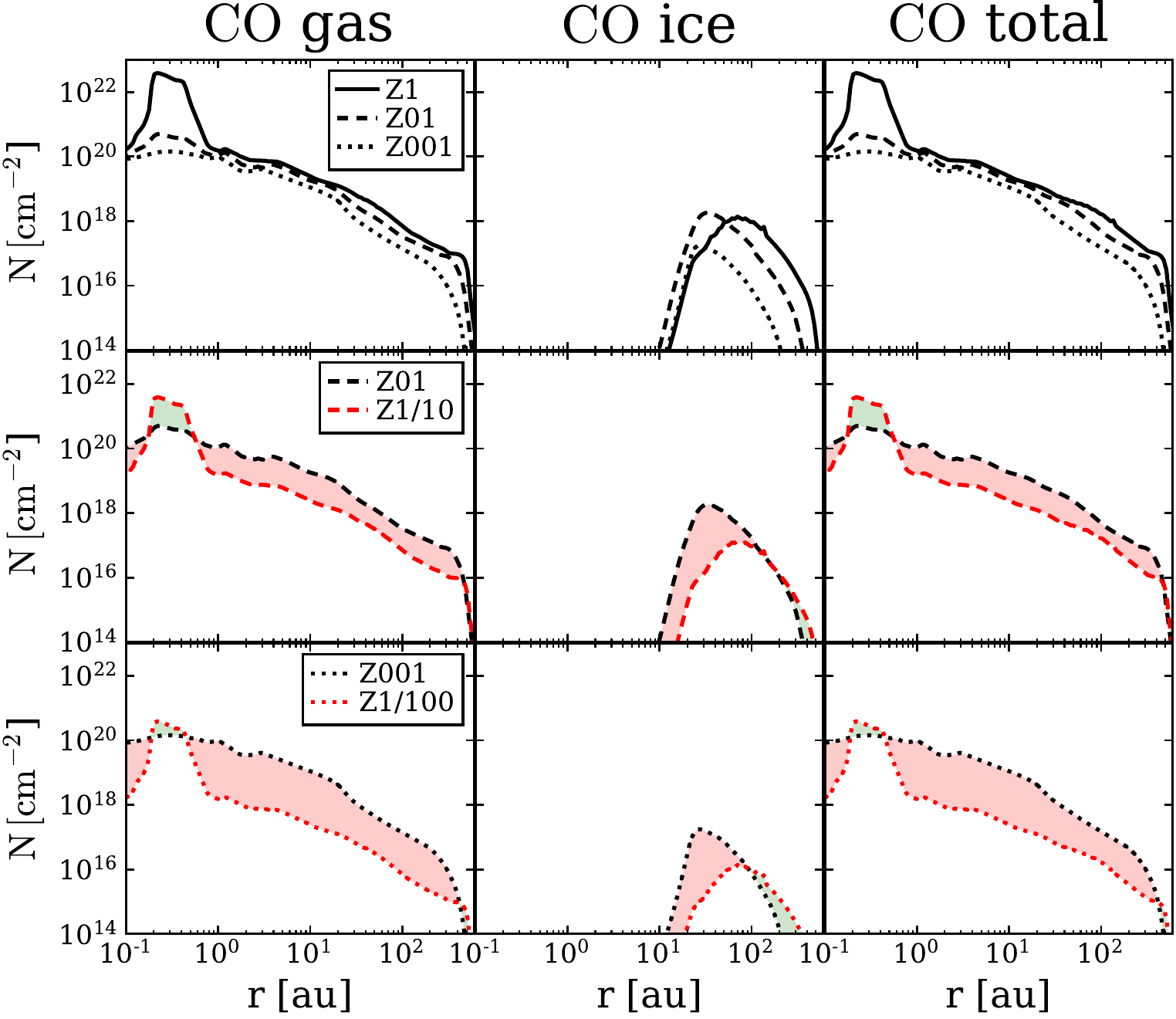}
  \end{subfigure}
\begin{subfigure}[b]{0.49\textwidth}
    \includegraphics[width=\textwidth]{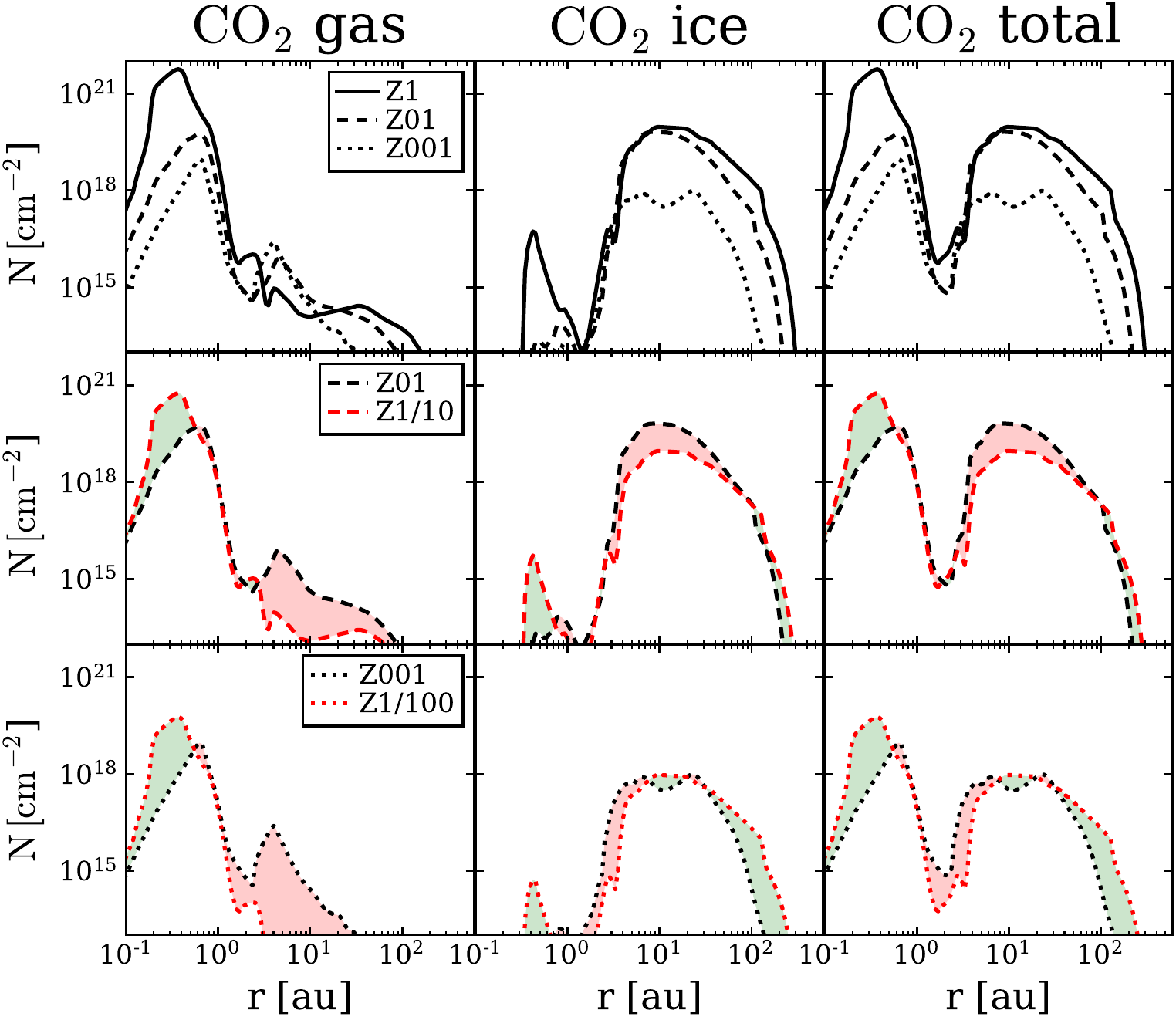}
  \end{subfigure}
  \begin{subfigure}[b]{0.49\textwidth}
    \includegraphics[width=\textwidth]{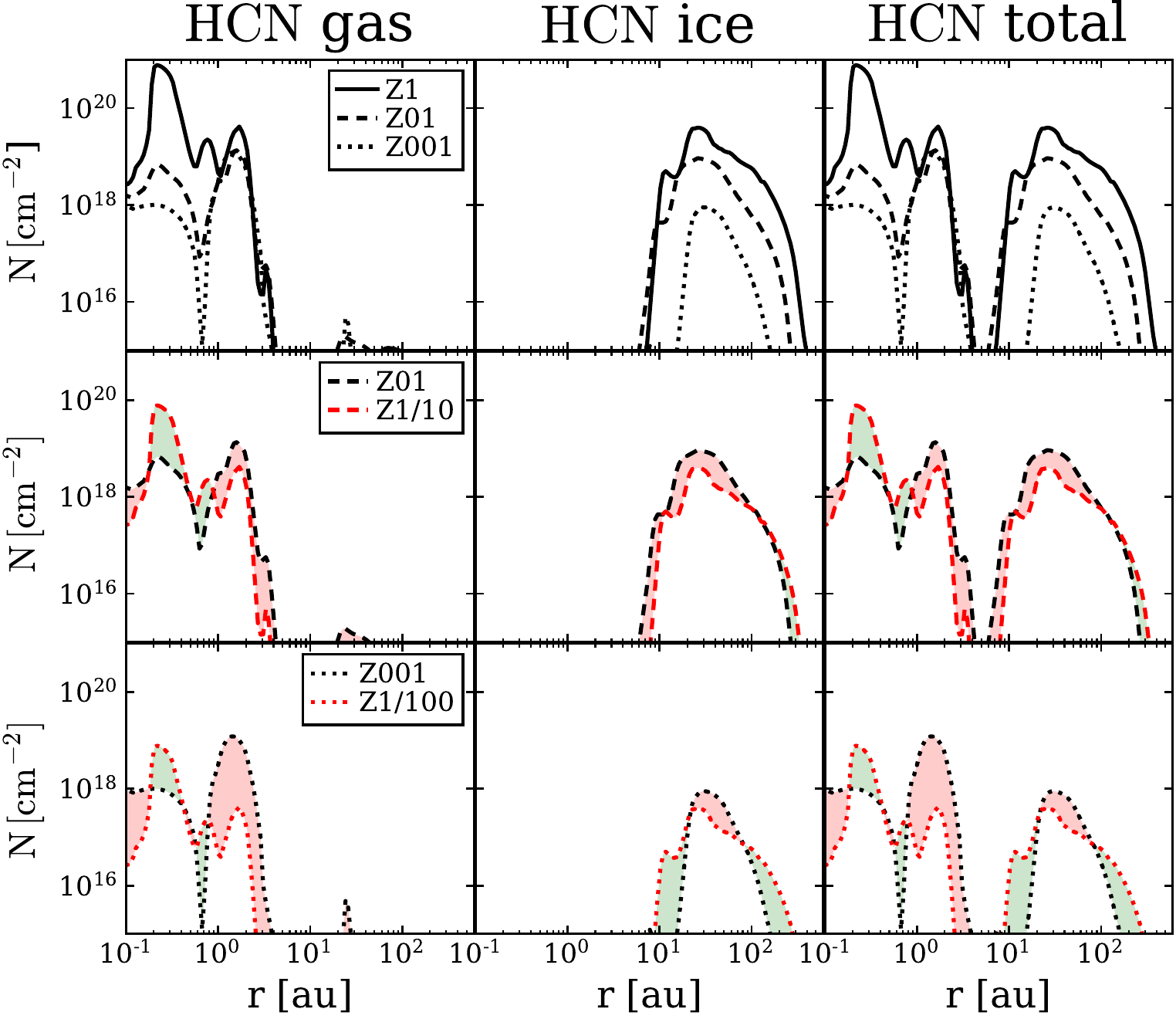}
  \end{subfigure}
  \begin{subfigure}[b]{0.49\textwidth}
    \includegraphics[width=\textwidth]{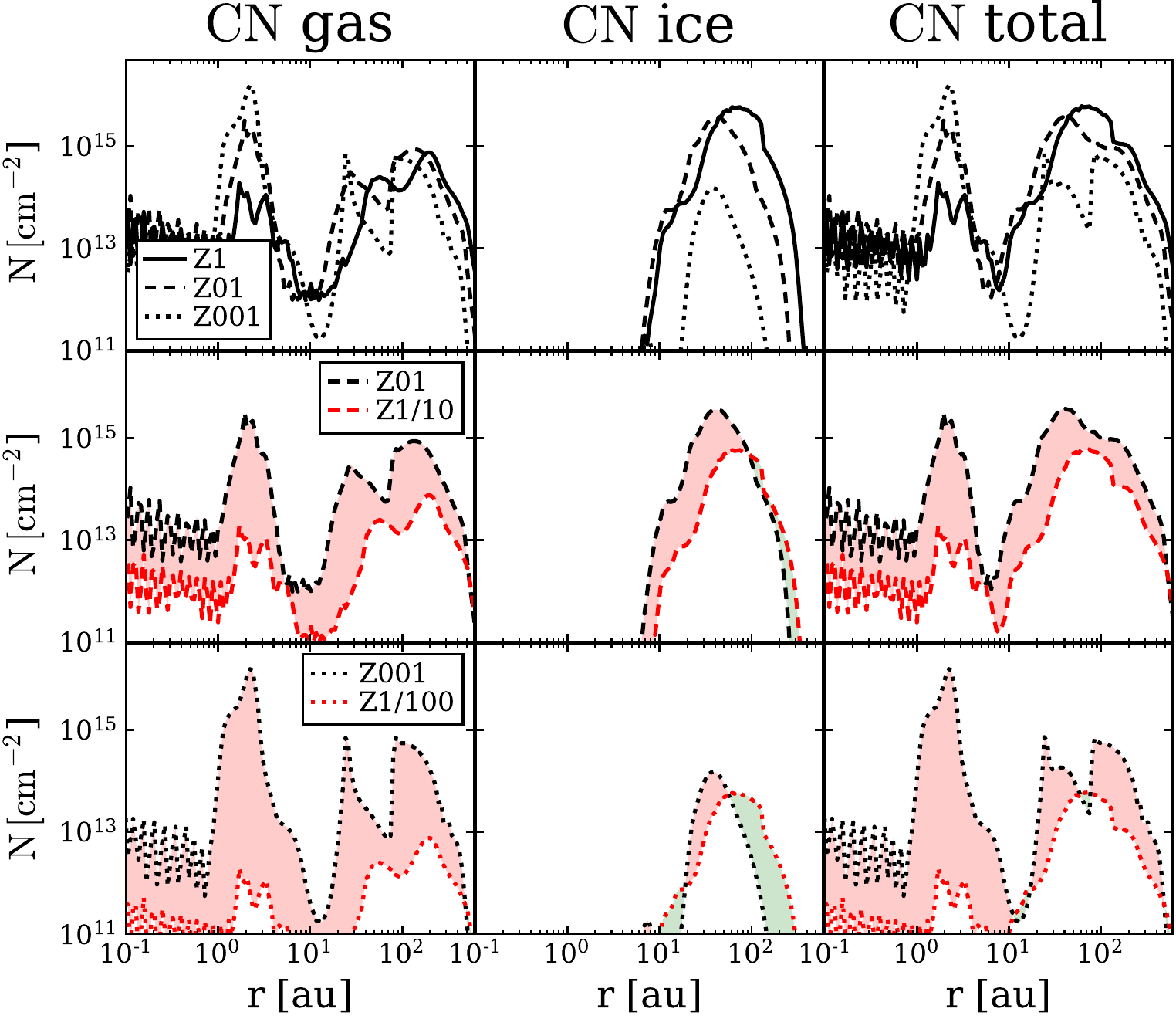}
  \end{subfigure}
  \caption{Same as Figs. \ref{fig:column_H2O}-\ref{fig:column_HCO+_N2H+} with the inclusion of the vertical column density of the gaseous (left column) and ice (middle column) phase.}
  \label{fig:all_column_dens}
\end{figure*}

In this appendix, we offer a more complete and detailed analysis of the behavior of the species in the sense that one can see if the ice or gas phase is more affected by the metallicity change. Fig. \ref{fig:all_column_dens} adds the behavior of the vertical column density for the gas and ice species as presented for the total column density in Figs. \ref{fig:column_H2O} to \ref{fig:column_HCO+_N2H+}. Table \ref{tab:react_eq} shows the specific chemical reactions {referred to} in Table \ref{table:form_destr} that take place at the positions where over-or underabundance is largest.

\begin{table}[ht!]
    \centering
     \caption{Reactions discussed in this study for different chemical species at specific points in the midplane from Table \ref{table:form_destr}.}
    \begin{adjustbox}{max width=0.49\textwidth}
    \begin{tabular}{l|c}
    \hline
    Species     & Reaction \\ 
\hline
\hline
    H$_2$O$\#$ & \vbox{\begin{equation}\label{eq:d_H2O_Z01_Z001}\mathrm{H_{2}O\# + photon \rightarrow OH\# + H\#}.\end{equation}}  \\
\hline
 CH$_4$ & \vbox{\begin{equation}\label{eq:f_CH4_Z01}
\mathrm{CH_{5}^{+} + e^{-} \rightarrow CH_{4} + H}.
\end{equation}}  
\\
    & \vbox{\begin{equation}\label{eq:f_CH4_Z001}\mathrm{CH_{5}^{+} + CO \rightarrow CH_{4} + HCO^{+}}.\end{equation}}  \\
\hline
CO & \vbox{\begin{equation}\label{eq:f_COb_Z1}\mathrm{H_{2}O + HCO^{+} \rightarrow CO + H_{3}O^{+}}\end{equation}}   \\
  & \vbox{\begin{equation}\label{eq:f_COb_Z01_Z001}
\mathrm{H + HCO \rightarrow CO + H_{2}}
\end{equation}}  \\ 
  &\vbox{\begin{equation}\label{eq:f_CO_Z1}
\mathrm{H_{2}CO^{+} + e^{-} \rightarrow CO + H + H}.
\end{equation}}  \\
  & \vbox{\begin{equation}\label{eq:f_CO_Z01}
\mathrm{H + HCO \rightarrow CO + H_{2}}.
\end{equation}}  \\
  & \vbox{\begin{equation}\label{eq:f_CO_Z001}
\mathrm{CO_2 + photon \rightarrow CO + O}.
\end{equation}}  \\
 \hline
CO$\#$ & \vbox{
\begin{equation}\label{eq:f_COi_Z1}
\mathrm{H\# + HCO\# \rightarrow CO\# + H_{2}\#}.
\end{equation}}  \\

\hline
CO$_2$ & 
\vbox{\begin{equation}\label{eq:f_CO2_Z001}
\mathrm{OH + CO \rightarrow CO_{2} + H}.
\end{equation}}  \\

\hline
CN & \vbox{
\begin{equation}\label{eq:f_CN_Z1_Z01}
\mathrm{HCNH^{+} + e^{-} \rightarrow CN + H + H}.
\end{equation}}  \\
 & \vbox{\begin{equation}\label{eq:f_CN_Z001}
\mathrm{HCN + photon \rightarrow CN + H}.
\end{equation}}  \\
\hline
HCN & \vbox{
\begin{equation}\label{eq:f_HCN_Z1_Z01}
\mathrm{HCNH^{+} + e^{-} \rightarrow HCN + H}.
\end{equation}}  \\
  & \vbox{\begin{equation}\label{eq:f_HCN_Z001}
\mathrm{CN + C_{2}H_{2} \rightarrow HCN + C_{2}H}.
\end{equation}}  \\
 \hline
 HCN$\#$ &  \vbox{\begin{equation}\label{eq:d_HCN_Z001}
\mathrm{HCN\# + photon \rightarrow CN\# + H\#}.
\end{equation}} \\
\hline
HCO$^+$ & \vbox{\begin{equation}\label{eq:f_HCO+_Z1}
\mathrm{H^{+} + H_{2}CO \rightarrow HCO^{+} + H_{2}}.
\end{equation}} \\
 & \vbox{\begin{equation}\label{eq:f_HCO+_Z01_Z001}
 \mathrm{H_{3}^{+} + CO \rightarrow HCO^{+} + H_{2}}
\end{equation}} \\
\hline
N$_2$H$^+$ & \vbox{\begin{equation}\label{eq:f_N2H+}
\mathrm{H_{3}^{+} + N_{2} \rightarrow N_{2}H^{+} + H_{2}}
\end{equation}} \\
\hline
\end{tabular}
\end{adjustbox}
    \label{tab:react_eq}
\end{table}

\section{Reducing the element abundance and the dust-to-gas ratio separately}
\label{appendix:b}
\begin{figure}[]
       \centering
    \begin{subfigure}{0.49\textwidth}
    \includegraphics[width=\textwidth]{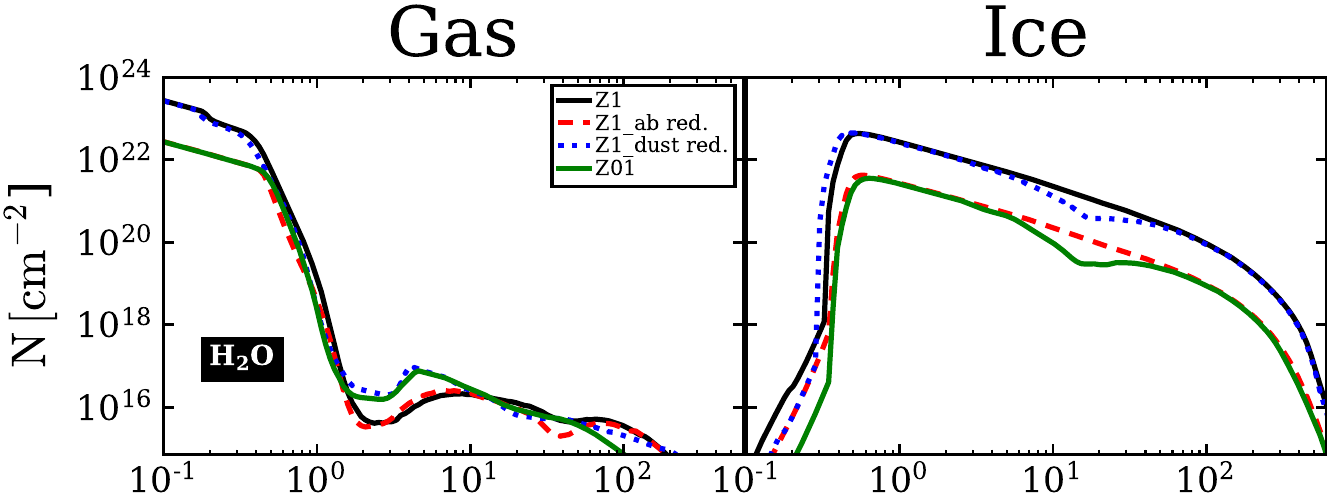}
    \end{subfigure}
    
    \begin{subfigure}{0.49\textwidth}
    \includegraphics[width=\textwidth]{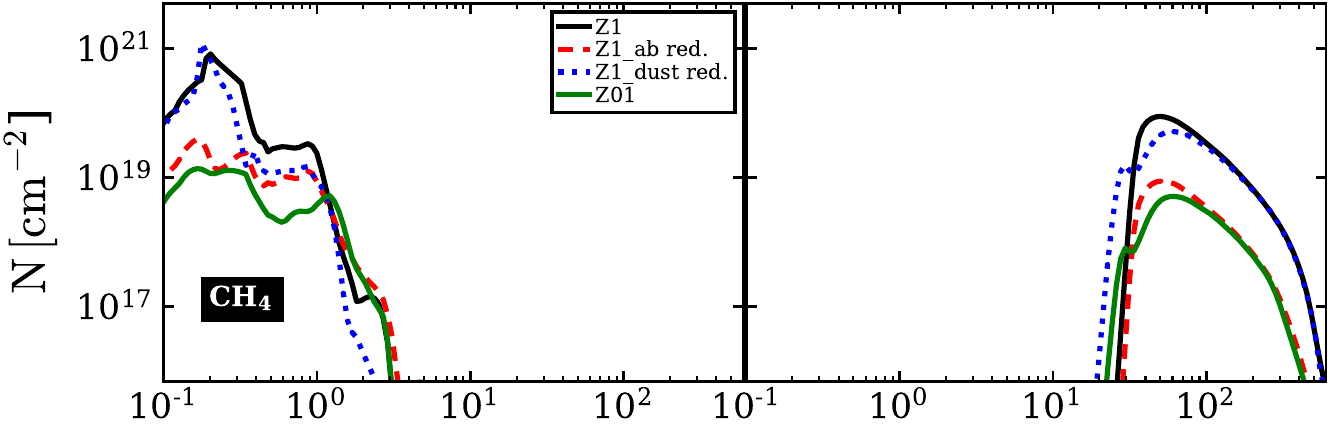}
    \end{subfigure}
    
    \begin{subfigure}{0.49\textwidth}
    \includegraphics[width=\textwidth]{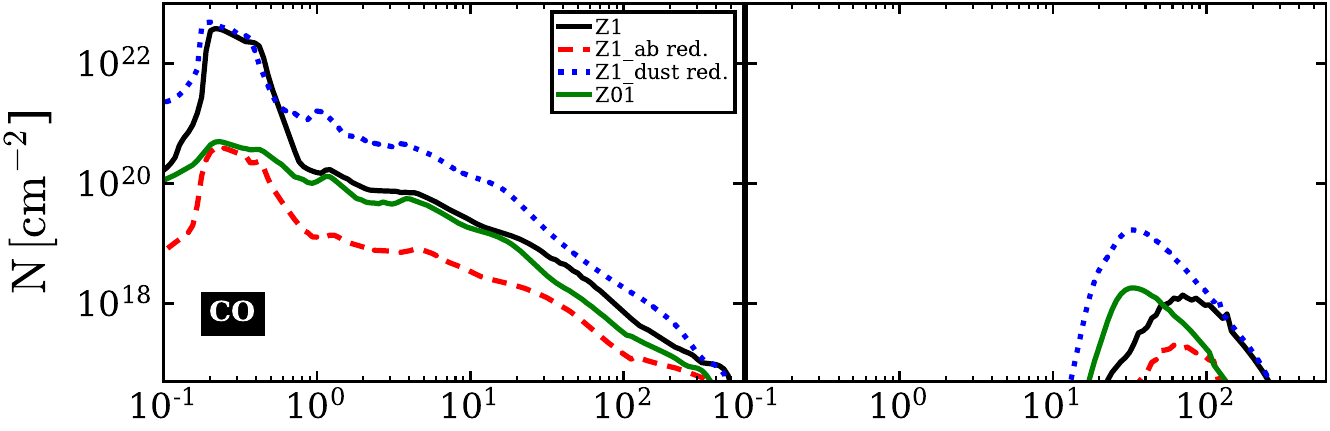}
    \end{subfigure}
    
    \begin{subfigure}{0.49\textwidth}
    \includegraphics[width=\textwidth]{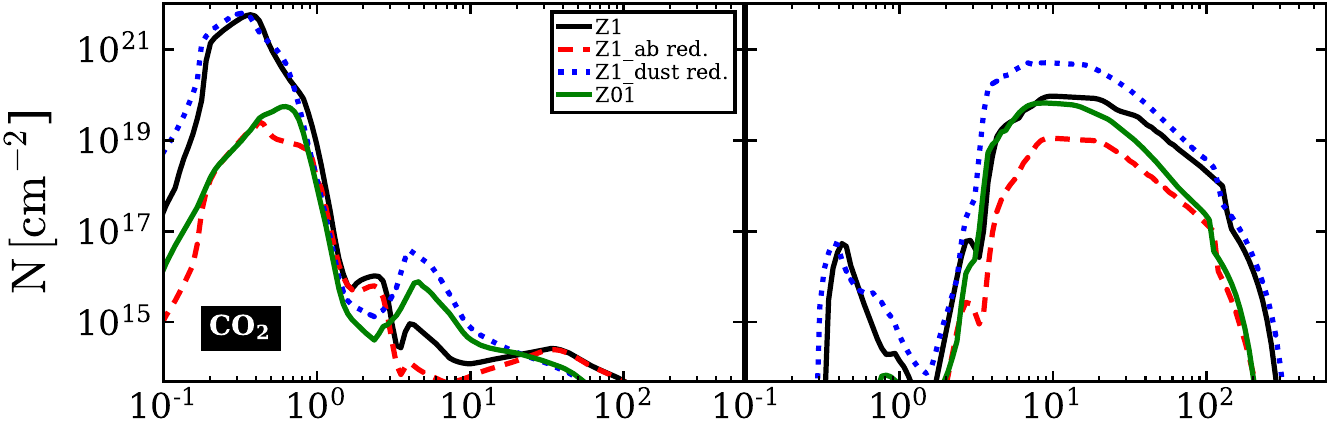}
    \end{subfigure}
    
    \begin{subfigure}{0.49\textwidth}
    \includegraphics[width=\textwidth]{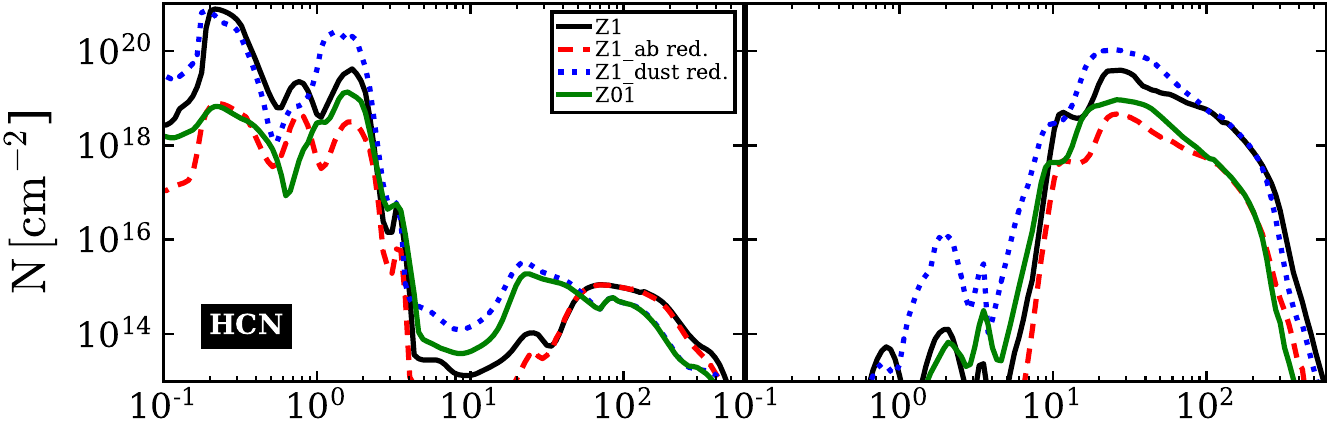}
    \end{subfigure}
    
    \begin{subfigure}{0.49\textwidth}
    \includegraphics[width=\textwidth]{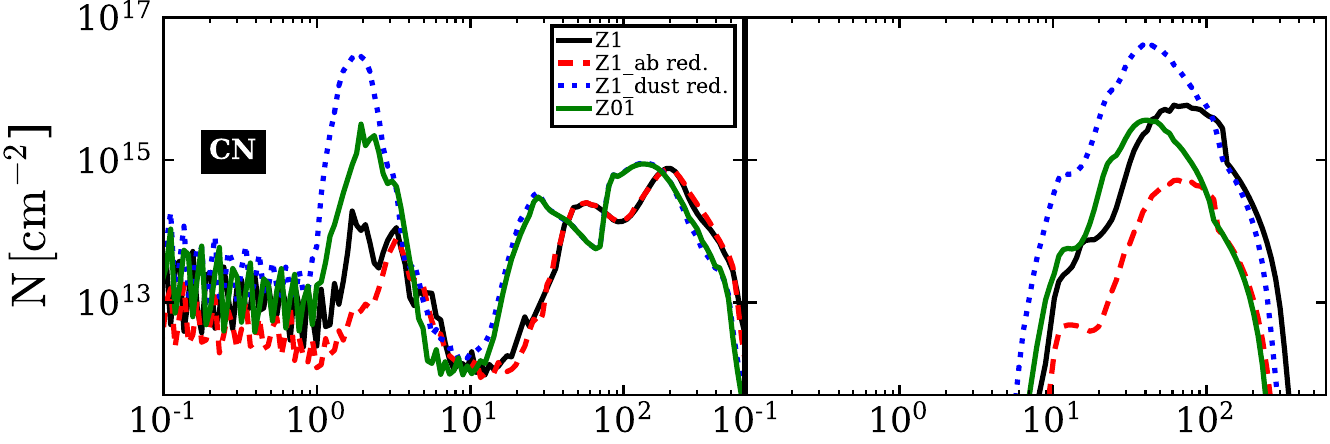}
    \end{subfigure}
    
    \begin{subfigure}{0.49\textwidth}
    \includegraphics[width=\textwidth]{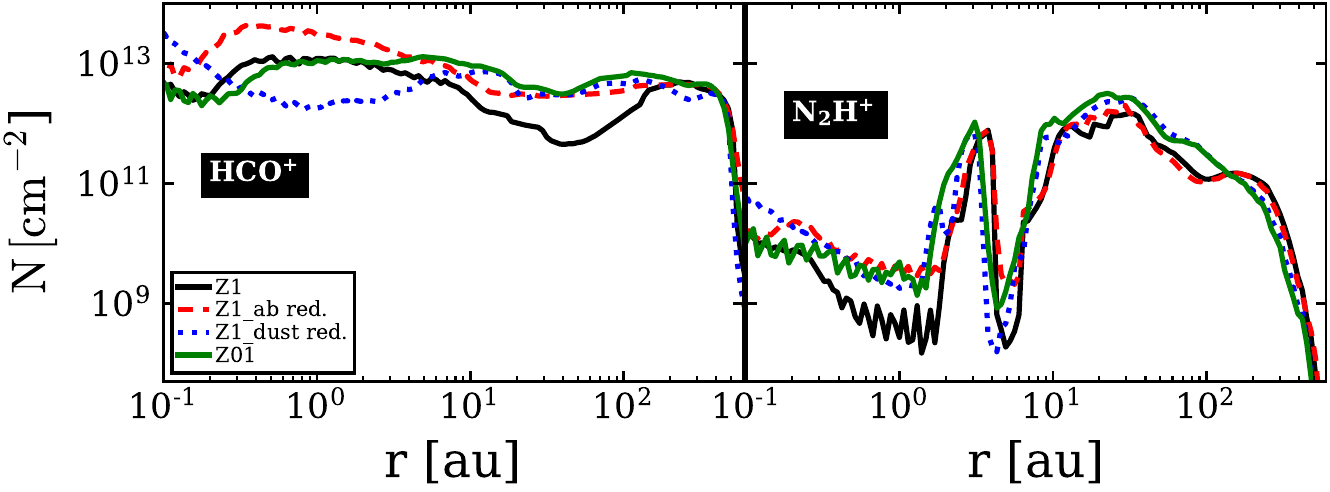}
    \end{subfigure}

    \caption{Vertical column density of chemical species. With the exception of the bottom row, the left column shows the gaseous species and the left column the frozen species. the last row shows the ions HCO$^+$ (left) and N$_2$H$^+$ (right). The reference model Z1 (solid black line), the model with only an elemental abundance reduction (dashed red line), the model with only dust-to-gas reduction (dotted blue line), and the Z01 (solid green line) show the effect of the two modifications done in this study separately.}
    \label{fig:red}
\end{figure}

To study the effect of the respective modifications we performed to simulate a lower metallicity we produced models where only the dust-to-gas ratio or the element abundances are lowered by a factor of 10. The resulting vertical column densities are shown in Fig.\ref{fig:red}. In most of the cases, the models where only the elemental abundance was reduced show a profile that resembles a simple scaled-down value from the Z1 model. {The models where only the dust-to-gas ratio was reduced exhibit different features for certain radii that match the profile of the lower metallicity model. A good example if this is water ice shown in the top right panel of Fig. \ref{fig:red}. The model where only the abundances were reduced shows a shift downwards that mostly matches the Z001 model. However, the drop in column density between 10 and 20 au in the Z001 model is absent for the abundance reduced model. The model with only dust reduction is the one that exhibits the drop at the same radius. This suggests that on the one hand, the reduction in the abundances is responsible for a general scaling down of the column density in the entire disk. On the other hand, dust reduction is responsible for the specific features (drop or rise) in certain parts of the disk.}

\end{document}